\documentclass[a4paper,11pt]{article}
\pdfoutput=1 

\usepackage{jcappub} 
\usepackage{orcidlink}
\usepackage[T1]{fontenc} 
\usepackage{xcolor}
\usepackage[normalem]{ulem}
\usepackage{bm}
 \usepackage{caption}

\newcommand{\fPBH}
{\ensuremath{f_\mathrm{PBH}}}

\newcommand{\csin}{\ensuremath{c_\mathrm{s, in }}}
\newcommand{\vin}{\ensuremath{v_\mathrm{in}}}
\newcommand{\vrel}{\ensuremath{v_\mathrm{rel}}}

\newcommand{\rhoin}{\ensuremath{\rho_\mathrm{in }}}

\def\beq{\begin{equation}}
\def\eeq{\end{equation}}
\def\bea{\begin{eqnarray}}
\def\eea{\end{eqnarray}}

\title{\boldmath Astrophysical uncertainties challenge 21-cm forecasts: A primordial black hole case study}

\author[a]{Dominic Agius\orcidlink{0009-0005-1454-5858},}
\author[b]{Rouven Essig\orcidlink{0000-0002-3066-0486},}
\author[c]{Daniele Gaggero\orcidlink{0000-0003-3534-1406},}
\author[a]{Sergio Palomares-Ruiz\orcidlink{0000-0001-9049-2288},}
\author[b]{Gregory Suczewski\orcidlink{0009-0007-7154-3589},}
\author[d]{Mauro Valli\orcidlink{0000-0002-0899-3735}}

\affiliation[a]{Instituto de Física Corpuscular (IFIC),  CSIC‐Universitat de València, \\  C/ Catedrático José Beltrán 2, E-46980 Paterna, Spain}
\affiliation[b]{C.N. Yang Institute for Theoretical Physics, Stony Brook University, NY 11794, USA}
\affiliation[c]{INFN Sezione di Pisa, Polo Fibonacci, \\ Largo B. Pontecorvo 3, 56127 Pisa, Italy}
\affiliation[d]{INFN Sezione di Roma, \\ Piazzale Aldo Moro 2, I-00185 Rome, Italy}

\emailAdd{dominic.agius@ific.uv.es}
\emailAdd{rouven.essig@stonybrook.edu}
\emailAdd{daniele.gaggero@pi.infn.it}
\emailAdd{sergiopr@ific.uv.es}
\emailAdd{gregory.suczewski@stonybrook.edu}
\emailAdd{mauro.valli@roma1.infn.it}

\abstract{
The 21-cm signal is a powerful probe of the early Universe’s thermal history and could provide a unique avenue for constraining exotic physics. Previous studies have forecasted stringent constraints on energy injections from exotic sources that heat, excite, and ionize the background gas and thereby modify the 21-cm signal. In this work, we quantify the substantial impact that astrophysical uncertainties have on the projected sensitivity to exotic energy injection. In particular, there are significant uncertainties in the minimum star-forming dark matter halo mass, the Lyman-$\alpha$ emission, and the X-ray emission, whose values characterize the fiducial astrophysical model when projecting bounds. As a case study, we investigate the energy injection of accreting primordial black holes of mass $\sim 1~M_\odot-10^3~M_\odot$, also taking into account uncertainties in the accretion model. We show that, depending on the chosen fiducial model and accretion uncertainties, the sensitivity of future 21-cm data could constrain the abundance of primordial black holes to be either slightly stronger, or significantly weaker, than current limits from the Cosmic Microwave Background. 
}

\begin{document}
\maketitle
\flushbottom
\section{Introduction}
Recent cosmological measurements have heralded a new era in physics, allowing both for precise measurements of the parameters of the $\Lambda$CDM model and for stringent constraints on new physics scenarios~\cite{Turner_2022}. Cosmology provides a particularly important avenue to test a wide range of dark matter (DM) candidates~\cite{Slatyer2024}, with interesting complementarity to other detection strategies, such as direct and indirect detection or collider searches~\cite{Cirelli2024}. 

The upcoming era of 21-cm cosmology promises to revolutionize our understanding of the early Universe, opening a new window into the dark ages ($30 \lesssim z \lesssim 200$), the cosmic dawn ($10 \lesssim z \lesssim 30$), and epoch of reionization ($ 6 \lesssim z \lesssim 10$)~\cite{Furlanetto:2006jb, Pritchard:2011xb, Furlanetto:2015apc, Villanueva-Domingo:2021vbi, Mondal:2023xjx}. By measuring the hyperfine transition of neutral hydrogen, global experiments like {\tt EDGES}~\cite{Bowman:2018yin}, {\tt SARAS}~\cite{Singh:2021mxo}, and {\tt REACH}~\cite{deLeraAcedo:2022kiu} alongside interferometers like {\tt HERA}~\cite{HERA:2016} and the ({\tt SKA})~\cite{Mellema:2012ht} aim to characterize the 21-cm global signal and power spectrum, providing unprecedented sensitivity to the thermal history of the Universe at redshifts $z \simeq 6 - 30$. Future lunar-based experiments such as {\tt LuSEE Night}~\cite{LuseeNight:2023}, {\tt FARSIDE}~\cite{Farside:2019}, {\tt Hongmeng}~\cite{Chen:2019xvd, Chen:2020lok, Shi:2022zdx, 2023ChJSS..43...43C},  {\tt DAPPER}~\cite{Burns_2017,Tauscher:2018uxi,Burns_2020}, {\tt LARAF}~\cite{Chen:2024tvn}, and {\tt FarView}~\cite{Polidan:2024kkh} may one day measure the 21-cm signal at even higher redshift (in the dark ages), where it could provide a clean cosmological probe free from astrophysical uncertainties~\cite{Jester:2009dw, Liu:2019awk}. Despite this promise, and the sentiment that ``\textit{the moon is our future}''~\cite{Silk:2025znp}, our focus here is on the two ground-based experiments, which will deliver data in the coming years. 

The new window that {\tt HERA} and {\tt SKA} will provide into the thermal and ionization properties of the intergalactic medium (IGM) makes the 21-cm signal a powerful upcoming tool for precision cosmology, and a promising avenue for constraining new physics beyond the Standard Model~\cite{Katz:2024ayw}. 
However, realizing this potential, especially in the context of exotic physics, faces a significant challenge: the dominant, yet uncertain, influence of the first astrophysical sources~\cite{Lopez-Honorez:2016sur, Decant:2024bpg}. The ultimate constraining power of the 21-cm signal is fundamentally dependent on the properties of the first stars, with some astrophysical scenarios being inherently less sensitive to exotic physics insofar as their constraining power.  Key properties of the first stars, such as their X-ray luminosity, Lyman-$\alpha$ emission, and the minimum halo mass required to form the first star-forming regions, are currently poorly constrained, and any exotic physics signature must be disentangled from this complex astrophysical signal. As such, a comprehensive assessment of these uncertainties is essential to determine the true constraining power of upcoming 21-cm experiments, and to assess whether they can probe a wider parameter space than existing observations of the Cosmic Microwave Background (CMB). 

The goal of this paper is to quantify how these astrophysical systematics, associated with the fiducial astrophysical scenario, impact forecasts for exotic energy injection. As a concrete and well-motivated case study, we investigate the effects of an exotic energy source: a monochromatic population of accreting primordial black holes (PBHs). 
PBHs are hypothetical compact objects that may have formed in the early Universe. A common formation mechanism involves the collapse of large overdensities at small scales that exceed a critical threshold~\cite{Carr:1975qj, Ivanov:1994pa, Niemeyer:1997mt, Yokoyama:1995ex, Yokoyama:1998qw, Niemeyer:1999ak, Shibata:1999zs, Musco:2004ak, Young:2014ana, Escriva:2020tak, Musco:2020jjb}. Interest in PBHs with masses above $\mathcal{O}(M_{\odot})$ has grown significantly following the gravitational wave detections by the LIGO/Virgo/KAGRA experiments~\cite{LIGOScientific:2016aoc, LIGOScientific:2016dsl, LIGOScientific:2016sjg, LIGOScientific:2017bnn, LIGOScientific:2017ycc} and speculation that a signal may have been of primordial origin~\cite{Bird:2016dcv}. Even a sub-dominant PBH population in this mass range would have important implications, potentially causing early structure formation~\cite{Inman:2019wvr, Delos:2024poq} and contributing to the formation of supermassive black holes at high redshifts~\cite{Bogdan:2023ilu, Sasaki:2018dmp}. In the mass range $(1 - 10^3) \, M_{\odot}$, PBHs would accrete baryonic matter, injecting radiation into the IGM~\cite{Carr:1981pko}. This process directly alters the IGM gas temperature and ionization fraction, leaving a distinct imprint on the 21-cm signal~\cite{Tashiro:2012qe, Hektor:2018qqw, LuisBernal:2017fmf, Mena:2019nhm,Yang:2021idt, Sun:2023acy,Sun:2025ksr}. The same physical mechanism of heating and ionizing the IGM also determines 21-cm forecasts for decaying~\cite{Shchekinov:2006eb, Furlanetto:2006wp, Valdes:2007cu, Poulin:2016anj, Clark:2018ghm, Liu:2018uzy, Mitridate:2018iag, Qin:2023kkk, Sun:2023acy, Novosyadlyj:2024bie, Zhao:2024jad, Zhao:2025ddy} or annihilating DM~\cite{Furlanetto:2006wp, Valdes:2007cu, Natarajan:2009bm, Evoli:2014pva, Lopez-Honorez:2016sur, DAmico:2018sxd, Liu:2018uzy, Qin:2023kkk, Facchinetti:2023slb, Novosyadlyj:2024bie, Zhao:2024jad, Bae:2025uqa, Natwariya:2025jlw, Sun:2025ksr}, and evaporating PBHs~\cite{Clark:2018ghm, Yang:2020egn, Halder:2021jiv, Halder:2021rbq, Mittal:2021egv, Natwariya:2021xki, Cang:2021owu, Saha:2021pqf, Mukhopadhyay:2022jqc, Yang:2022puh, Zhao:2024jad, Zhao:2025ddy,Sun:2025ksr}, and our PBH case study can be generalized to these results.\footnote{We note that depositing energy into Lyman-$\alpha$ photons without additional heating or ionization can also provide some constraining power~\cite{Agius:2025nfz}.}

The case study of PBHs is particularly relevant, since a previous study has shown that the 21-cm signal has better sensitivity than other probes to PBHs in the mass range $(1-10^3) \, M_\odot$~\cite{Mena:2019nhm}. In addition, we further investigate a second theoretical uncertainty associated with the accretion physics model. Constraints on accreting PBHs from CMB observations are already known to be strongly dependent on how the accretion is modeled~\cite{Ricotti:2007au, Ali-Haimoud:2016mbv, Horowitz:2016lib, Poulin:2017bwe, Serpico2020, Piga:2022ysp, Facchinetti:2022kbg, Agius:2024ecw}. Several effects have been explored in this context, including modifications to the accretion rate from DM minihalos around PBHs~\cite{Mack:2006gz, Serpico2020}, accretion outflows~\cite{Piga:2022ysp}, black hole spin~\cite{DeLuca:2023bcr}, and radiative feedback~\cite{Facchinetti:2022kbg, Agius:2024ecw}. In this work, we extend this investigation to the 21-cm signal by comparing the widely used Bondi-Hoyle-Lyttleton (BHL) accretion model~\cite{Hoyle1939effect, Hoyle1940accretion, Hoyle1940physical, Hoyle1941accretion, Bondi:1944rnk, Bondi:1952ni} with the state-of-the-art Park-Ricotti (PR) model, which includes the important effect of radiative feedback~\cite{Park:2010yh, Park:2011rf, Park:2012cr}. This case study allows us to explore the interplay between uncertainties from a scenario with exotic energy injection, and the much larger uncertainties from standard astrophysics.    

To summarize, the goal of this paper is twofold:  a) to characterize how astrophysical uncertainties in the  fiducial 21-cm model can alter the forecasted sensitivity to an exotic energy injection, and b) using PBHs as our case study, to provide updated 21-cm forecasts that account for systematics in the accretion model, in light of the astrophysical uncertainties. This will complement the study presented in ref.~\cite{Agius:2024ecw} and assess the impact of the PR model in the context of 21-cm cosmology.

This paper is structured as follows. In \autoref{sec:21-cmintro}, we provide an overview of 21-cm basics, highlighting the key ingredients that are altered by the presence of DM, with particular reference to accreting PBHs. In \autoref{sec: methodology}, we describe our methodology for providing 21-cm sensitivity forecasts, with specific emphasis on the impact of astrophysical uncertainties entering the fiducial model. We go on to quantify our results in \autoref{sec:Results}, where we present forecasts for the upcoming {\tt SKA} interferometer. Finally, we conclude and discuss the  outlook in \autoref{sec:conclusions}. An appendix provides a comparison to previous work, and lists astrophysical parameters that are held fixed in our study.

\section{The 21-cm signal and exotic energy injection}
\label{sec:21-cmintro}

In this section, we introduce the basics of 21-cm phenomenology, and describe the standard astrophysics scenario governing the signal and its uncertainties.  We then discuss the impact of PBH accretion on the 21-cm observables: the brightness temperature and the power spectrum.

\subsection{Overview of 21-cm basics}
\label{sec:overview_of_21cm_pheno}

The 21-cm signal is often parametrized by the so-called ``spin temperature'' ($T_{\rm S}$), defined by the fraction of neutral hydrogen in the triplet state compared to the singlet state,
\begin{align}\label{eqn:spintemp}
    \frac{n_1}{n_0} = 3 \, e^{-\frac{T_0}{T_{\rm S}}}\ ,
\end{align}
where $n_1$ and $n_0$ are the number densities of neutral hydrogen in the triplet (excited) and singlet (ground) states, respectively. The factor of $3$ comes from the degeneracy of the triplet state compared to the singlet, and $k_{\rm B} \, T_0 = h \, \nu_0$ is the energy of the 21-cm photons. By definition, higher spin temperatures correspond to a higher proportion of excited states (with $n_1 \approx 3 \, n_0$ for $T_{\rm S} \gg T_0$), while lower temperatures correspond to a smaller proportion of excited states (with $n_1 \ll n_0$ for $T_{\rm S} \ll T_0$). The ratio of these densities (and therefore the spin temperature) depends on the interactions that can induce hyperfine transitions. There are three dominant ingredients that determine the spin temperature, which can be understood as the weighted average of the temperatures associated with each process: 
\begin{align}\label{eqn:spintempinteractions}
    T_{\rm S}^{-1}  = \frac{T_{\text{CMB}}^{-1} + x_{\alpha} \,  T_{\rm c}^{-1} + x_c \, T_k^{-1}}{1+x_{\alpha} + x_c} ~.
\end{align}
Each term corresponds to a key interaction. The first corresponds to the stimulated emission and absorption caused by a background radiation field (which we take to be the CMB), a process characterized by the CMB temperature, $T_{\rm CMB}$. The second term accounts for the coupling between the spin temperature and the color temperature, $T_c$, mediated by Lyman-{$\alpha$} photons through the Wouthuysen–Field (WF) effect~\cite{Wouthuysen:1952, Field:1958}. The color temperature of the Lyman-$\alpha$ radiation field can be understood as the slope of the spectrum near the Lyman-$\alpha$ frequency, and when the radiation spectrum reaches a steady state, this color temperature is equal to the gas kinetic temperature field at the Lyman-$\alpha$ frequency~\cite{Chen:2003gc}. The strength of this interaction is determined by the coupling coefficient $x_\alpha$, which is proportional to the flux of Lyman-$\alpha$ photons. Physically, this coupling arises because Lyman-$\alpha$ photons excite electrons from their original hyperfine ground state (with number densities $n_0$ or $n_1$). Upon relaxation, the electron can decay to either of the ground state levels, therefore altering the original population between the singlet and triplet states. The third interaction in eq.~\eqref{eqn:spintempinteractions} is due to hydrogen--hydrogen collisions, whose strength is determined by the collisional coupling, $x_c$. This final term becomes negligible after $z \simeq 30$, as the expansion of the Universe dilutes the density of hydrogen atoms. 

Upcoming 21-cm experiments, such as the {\tt SKA} telescope, are designed to measure the frequencies probing the cosmic dawn and the epoch of reionization (corresponding to $z \simeq 6 - 30$). At these times, the collisional coupling is negligible ($x_c \ll 1$). Lastly, for all scenarios of interest, the color temperature is approximately equal (within 5\%~\cite{Hirata:2005mz}) to the kinetic temperature of the baryons, $T_c \simeq T_k$, so we use them interchangeably in this work.

In the Rayleigh-Jeans limit ($h \, \nu \ll k_{\rm B} \, T_b(\nu)$), the specific intensity of a blackbody is proportional to the brightness temperature, $T_b$. The primary 21-cm observable is the differential brightness temperature, $\delta T_b$, which measures the intensity of the 21-cm line relative to the CMB background (or in general, the background radiation), and depends on both the spin temperature and the optical depth of the 21-cm line, $\tau_{21}$, 
\begin{equation}
\label{eq:T21}
    \delta T_b = \frac{T_{\rm S} - T_{{\rm CMB}}}{1 + z} \, \left(1 - e^{-\tau_{21}}\right) ~.
\end{equation}

When the spin temperature is greater (less) than the CMB temperature, the signal is in emission (absorption). There is no signal when the spin temperature is in equilibrium with the CMB. For the frequencies (redshifts) of interest, the optical depth is small ($\tau_{21} \ll 1$) and eq.~\eqref{eq:T21} can be approximated as~\cite{Madau:1996cs, Furlanetto:2006jb, Barkana:2016nyr}
\begin{equation}
\label{eq:T21_simplified}
\delta T_b \simeq 27 \, x_{\rm HI} \, (1 + \delta_b) \left(1 - \frac{T_{\rm CMB}}{T_{\rm S}}\right) \left(\frac{1}{1 + H^{-1} \, \partial v_r / \partial r}\right)  \left(\frac{1 + z}{10}\right)^{1/2}  \, \text{mK} ~,
\end{equation}
where $H$ is the Hubble constant, $x_{\rm HI}$ is the fraction of neutral hydrogen, $\delta_b$ is the fractional baryon density perturbation, and $\partial v_r/\partial r$ is the line-of-sight gradient of the neutral hydrogen velocity. Eq.~\eqref{eq:T21_simplified} captures the signal at a specific spatial point, and its sky average, $\overline{\delta T_b}$, is what is referred to as the global 21-cm signal. 

Since the properties of the IGM are not completely homogeneous, any fluctuations in the 21-cm signal in space and time, can be characterized using the statistical properties of the signal. Particularly useful is the 21-cm power spectrum, $P_{21}(\mathbf{k}, z)$, which quantifies the variance of the brightness temperature as a function of spatial scale (or wavenumber $\mathbf{k}$), at a given redshift. It is defined via the two-point correlation of the signal in Fourier space, 
\begin{equation}
	\langle \delta T_{21}(\mathbf{k}, z) \delta T_{21}^*(\mathbf{k}', z) \rangle = (2\pi)^3 \delta_D(\mathbf{k} - \mathbf{k}') P_{21}(\mathbf{k}, z) ~,
\end{equation}
where $\delta_D$ is the Dirac delta function, and $\delta T_{21}(\mathbf{k}, z)$ corresponds to the Fourier transform of $\delta T_{b}(\mathbf{x}, z) - \overline{\delta T_{b}}(\mathbf{x}, z)$. Throughout this work we use the dimensional reduced power spectrum,
\begin{equation}
	\Delta_{21}^2(\mathbf{k}, z) = \frac{k^3 P_{21}(\mathbf{k}, z)}{2\pi^2} \, [\mathrm{mK}^2] ~.
\end{equation}

\subsection{Shape of the signal from the dark ages to the cosmic dawn: the crucial role of astrophysical parameters}

The 21-cm signal is dictated by the changing thermal and ionization properties of the Universe around the epoch of reionization, caused by the formation of the first astrophysical sources and their emission. 
These first stars determine the 21-cm signal through three key physical effects: i) through their impact on the kinetic temperature of the gas, via X-ray heating, ii) via the strength of WF coupling $x_\alpha$, induced by Lyman-$\alpha$ pumping, and iii) via changes in the fraction of neutral hydrogen $x_{\rm HI}$, shown explicitly in eqs. \eqref{eqn:spintempinteractions} and \eqref{eq:T21_simplified}. 

The epoch of interest for the present study is the redshift range $z\simeq 10 - 25$, and begins in the last portion of the dark ages, when no stars have formed yet, and the IGM is neutral. At $z \sim 25$, the Universe has expanded sufficiently such that the collisional coupling between the spin temperature and the kinetic temperature of the gas is no longer effective, and $\delta T_b=0$ is expected because $T_{\rm S} = T_{\rm CMB}$. When the first astrophysical sources switch on, they emit both Lyman-$\alpha$ and X-ray photons. Initially, Lyman-$\alpha$ emission is expected to couple the spin temperature to the gas temperature, generating an absorption signal, since the gas temperature is below that of the CMB. Then, as sources begin to emit more strongly in X-rays, the gas temperature increases, damping the absorption signal. Eventually, $T_k$ (which is $\sim T_S$) increases above $T_{\rm CMB}$, resulting in an emission signal, which is damped as the fraction of neutral hydrogen vanishes, and the Universe becomes completely ionized.  

The sources responsible for the emission of Lyman-$\alpha$ photons (i.e., the first stars and galaxies) are discrete and clustered, and hence the Lyman-$\alpha$ background responsible for the WF coupling builds up inhomogeneously in space. 
Later, the X-ray flux originating from the first X-ray binaries is also expected to be patchy. Therefore, the temperature contrast between the heated and cold regions leads to spatial variations in $T_k$ and thus in the brightness temperature, and these spatial variations can be captured with the power spectrum. 

Both the global averaged absorption/emission signal and the power spectrum that characterizes the spatial variations are shaped by the details of the star formation process, and eventually depend on some key parameters that are crucial in the current study. In particular, three quantities are certainly relevant: {\it 1)} the number of Lyman-$\alpha$ photons emitted per baryon, $N_{\alpha}$; {\it 2)} the X-ray luminosity, $L_X$, normalized to the star-formation rate, $\dot{M}_*$; {\it 3)} the minimum mass of star-forming DM halos, $M_{\rm turn}$.

The first quantity is responsible for the WF coupling described above, and therefore impacts the coupling of the spin temperature and the kinetic temperature of the gas. The second parameter controls the heating of the cold gas. Both parameters depend on the Star Formation Rate Density (SFRD), which quantifies how much stellar mass is formed per unit volume and per unit time at each redshift. It plays a central role in the 21-cm modeling and determines the normalization of the radiation fields mentioned above that shape the signal and its fluctuations. In \autoref{sec: methodology}, we will address how the SFRD  evolves over cosmic history and how it is modeled with the numerical tools that we use.
The third quantity, instead, sets the threshold mass for DM halos to cool and form stars: larger values of $M_{\rm turn}$ restrict star formation to rarer massive halos, delaying the onset of radiation backgrounds. In \autoref{sec: methodology}, we will also provide more details on  $M_{\rm turn}$.

These three astrophysical parameters are uncertain. As a consequence, both the shape of the absorption signal and the power spectrum that future experiments aim to probe are uncertain, and our current knowledge prevents us from characterizing it accurately. Therefore, the sensitivity of these upcoming data to new physics scenarios depends strongly on the values chosen for the astrophysical parameters, which we will illustrate in this paper by forecasting the sensitivity to PBHs.

\subsection{The case study of Primordial Black Holes: impact on the signal}

Additional physics phenomena beyond the Standard Model can alter the 21-cm signal described in the previous subsection. In particular, we consider a hypothetical population of massive PBHs that accrete baryonic matter. These objects can affect the 21-cm brightness temperature by altering the spin temperature and the 21-cm optical depth. 
Through the process of accretion, PBHs would inject high-energy photons, which can deposit energy in the IGM, heating, exciting, and ionizing the gas. Therefore, PBHs would alter the 21-cm optical depth by reducing the neutral hydrogen fraction and also alter the spin temperature by increasing the kinetic temperature from heating. 

The accretion process by PBHs would also produce Lyman-$\alpha$ photons, which would directly impact the efficiency of the WF-effect through the dimensionless coupling $x_{\alpha}$, by increasing the flux of Lyman-$\alpha$ photons, $J_{\alpha}$, 
\begin{equation}
\label{eq:lyalpha_coupling}
    x_{\alpha} = 0.416 \times \frac{16 \pi^2 e^2}{27 \, m_e \, c \, A_{10}}\frac{h \, \nu_o}{k_{\rm B} \, T_k} J_{\alpha} ~,
\end{equation}
where $A_{10}$ is the Einstein coefficient for spontaneous emission, and $e$ and $m_e$ are the charge and mass of the electron.
In particular, the Lyman-$\alpha$ flux can be split into two contributions, 
\begin{equation}
\label{eq:Jalpha}
    J_{\alpha} = J_{\alpha}^{\rm astro} + J_{\alpha}^{\rm PBH} ~,
\end{equation}
where $J_{\alpha}^{\rm astro} \propto N_{\alpha}$ is the astrophysical contribution from the first stars~\cite{Munoz:2023kkg},\footnote{Note that \texttt{Zeus21} does not include the contribution from X-rays excitations of neutral hydrogen~\cite{Chen:2006zr}, which is subdominant for the astrophysical parameters and redshifts we consider~\cite{Lopez-Honorez:2020lno}.} and $J_{\alpha}^{\rm PBH}$ denotes the contribution from accreting primordial black holes. 

Additionally, injection of energy from accreting PBHs also alters the free electron fraction, $x_e$, and the baryon temperature, $T_k$, of the IGM. To account for this, we use the modified evolution equations described in ref.~\cite{Stocker:2018avm} and given by 
\begin{align}
\label{eq:evolution equations}
\nonumber
\frac{dx_e(z)}{dz} &= \frac{1}{(1+z)\, H(z)}\left(R(z) - I(z) - I_X(z)\right) ~, \\
\frac{dT_k}{dz} &= \frac{1}{1+z}\left[2 \, T_k + \gamma\, \left(T_k - T_{\rm CMB}\right)\right] + K_h ~, 
\end{align} 
where $R(z)$ and $I(z)$ are the standard recombination and ionization rates, and $\gamma$ is the dimensionless opacity of the gas. $I_X(z)$ and $K_h$ denote the additional respective ionization and heating rates from exotic injections, which are proportional to the deposited energy due to exotic sources. We will define the deposited energy in \autoref{sssec:Deposition}. We refer the reader to refs.~\cite{Poulin:2015pna, Poulin:2016anj, Stocker:2018avm} for further details on these coefficients. 

\begin{figure}[t]
    \centering
    \includegraphics[width=1\linewidth]{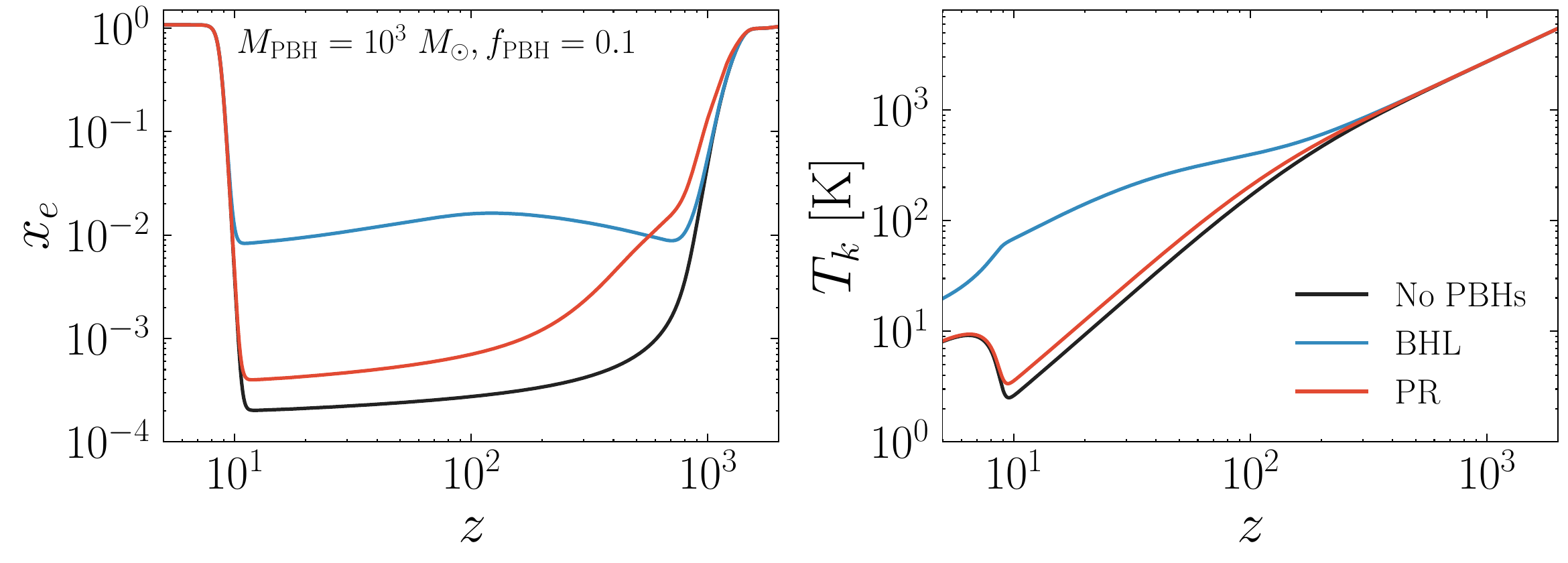}
    \caption{Impact of accreting PBHs on the free-electron fraction (left panel), $x_e$, and on the gas temperature (right panel), $T_k$. Results are shown for two different accretion models, the Park-Ricotti and Bondi-Hoyle-Lyttleton models. For both panels we assume a PBH mass of $M_{\rm PBH} = 10^3 \ M_{\odot}$ and a fractional PBH abundance of $f_{\rm PBH} = 0.1$.}
    \label{fig:xe_Tk}
\end{figure}

We show in \autoref{fig:xe_Tk} how the quantities in eq.~\eqref{eq:evolution equations} are affected by the presence of accreting PBHs.  We show the impact for the two different accretion models: the Bondi-Hoyle-Lyttleton (BHL) model described in \autoref{sssec:BHL_model}; and the Park-Ricotti (PR) model, described in \autoref{sssec:PR_model}. We set the PBH mass to $M_{\rm PBH }=10^3~M_\odot$, and the fraction of dark matter contained in PBHs, $f_{\rm PBH}= \Omega_{\rm PBH} / \Omega_{\rm DM}$, to $ 0.1$, in this figure. As can be clearly seen, the BHL accretion model has a much larger impact than the PR accretion model on $x_e$ and $T_k$, especially at late times, changing the free electron fraction and the baryon temperature by more than one order of magnitude. Hence, the modeling of accretion strongly influences the ionization fraction and gas temperature, and thus the predicted 21-cm signal. The key quantity governing the impact of the accreting PBHs on the 21-cm signal is the {\it accretion rate}, $\dot{M} \equiv \frac{{\rm d}M}{ {\rm d} t}$, which quantifies the rate at which baryonic matter is captured by a black hole. This quantity is typically a function of the PBH mass and PBH speed, and also depends on the properties of the medium. It is used to calculate the amount of energy that is injected into the IGM, and thus the impact on the observables.

In the next sections we will compare the two accretion models, BHL and PR. We remark here that we do not study the impact of dark matter mini-halos on the accretion rate in this work.\footnote{We refer the interested reader to refs.~\cite{Serpico2020, Agius:2024ecw} for an analysis with the CMB, and ref.~\cite{Sun:2025ksr} for a 21-cm analysis.} Our conclusions---that astrophysical uncertainties associated with the choice of fiducial model can significantly impact the derived constraints---are expected to remain unchanged in this case.

\subsubsection{The Bondi--Hoyle--Lyttleton model}
\label{sssec:BHL_model}

The Bondi-Hoyle-Lyttleton (BHL) model was developed as an interpolation between two limiting cases: the ballistic regime, which describes accretion onto a point mass moving at constant velocity through a uniform-density medium under purely gravitational influence~\cite{Hoyle1939effect, Hoyle1940accretion, Hoyle1940physical, Hoyle1941accretion}, and spherical accretion, which considers a stationary, spherically symmetric object accreting matter, while accounting for both pressure and gravity~\cite{Bondi:1944rnk, Bondi:1952ni}. With these simplifying assumptions, the resulting interpolation formula can provide an order-of-magnitude estimate of the accretion rate~\cite{Bondi:1952ni}, without capturing complicated hydrodynamical effects such as radiation feedback. Despite these simplifications, it is commonly used to derive bounds and forecasts in the cosmological setting. We provide a summary of the model below.

Under the BHL accretion model~\cite{Hoyle1939effect, Hoyle1940accretion, Hoyle1940physical, Hoyle1941accretion, Bondi:1944rnk, Bondi:1952ni}, the mass accretion rate onto an isolated compact object with mass $M$ and with a velocity \ensuremath{v_\mathrm{rel}} relative to the ambient gas, is given by
\begin{equation}
\label{eq:bondi}
\dot{M}_\mathrm{BHL} = 4 \pi \lambda \frac {(GM)^2 \rho_{\rm b}} {(\ensuremath{v_\mathrm{rel}}^2 + c_{\mathrm s}^2)^{3/2}} ~,
\end{equation}
where $\rho_{\rm b}$ and $c_{\mathrm s}$ denote the density and sound speed of the surrounding medium, respectively, and $\lambda$ is a dimensionless parameter determining the normalization of the accretion rate. Bondi originally computed the maximal value of $\lambda$ as a function of the equation of state of the gas under simplifying assumptions, finding $\lambda \sim \mathcal{O}(1)$ \cite{Bondi:1952ni}. Later studies corrected this value of $\lambda$, specifically computing it in cosmological scenarios, where the effects of cosmic expansion and DM overdensities were included~\cite{Ricotti:2007jk, Ricotti:2007au, Ali-Haimoud:2016mbv}. These studies generally assume spherical symmetry for the accretion flow and identify bremsstrahlung (free-free) emission near the Schwarzschild radius as the dominant cooling mechanism, while also accounting for the Hubble flow.

However, the calculation of $\lambda$ from first principles involves simplifying complicated accretion processes, and an alternative phenomenological approach for quantifying $\lambda$ exists in the literature. This approach involves setting the value of $\lambda$ to be consistent with astrophysical observations. Indeed, values of $\lambda \simeq 10^{-2}-10^{-3}$ are frequently adopted to align theoretical predictions with observations, such as the absence of a large population of isolated neutron stars~\cite{Perna:2003ck} or stellar-mass black holes~\cite{Fender:2013ei} in the local Universe. This correction is also motivated by studies of nearby active galactic nuclei~\cite{Pellegrini:2005pi} and the accretion environment around the central supermassive black hole of the Milky Way~\cite{Wang2013}. The suppression factor $\lambda$ is intended to account for a variety of non-gravitational effects (including pressure forces, viscosity, and radiation feedback) that can act to reduce the accretion rate. In what follows, we adopt $\lambda = 0.01$ as a representative value for disk accretion scenarios.

\subsubsection{The Park--Ricotti model}
\label{sssec:PR_model}

The role of {radiative feedback} in the accretion process was investigated by Park and Ricotti through hydrodynamical simulations of accretion from a homogeneous medium onto a compact object~\cite{Park:2010yh, Park:2011rf, Park:2012cr}. These simulations revealed the formation of an ionization front (sometimes preceded by a shock wave depending on the velocity regime) alongside a sharp suppression of the accretion rate at low speed. The authors developed a simplified analytical prescription, the PR model, which successfully reproduces the accretion rates observed in the simulations, and which we briefly outline in the following.

The high-energy radiation generated during accretion ionizes the surrounding gas, altering its temperature, density, and flow velocity relative to the black hole. These changes, in turn, impact the accretion dynamics. The PR model finds a semi-analytic fit to their simulation that agrees with the functional form of the BHL accretion formula, eq.~\eqref{eq:bondi}, but in this case accreting from the ionized gas. Hence, the PR accretion rate is
\begin{equation}
\label{eq:ricotti}
\dot{M}_\mathrm{PR} = 4 \pi \frac{(GM)^2 \rhoin}{(\vin^2 + \csin^2)^{3/2}} ~,
\end{equation}
where $\rhoin$, $\vin$, and $\csin$ are the density, relative velocity, and sound speed of the ionized gas, respectively. The sound speed $\csin$ is treated as a constant, parameterizing the temperature of the ionized gas. The values of $\rhoin$ and $\vin$ are determined from the ambient gas properties $\rho_{\rm b}$ and $v_{\rm rel}$ by enforcing one-dimensional mass conservation and force equilibrium across the ionization front, as detailed in ref.~\cite{Agius:2024ecw} and references therein.

At high relative velocities, $\vrel \gtrsim  2 \, \csin$, the PR and BHL models converge, since the ram pressure dominates over the ionization pressure. However, the BHL rate is typically reduced by the normalization parameter $\lambda$, resulting in an accretion rate lower than that of the PR model. At lower velocities, $\vrel \lesssim 2 \, \csin $, the formation of a shock front leads to significant deviations: whereas the BHL rate increases, the PR rate declines and becomes strongly suppressed relative to BHL. It is this regime that is most relevant in the cosmological setting, and has the largest phenomenological impact.

The value of $\csin$, or equivalently the gas temperature within the ionized region, sets the characteristic velocity and hence the peak of the PR accretion rate. While $\csin$ is determined, in principle, by the balance between radiative heating and cooling, the PR model treats it as a free parameter. A commonly adopted benchmark, that we use in this work is $\csin = 23~\mathrm{km/s}$, corresponding to a temperature of $4 \times 10^4$~K~\cite{Park:2012cr}. A more detailed analysis of this parameter and its implications for cosmological constraints is provided in ref.~\cite{Agius:2024ecw}.

Once the accretion rate has been specified, it can be applied to a specific cosmological setup characterized by a redshift-evolving density and relative speed between PBHs and baryons.  It can subsequently be used to compute the PBHs luminosity, which plays a central role in determining the rate of energy injection. In the following section, we describe in more detail our modeling of the cosmological medium.

\subsubsection{Environmental properties}
\label{sssec:Environment_properties}

In this work, we adopt both the BHL and PR models to describe the accretion of a gas with a homogeneous density distribution onto PBHs.\footnote{Under this assumption, the energy injection is modeled in this work as homogeneous. The effects of inhomogeneous energy injection have been shown to have a small impact in the case of decaying dark matter~\cite{Sun:2023acy}. See also ref.~\cite{Sun:2025ksr} for a treatment of inhomogeneous energy injection in the case of evaporating and accreting PBHs.} As mentioned above, the Universe is starting to develop virialized halos in the epoch of interest in this study, and some of these halos are forming stars, giving rise to the cosmic dawn. However, prior to the end of the cosmic dawn and the onset of reionization, the vast majority of PBHs are still expected to be isolated in the cosmological medium rather than in virialized halos (see for instance Figure~14 in ref.~\cite{Jangra:2024sif} and ref.~\cite{Ricotti:2007au}).\footnote{See, however, ref.~\cite{Sun:2025ksr} for an analysis that also considers the contribution from some PBHs that are contained in virialized halos.} Hence, similarly to our previous work~\cite{Agius:2024ecw}, we take the cosmological background density of baryons to evolve with redshift as 
\begin{equation}
    \rho_{\mathrm{b}} = 200 \, m_{\rm p} \left(\frac{  1 + z}{ 1 + z_{\rm rec}} \right)^3 \,\, {\rm g \, cm^{-3}} ~,
\end{equation} 
where $m_{\rm p}$ is the proton mass in grams and $z_{\rm rec} \simeq 1100$ is the redshift at recombination. The baryon sound speed is given by 
\begin{equation}
    c_{\mathrm{s}} = \sqrt{\frac{\gamma \, (1 + x_{\mathrm{e}}) \, T_k}{ m_{\rm p}}} \,\, {\rm km\,s^{-1}} ~,
\end{equation}
where $T_k$ is the gas temperature, $x_{\mathrm{e}}$ is the ionization fraction, and $\gamma$ is the adiabatic index. 

We also adopt the linear cosmological relative velocity between baryons and DM as the relative velocity between PBHs and the background gas, $v_{\rm rel}$, with the RMS value given by~\cite{Tseliakhovich:2010bj, Dvorkin:2013cea}
\begin{equation}\label{eqn:v_rms}
    \sqrt{\langle v_{\rm rel}^2 \rangle} = \min \left[ 1, (1+z)/1000 \right] \times 30 \,\, {\rm km\,s^{-1}} ~.
\end{equation}
The PBH velocity distribution is assumed to follow a Maxwell--Boltzmann profile, where we relate the root-mean-square velocity in eq.~\eqref{eqn:v_rms} to the temperature, to uniquely define the distribution.\footnote{As pointed out in ref.~\cite{Mena:2019nhm}, previous works, such as refs.~\cite{Ali-Haimoud:2016mbv, Poulin:2017bwe} have (incorrectly) used a proxy for this averaging, $v_{\rm eff} $, corresponding to a limiting case where the accretion luminosity has the proportionality $L_{\rm acc} \propto \dot{M}^2$. We further discuss the effect of varying the velocity distribution in \autoref{appendix:comparison_to_mena}.}

We remark here that any contribution to the power spectrum in the form of Poisson noise~\cite{Afshordi:2003zb}, due to the discreteness of PBHs is not accounted for here. This has been studied previously (e.g., in refs.~\cite{Mena:2019nhm, Cole:2019zhu}), where it was found that for the scales probed by ${\tt SKA}$ and {\tt HERA} with wavenumbers $k \simeq 0.1 - 1$ Mpc$^{-1}$, the effect is subdominant. Furthermore, we have also not modeled the potential impact of PBHs on early structure formation \cite{Delos:2024poq, Inman:2019wvr} or on the formation of the first stars themselves~\cite{Koulen:2025xjq}, which remain interesting avenues for future investigation.

\section{Methodology}
\label{sec: methodology}

Our goal is to provide 21-cm sensitivity forecasts for PBHs, highlighting the impact of choice of fiducial model on the forecasts. To this end, we implement the exotic physics of accreting PBHs into the {\tt Zeus21} code~\cite{Munoz:2023kkg, Cruz:2024fsv}, modeled with both the BHL and PR models. In \autoref{ssec:Zeus21}, we first describe the numerical modeling of the first stars, as accounted for in {\tt Zeus21}. We go on to describe our modeling of injected and deposited energy from accreting matter around PBHs in \autoref{sec:Einjection_and_deposition}. In \autoref{ssec:Null_hypothesis}, we introduce three plausible astrophysical scenarios as our fiducial models for mock data generation. In \autoref{ssec:statistic}, we outline our statistical analysis pipeline, including details on the expected sensitivities of the {\tt HERA} and {\tt SKA} telescope configurations. Finally, in \autoref{ssec:Marginalization}, we marginalize over astrophysical nuisance parameters and present 21-cm sensitivities to PBHs for three astrophysical scenarios.

\subsection{{\tt Zeus21} calculation of the 21-cm signal}
\label{ssec:Zeus21}

The 21-cm signal during the cosmic dawn is principally determined by the effects of the radiation fields produced by the first stars, whose formation and emission properties are uncertain. Modeling the formation and associated emission of the first stars is very challenging, especially on cosmological scales. There have been dedicated efforts using hydrodynamical simulations (see, e.g., refs.~\cite{Trac:2006vr, Gnedin:2014uta}), but these are computationally very expensive. A faster \textit{semi-numerical} approach also exists, where 3D realizations of evolved density, temperature, ionization, velocity, and radiation fields are computed using Lagrangian perturbation theory (see, e.g., the popular code {\tt 21cmFAST}~\cite{Mesinger:2010ne, Murray:2020trn}). These Lagrangian methods produce results comparable to full hydrodynamical simulations~\cite{Zahn:2010yw}, at a fraction of the computational cost. Despite this speed-up of this code, it still is computationally expensive to run parameter scans using these methods.

In this work, we instead opt to use the fully analytic code, {\tt Zeus21}~\cite{Munoz:2023kkg, Cruz:2024fsv}, where the SFRD is the main quantity determining the 21-cm signal, and both the global signal and  power spectrum are computed approximately two orders of magnitude faster than with {\tt 21cmFAST}. Comparisons have shown that {\tt Zeus21} and {\tt 21cmFAST} agree at the $10 \%$ level for both the 21-cm global signal and power spectrum~\cite{Munoz:2023kkg, Cruz:2024fsv}. We summarize below how we use the {\tt Zeus21} code in our analysis.

Following ref.~\cite{Munoz:2023kkg}, we take the SFRD to be an approximately log normal variable at cosmic dawn. This allows for the fully analytic computation of the global signal and power spectrum. This analytic calculation relies on the assumption that the SFRD scales exponentially with the over/underdensities $\delta_R$, where these $\delta_R$ are assumed from the cosmological initial conditions. The 21-cm signal thus depends on the sum of SFRDs, averaged over different comoving radii $R$. Under these assumptions, the SFRD in a region of comoving radius $R$, and with density contrast $\delta_R$, is given by~\cite{Barkana:2004vb, Munoz:2021psm}
\begin{equation}\label{eq:SFRD}
     {\rm SFRD}(z|\delta_R) = (1 + \delta_R) \int dM_h \frac{dn}{dM_h}(\delta_R)\dot{M}_*(M_h) ~,
\end{equation}
where $\frac{dn}{dM_h}(\delta_R)$ is the density-modulated halo mass function (HMF), and $M_*(M_h) \equiv M_*(M_h,z)$ is the star formation rate (SFR) of a galaxy hosted in a halo of mass $M_h$ (see ref.~\cite{Munoz:2021psm} for more details).  We use a Sheth-Tormen halo mass function~\cite{Sheth:1999su}, and consider a SFR that assumes that some fraction of baryonic matter accreted by galaxies is converted into stars
\begin{equation}
    \dot{M}_* = f_* \, f_b \, \dot{M}_h ~,
\end{equation}
where $f_b = \Omega_b/\Omega_m$ is the baryon fraction, $\dot{M_h}$ is the mass accretion rate of a galaxy, and $f_*\equiv f_*(M_h)$ is the SFR efficiency.

The efficiency $f_*(M_h)$ is regulated by an exponentially suppressed duty fraction, ensuring that stars do not form efficiently below some threshold $M_{\rm turn}$, as given by 
\begin{equation}
   f_*(M_h) = \frac{2 \epsilon_* }{(M_h/M_{\text{pivot}})^{-\alpha_*} + (M_h/M_{\text{pivot}})^{-\beta_*}}   \operatorname{exp} \left( - \frac{M_{\rm turn}}{M_h}\right) ~, 
\end{equation} 
where $\epsilon_*$, $M_{\rm pivot}$, $\alpha_*$, and $\beta_*$ are constant parameters, fixed to the values shown in \autoref{tab:All_Astroparams}. 
In {\tt Zeus21}, $M_{\rm turn}$ is taken to be the atomic cooling threshold, corresponding to a fixed virial temperature of $T_{\rm vir} = 10^4$ K. We generalize this implementation following ref.~\cite{Barkana:2000fd}, and modify {\tt Zeus21} by taking $M_{\rm turn}$ to depend on the virial temperature as 
\begin{equation}
\label{eq:Mturn}
    M_{\text{turn}} = 100.58 \times  T_{\text{vir}}^{3/2} \left(\frac{1+z}{10}\right)^{-3/2} ~.
\end{equation}
$\tt Zeus21$ additionally uses a broadcasting methodology to compute the effective biases $\gamma_R$, by positing that the SFRD in eq.~\eqref{eq:SFRD} is related to its average value 
\begin{equation}
\label{eq:calculate_effective_biases}
    {\rm SFRD}(z|\delta_R) \approx \overline{{\rm SFRD}(z)} \, e^{\gamma_R \tilde{\delta}_R} ~, 
\end{equation}
where $\overline{\rm SFRD}(z)$ is computed following~\cite{Madau:1996yh}, and $\tilde{\delta}_R = \delta_R - \gamma_R \, \sigma_R^2 / 2$, where $\delta_R$ and $\sigma_R$ is the density contrast and variance in a region of comoving radius $R$, respectively (see ref.~\cite{Munoz:2023kkg} for details). The  approximation as an exponential allows an analytic calculation of the correlation function of the SFRD given $\delta_R$, which is a well-known cosmological output of the {\tt CLASS} code~\cite{Blas:2011rf}. The nonlinearities associated with structure formation are therefore encoded in the {\it effective bias} $\gamma_R$. 

Using the above prescription for the SFRD, the  Lyman-$\alpha$ radiation field $J_{\alpha}^{\rm astro}$ and the X-ray radiation field $J_X^{\rm astro}$ can be written as integrals of the form
\begin{equation}\label{eq:lyalpha_x_ray_from_SFRD}
    J_{\alpha/X}^{\rm astro} \propto \int dR\, c_{\alpha/x}(R)\, \text{SFRD}(R) ~,
\end{equation}
where the integral is performed over the comoving  radius $R$, and $c_{\alpha/x}$ are coefficients accounting for photon propagation, defined explicitly in ref.~\cite{Munoz:2023kkg}. We remark here explicitly that the Lyman-$\alpha$ normalization in eq.~\eqref{eq:lyalpha_x_ray_from_SFRD} is $J_{\alpha}^{\rm astro} \propto N_\alpha$, the number of Lyman-$\alpha$ photons per baryon.  Additionally, the X-ray term that contributes to heating is normalized by $ J_{X}^{\rm astro} \propto L_X / \dot{M}_*$, where $L_X / \dot{M}_*$ is the soft-band ($E<2 {\rm \, keV}$) luminosity/SFR in X-rays in units of ${\rm erg} \, {\rm s}^{-1} \,M_\odot^{-1} \,{\rm yr}$~\cite{Greig:2017jdj,Munoz:2023kkg}. These normalizations, and their associated uncertainties, are a focus of this paper, and will be discussed in detail \autoref{ssec:Null_hypothesis}.

{\tt Zeus21} uses the background cosmological evolution of the ionization fraction and IGM temperature as computed by  \texttt{CLASS}~\cite{Blas:2011rf}, and calculates the stellar contribution to these quantities throughout cosmic dawn using the prescription described above. To include the contribution from PBHs, we modify both {\tt Zeus21} and \texttt{CLASS}.  In {\tt Zeus21}, we include the Lyman-$\alpha$ contribution from PBHs as in eq.~\eqref{eq:lyalpha_coupling}, and in {\tt CLASS}, we use the modified evolution equations described in ref.~\cite{Stocker:2018avm} and given in eq.~\eqref{eq:evolution equations}, evolved using {\tt HyRec}~\cite{AliHamoud:2011}.

\subsection{Energy injection and energy deposition from accreting matter}
\label{sec:Einjection_and_deposition}
\subsubsection{Energy injection}\label{sec:Injection}

The total energy injected into the medium per unit volume is given by
\begin{equation}
\left. \frac{\mathrm{d}^2 E}{\mathrm{d} V\, \mathrm{d} t} \right|_{\mathrm{inj}} = L\, n_{\mathrm{PBH}} = L\, f_{\mathrm{PBH}}\, \frac{\rho_{\mathrm{DM}}}{M_{\rm PBH}} ~,
\end{equation}
where $L$ is the bolometric luminosity of a PBH with mass $M_{\rm PBH}$, $n_{\rm PBH}$ is the number density of PBHs, $f_{\rm PBH}$ is the fraction of DM in the form of PBHs, and $\rho_{\rm DM}$ is the DM density today.  The bolometric luminosity of an accreting PBH depends on the accretion rate, $\dot{M}$, and is connected to the total rest-mass energy inflow of the accreted material through the radiative efficiency parameter $\epsilon$, defined as
\begin{equation}
L = \epsilon\, \dot{M} c^2 ~.
\end{equation}
The efficiency $\epsilon$ is generally a function of the accretion rate, often modeled as a power law, $\epsilon(\dot{M}) \propto \dot{M}^a$. The exponent $a$, along with the normalization, depends on the nature of the accretion flow, particularly on whether an accretion disk forms and on its specific properties.

A key criterion for the formation of an accretion disk is whether the angular momentum of the baryons is sufficient to maintain Keplerian motion at radii larger than the innermost stable circular orbit~\cite{Shapiro1976}. Following refs.~\cite{Poulin:2017bwe, Mena:2019nhm, Agius:2024ecw}, we assume that an accretion disk always forms.

The modeling of the transport of angular momentum and energy (through turbulence, viscosity, shear, and magnetic fields) has significant uncertainties, which translates into uncertainties in forecasting cosmological constraints. The angular momentum and energy transport and its influence on the radiative output have been explored in depth in prior studies~\cite{Poulin:2017bwe, Agius:2024ecw}. Nonetheless, in line with refs.~\cite{Poulin:2017bwe, Mena:2019nhm, Agius:2024ecw}, we adopt a fiducial model in which a hot, geometrically thick accretion disk develops, known as advection-dominated accretion flow (ADAF). In this regime, the radiative efficiency $\epsilon$ decreases with decreasing accretion rate.
In the ADAF scenario, $\epsilon$ is modeled as
\begin{equation} 
\label{eq:epsilon}
\epsilon (\dot{M}) = \epsilon_0 \left( \frac{\dot{M}}{0.01\, \dot{M}_{\rm Edd}} \right)^a  ~,
\end{equation}
where $\dot{M}_{\rm Edd}$ denotes the Eddington accretion rate. The values of $\epsilon_0$ and $a$ depend on the accretion rate itself and follow a piecewise definition, provided in \autoref{tab:deltas} (taken from ref.~\cite{Xie:2012rs}). The parameter $\delta$ in \autoref{tab:deltas} quantifies the fraction of turbulent energy in the accretion disk that heats the electrons directly, and we assume a benchmark value $\delta = 0.1$ in this work.\footnote{See ref.~\cite{Agius:2024ecw} for a study on the effect of varying delta on cosmological bounds, and ref.~\cite{Xie:2012rs} for more details on the ADAF parameterization.} For the typical accretion rates arising in both the BHL and PR model and assuming $\delta = 0.1$, then usually we fall in the regime where $\epsilon_0 = 0.12$ and $a = 0.59$. The other values of $\epsilon_0$ and $a$ that occur for larger values of the accretion rate are given in \autoref{tab:deltas}. The functional form in eq.~\eqref{eq:epsilon} encodes the necessary information required to compute the energy injection into the IGM.

\begin{table}[t]
\centering
\begin{tabular}{|c|c|c|c|}
\hline
$\delta$ & $\dot{M} / \dot{M}_{\text{Edd}}$ range & $\epsilon_0$ & $a$ \\
\hline \hline
& $< 9.4 \times 10^{-5}$ & 0.12 & 0.59 \\
$0.1$ & $9.4 \times 10^{-5} - 5.0 \times 10^{-3}$ & 0.026 & 0.27 \\
& $5.0 \times 10^{-3} - 6.6 \times 10^{-3}$ & 0.50 & 4.53 \\
\hline
\end{tabular}
\caption{Piecewise power-law fitting parameters for the radiative efficiency $\epsilon$, used in eq.~\eqref{eq:epsilon}. Values are taken from ref.~\cite{Xie:2012rs}.}
\label{tab:deltas}
\end{table}

\subsubsection{Energy deposition}
\label{sssec:Deposition}

It is well established that energy injected into the medium during the dark ages is not deposited immediately~\cite{Shull:1985, Chen:2003gz, Slatyer:2009yq}. Instead, energy is deposited at later times. Initially, on short time-scales, injected particles initiate an electromagnetic cascade through the interaction with thermal photons, where there is an increase in the number of non-thermal particles at the expense of a decrease in their average energy. These non-thermal particles then cool slowly with redshift, on cosmological times scales, and when these particles have energies of the order of keV, they start interacting strongly with the hydrogen atoms of the IGM, thus depositing their energy~\cite{Chen:2003gz, Stocker:2018avm}. This delayed deposition is quantified by the energy deposition functions $f_c(z, x_e)$, which describe the fraction of injected energy deposited at redshift $z$ into a specific channel $c$. The most relevant deposition channels are heating, ionization, and excitation of atoms,
\begin{equation}
\label{eqn:dep_fn_defn}
\left. \frac{\mathrm{d}^2 E}{\mathrm{d}V\, \mathrm{d}t} \right|_{\mathrm{dep}, \, c} 
= f_c(z, x_e) \left. \frac{\mathrm{d}^2 E}{\mathrm{d}V\, \mathrm{d}t} \right|_{\mathrm{inj}} ~.
\end{equation}
The energy deposition functions $f_c(z, x_e)$ can be computed directly from a given energy-differential luminosity spectrum $L_\omega$ using 
\begin{equation}\label{eq:f_c integral}
f_c(z, x_e) = H(z)\, \frac{\int \frac{\mathrm{d} \ln (1+z')}{H(z')} \int T(z', z, \omega)\, L_\omega\, \mathrm{d}\omega}{\int L_\omega\, \mathrm{d}\omega} ~,
\end{equation}
where $T(z', z, \omega)$ is are the energy and redshift-dependent transfer functions tabulated in refs.~\cite{Slatyer:2012yq, Slatyer:2015kla}, and $L_\omega $ is given as the disk accretion scenario from ref.~\cite{Poulin:2017bwe}, explicitly defined in the  \texttt{DarkAges} module~\cite{Stocker:2018avm}. 

Moreover, we use the \texttt{DarkAges} module~\cite{Stocker:2018avm} to perform the integrals in eq.~\eqref{eq:f_c integral}. One of the assumptions underlying this approach is that changes to the free electron fraction $x_e$ from any additional energy injection do not significantly alter the evolution of the electromagnetic cascade~\cite{Stocker:2018avm}. Accordingly, the ionization fraction $x_e$ is taken to follow its standard evolution in the absence of energy injection. 

\begin{figure}[t]
    \centering
    \includegraphics[width=.7\linewidth]{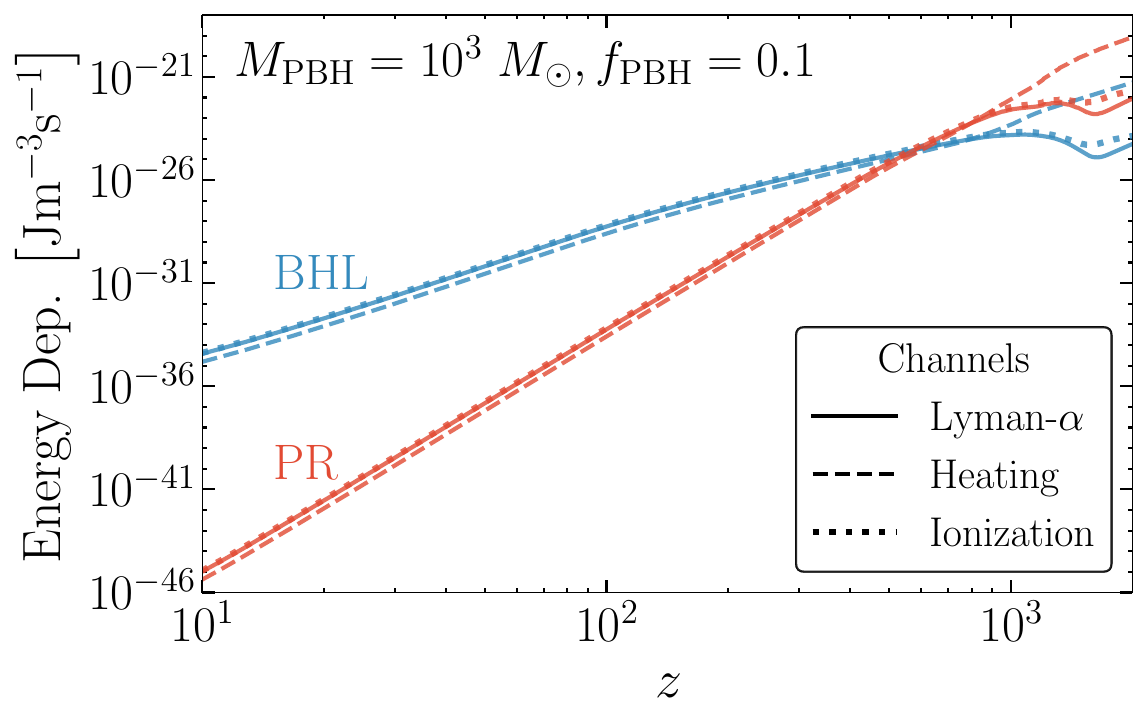}
    \caption{Energy deposition into the IGM from PBHs into Lyman-$\alpha$ (solid lines), heating (dashed lines), and ionization (dotted lines) for the BHL (red lines) and PR (blue lines) accretion models, as a function of redshift. We assume a PBH mass of $M_{\rm PBH} = 10^3~M_\odot$ and a fractional PBH abundance of $f_{\rm PBH} = 0.1$.}
    \label{fig:energy_dep}
\end{figure}

In \autoref{fig:energy_dep}, we show the energy deposition from PBHs into heating, ionization and excitation (Lyman-$\alpha$) for both accretion models. The energy deposition from heating, ionization and through Lyman-$\alpha$ photons decreases monotonically after $z\sim 1000$, largely a consequence of the dilution of the background medium. However, the two accretion models differ substantially in how quickly they decrease. Much more energy is deposited in the early Universe in the PR model than in the BHL model, but the PR model shuts off more rapidly at late times than the BHL model. The implications of these differences for the 21-cm signal will be discussed in the next sections.

With the ingredients described above, we can compute the Lyman-$\alpha$ contribution from accreting PBHs as
\begin{equation}
    J_{\alpha}^{\rm PBH} = \frac{c \, }{4\pi \, H(z) \, \nu_{\alpha}^2 \, h}  \left. \frac{\mathrm{d}^2 E}{\mathrm{d}V\, \mathrm{d}t} \right|_{\mathrm{dep}, \, \text{Ly}\alpha} ~,
\end{equation}
where $\nu_\alpha$ is the Lyman-$\alpha$ frequency, and we have explicitly written the Lyman-$\alpha$ contribution from eq.~\eqref{eqn:dep_fn_defn}.

Finally, we note that the \texttt{DarkHistory} code~\cite{Liu2019} improves upon the {\tt DarkAges} module by  self-consistently accounting for the backreaction of changes to $x_e$ by exotic injection on the evolution of the electromagnetic cascade, by recomputing $x_e$-dependent transfer functions. This correction becomes particularly relevant when the ionization fraction rises rapidly during the epoch of reionization, or when the ionization fraction is modified substantially by exotic injections in the ionization channel. The first effect becomes most relevant at $z \lesssim 12$ (see left panel of \autoref{fig:xe_Tk}). Since our analysis is restricted to $z > 10$, we do not expect substantial deviations from our results stemming from this narrow redshift range. This is especially the case, since at these redshifts heating from astrophysical sources dominates over any additional heating contributions from exotic sources. The second effect could be relevant, especially in less-constraining astrophysical scenarios, where the 21-cm signal does not rule out significant energy injections. In particular, as shown in Figure~6 of ref.~\cite{Liu2019}, constraints become stronger when considering the backreaction effect of modifications to the ionization and thermal history induced by exotic sources, by 10\%--50\% stronger for the cases of interest. This is attributed to larger temperatures caused by including backreaction effects, as shown in Figure~4 of ref.~\cite{Liu2019}. We do not attempt to quantify this effect further, given that the systematics that we will present in this paper correspond to orders of magnitude effects on the bound, and will be most relevant in a region of the parameter space corresponding to large energy injections, that are constrained by other observations (see, e.g., \autoref{fig:bound} below and references in the caption).

\subsection{Mock data generation}
\label{ssec:Null_hypothesis}

To forecast constraints using the 21-cm signal, a fiducial set of astrophysics parameters for mock data generation must be chosen. However, the properties of the first stars are poorly known, and consequently, the fiducial 21-cm signal expected to be measured by experiments represents a significant uncertainty. Current theoretical predictions of the power spectrum during cosmic dawn vary by at least one order of magnitude, with similar uncertainties in the astrophysical parameters underlying the fiducial theory model (see, e.g., refs.~\cite{Fialkov:2016zyq, Lopez-Honorez:2016sur, Cohen:2016jbh, Park:2018ljd, Munoz:2021psm, Pochinda:2023uom, Katz:2024ayw, Decant:2024bpg, Sims:2025hfm, Dhandha:2025dtn}). 

Several fiducial astrophysical scenarios have been studied in the literature. Some include only atomically cooled Pop II stars (e.g., refs.~\cite{Lopez-Honorez:2016sur, Park:2018ljd, Mena:2019nhm}), while others also consider the effects of an additional population of molecularly cooled halos forming Pop III stars (e.g., refs.~\cite{Munoz:2021psm, Facchinetti:2023slb, Sun:2023acy}). However, when confronted with existing data, the allowed parameter space for each model widens substantially, such that the theory prediction for the global and power spectrum signals spans roughly two orders of magnitude~\cite{Fialkov:2016zyq, Cohen:2016jbh, Pochinda:2023uom, Katz:2024ayw, Katz:2025sie, Sims:2025hfm, Dhandha:2025dtn}.  

To account for this uncertainty, we consider three distinct fiducial astrophysical scenarios for mock data generation, parametrized similarly to ref.~\cite{Mena:2019nhm}. In particular, we consider an astrophysical model consisting of three parameters that we vary, and several additional parameters that we set to the fiducial values given in \autoref{tab:All_Astroparams} in \autoref{App:Fixed Astro Params}. The value (or range of values) of each parameter is chosen to be consistent with existing independent observational constraints from ultraviolet luminosity functions (UVLFs)~\cite{Bouwens:2014fua, Bouwens:2021, Park:2018ljd}, spectra from Chandra X-ray observatory~\cite{Cappelluti:2012rd, Lehmer:2012ak, Fialkov:2016zyq}, and X-ray limits from {\tt HERA}~\cite{HERA:2021noe, HERA:2022wmy}.\footnote{These observational constraints necessarily depend on additional modeling assumptions about the star formation history and emission properties of early galaxies. We leave a fully self-consistent study of all independent constraints to future work.} For simplicity, we consider an astrophysical model containing only  Pop II stars, but remark that the properties of Pop III stars are also poorly understood, so the uncertainties associated with choosing a fiducial model would remain if we include Pop III stars~\cite{Klessen:2023qmc}.\footnote{Pop II stars have been found in surveys looking at metal-poor stars in our galaxy and neighboring satellites. However, Pop III stars have not yet been detected~\cite{Klessen:2023qmc}.} 

The three parameters of the astrophysical model that we vary over in this work are:
\begin{itemize}
    \item $N_\alpha$: The number of Lyman-$\alpha$ photons between $10.2$ eV and $13.6$ eV per baryon.
    \item $L_{X}$: The luminosity/$\dot{M}_*$ in soft X-rays ($0.5~{\rm keV} \leq E \leq 2~{\rm keV}$) in units of ${\rm erg} \, {\rm s}^{-1} \,M_\odot^{-1} \,{\rm yr}$.
    \item  $T_{\rm vir}$: The minimum virial temperature of galaxy forming halos. 
\end{itemize}
In particular, we vary the astrophysical parameters between three scenarios for mock data generation. We label these as \textsc{benchmark}, \textsc{more-constraining}, and \textsc{less-constraining}, with the astrophysics parameter choices for each of these scenarios given in \autoref{tab:Astroparameters}. Each of these three scenarios are chosen to be consistent with current observational and theory constraints, and named according to their constraining power for exotic scenarios. All scenarios are consistent with the $2\sigma$ upper limits on the power spectrum reported by {\tt HERA}~\cite{HERA:2021noe, HERA:2022wmy}. We justify the parameter choices for our three scenarios below. As far as the constraining power associated to each scenario is concerned, the full discussion will be presented in \autoref{sec:Results}.

\begin{table}[t]
\centering
\begin{tabular}{|c|c|c|c|}
\hline
Astrophysical parameter   & {\sc Benchmark} & { \sc Less-Constraining}  & {\sc More-Constraining } \\
\hline \hline 
$N_{\alpha}$ &       9690     & $3\times 10^3$ &  $4\times 10^4$   \\[1ex]
$\log_{10} \, L_{X} / \dot{M}_* \ [ \frac{\text{erg}}{\text{s}} \frac{\rm yr}{M_{\odot}}] $         &     $40.5$          & $41$  &$ 39.5$       \\[1ex]
$T_{\rm vir} \ [K] $         &  $10^4$ & $10^4$ & $ 10 ^5$  \\
\hline
\end{tabular}
\caption{The three astrophysical parameters that we vary over for forecasting the sensitivity of the 21-cm signal to PBHs. These parameters are: the number of photons between Ly-$\alpha$ and Ly continuum per baryon, $N_{\alpha}$; the soft-band ($E<2 {\rm \, keV}$) luminosity/SFR in X-rays, $L_{X}$; the minimum virial temperature of halos hosting galaxies, $T_{\rm vir}$. The second column shows our \textsc{benchmark} scenario, while the third and fourth columns shows the set of parameters leading to less sensitivity to PBHs and more sensitivity to PBHs, respectively.}
\label{tab:Astroparameters}
\end{table}

To specify the ranges on $N_\alpha$, we begin by remarking that the typical benchmark scenario of $N_\alpha = 9690$ Lyman-$\alpha$ photons per baryon is given in ref.~\cite{Barkana:2004vb}, and computed from the spectral energy distributions provided in the {\tt Starburst99} simulations~\cite{Leitherer:1999rq}. To remain consistent with the existing literature, we also use this as our benchmark value. However, the {\tt Starburst99} model predictions include data for different metallicities, initial mass functions, ages of the stellar population, and whether the stars formed at a continuous rate, or instantaneously. Integrating the {\tt Starburst99} spectra for different combinations of these parameters to derive a value of $N_\alpha$ gives different results ranging from several hundreds to tens of thousands of photons per baryon. Additionally, more recent additions to the {\tt Starburst99} have highlighted further theoretical uncertainties affecting the emission properties of stellar sources, such as stellar rotation, which can change the UV luminosity by up to a factor of 5~\cite{Leitherer:2010, Leitherer:2014}. Given the absence of any robust theoretical model of the UV spectrum of the first stars, we bracket this uncertainty through the parameter choices for $N_\alpha$. In particular, we choose  the \textsc{less-constraining} and \textsc{more-constraining} values of $N_\alpha$ to be $3\times 10^3$ and $4 \times 10^4 $ photons per baryon. These values are contained within ranges on $N_\alpha$ implicitly considered in refs.~\cite{Greig:2015qca, Greig:2017jdj, Greig:2018hja, Witte:2018itc, Mena:2019nhm}.

For $L_{X}$, an upper bound can be derived using stacked X-ray observations from Chandra, giving $L_{X} / \dot{M}_* \lesssim 10^{42}$ ${\rm erg} \, {\rm s}^{-1} \,M_\odot^{-1} \,{\rm yr}$~\cite{Lehmer:2012ak, Lehmer:2016dxd}. Recently, the {\tt HERA} collaboration released a lower limit, finding that $L_{X} / \dot{M}_*  > 10^{39.9}$ ${\rm erg} \, {\rm s}^{-1} \,M_\odot^{-1} \,{\rm yr}$ at 2$\sigma$ confidence~\cite{HERA:2021noe, HERA:2022wmy}. This lower limit is derived from a lower bound on the gas temperature, from the non-detection of the 21-cm power spectrum by {\tt HERA}, leading to the conclusion that some heating of the gas must be present in the early Universe. However, the translation of this gas temperature limit to an X-ray limit requires assumptions on the modeling of astrophysical sources. To include the uncertainties associated with the astrophysical modeling, we conservatively choose our lower limit to be $L_{X} / \dot{M}_* = 10^{39.5}$ ${\rm erg} \, {\rm s}^{-1} \,M_\odot^{-1} \,{\rm yr}$. We choose a \textsc{benchmark} scenario contained within these limits, assuming high-mass X-ray binaries consistent with ref.~\cite{Fragos:2012vf}, with $L_{X} / \dot{M}_*=10^{40.5}$ ${\rm erg} \, {\rm s}^{-1} \,M_\odot^{-1} \,{\rm yr}$,\footnote{This benchmark value should be understood as an order of magnitude estimate on the X-ray luminosity. See, e.g., ref.~\cite{Oh:2001}, and references therein.} and we choose an upper limit of $L_{X} / \dot{M}_* = 10^{41}$ ${\rm erg} \, {\rm s}^{-1} \,M_\odot^{-1} \,{\rm yr}$.\footnote{We remark here that using a larger upper limit for $L_X$ would further increase heating to the IGM from astrophysical sources, making distinguishability from exotic sources even more difficult (and can be roughly understood as an even less-constraining scenario). Similarly, using a lower limit would have the opposite effect, allowing for a more-constraining scenario.}

For $T_{\rm vir}$, a lower limit can be set by the atomic cooling threshold, and an upper limit can be set by requiring consistency with observed
high-$z$ Lyman break galaxies~\cite{Bouwens:2014fua}, and from observations of the Lyman-$\alpha$ forest~\cite{Tegmark:1996yt, Mesinger:2012ys, Greig:2015qca, Lopez-Honorez:2016sur}.\footnote{Attempts to constrain $M_{\rm turn}$ are given, e.g., in ref.~\cite{Park:2018ljd}.} We consider our \textsc{benchmark} scenario at the atomic cooling threshold $T_{\rm vir} = 10^4$ K, and choose our \textsc{less-constraining} and \textsc{more-constraining} values at $10^4$~K and $10^5$~K, respectively.

The substantial variation in the astrophysics parameters reflects the absence of powerful observational probes constraining cosmic dawn. Ongoing efforts to constrain these parameters using independent observables include UVLFs~\cite{Park:2018ljd, Qin:2020xyh, Munoz:2021psm, Katz:2024ayw, Katz:2025sie}, quasar dark fraction measurements~\cite{McGreer:2014qwa, Greig:2016wjs}, cosmic X-ray and radio background measurements~\cite{Cappelluti:2012rd, Lehmer:2012ak, Fialkov:2016zyq, Pochinda:2023uom, Dhandha:2025dtn}, and the high redshift Lyman-$\alpha$ forest~\cite{Qin:2021gkn, Sims:2025hfm}. As the James Webb Space Telescope ({\tt JWST}) continues to shed light on the early Universe~\cite{Gardner:2023}, such approaches may further narrow the allowed parameter space~\cite{Finkelstein2023}. In addition, upcoming missions such as The Advanced Telescope for High Energy Astrophysics ({\tt ATHENA}) X-ray telescope, will work in synergy with 21-cm observatories to constrain the cosmic dawn~\cite{Cassano:2018zwm}. 

However, until the 21-cm signal is measured, and independent observatories constrain the UV and X-ray properties of the first stars, a significant systematic will remain in the mock data generation for forecasts. Other groups have considered these astrophysical uncertainties on the 21-cm signal, studying the impact of the first galaxies being bright or faint~\cite{Mesinger:2016ddl, Munoz:2021psm}, or how altering astrophysical scenarios would have an impact on the constraints on annihilating DM~\cite{Lopez-Honorez:2016sur} or warm DM~\cite{Decant:2024bpg}.

\subsection{Statistics}
\label{ssec:statistic}

Following ref.~\cite{Mena:2019nhm}, we choose a multivariate Gaussian likelihood of the form
\begin{equation}
\label{eq:likelihood}
    \operatorname{log} \mathcal{L} = -\frac{1}{2} \sum_{i_z}^{N_z} \sum_{i_k}^{N_k} \frac{\left[ \Delta_{21}^2  \left( z_{i_z},k_{i_k} \right)_{\rm test} - \Delta_{21}^2  \left( z_{i_z},k_{i_k} \right)_{\rm fid} \right]^2}{\sigma^2_{\rm tot}\left(z_{i_z},k_{i_k} \right)} ~, 
\end{equation}
where $\Delta_{21}^2  \left( z_{i_z},k_{i_k} \right)_{\rm test / fid}$ are the 21-cm power spectrum evaluated at a given $z$ and $k$, for either the test or fiducial model, and $\sigma_{\rm tot}\left(z_{i_z},k_{i_k} \right)$ is the total measurement error, defined in the next section. The sums run over redshift $z$ and through $k$-space (in ${\rm Mpc}^{-1}$), with the range we consider explicitly including  $k$ = $\{ $0.13, 0.18, 0.23, 0.29, 0.34, 0.39, 0.45, 0.5, 0.55, 0.61, 0.66, 0.71, 0.77, 0.82, 0.87, 0.93$\}$ Mpc$^{-1}$, consistent with those used in ref.~\cite{Facchinetti:2023slb}, and ten log-spaced $z$ measurements  $z$ = $\{ $10.27, 11.04, 11.91, 12.93, 14.11, 15.52, 17.21, 19.29, 21.91, 25.30$\}$, matching the 8~MHz bandwidth of the  experiments we consider.\footnote{We consider only $z\gtrsim 10$, since {\tt Zeus21} is designed to be valid until reionization at $z\simeq 10$~\cite{Munoz:2023kkg}.}

We adopt the three-parametric model outlined in \autoref{ssec:Null_hypothesis}, and explicitly defined in \autoref{tab:Astroparameters}. We compute the likelihood when varying $f_{\rm PBH}$ under the three different astrophysical scenarios defined in \autoref{tab:Astroparameters}.

\subsubsection{Sensitivities}
\label{sssec:HERAandSKAsensitivities}

The total measurement error in the power spectrum is given by~\cite{Facchinetti:2023slb,Park:2018ljd,Mason:2022obt}
\begin{equation}\label{eq:sensitivity}
\sigma^2_{\rm tot} \left( z , k \right)  = \sigma^2_{\rm exp}\left( z , k \right)+ \sigma^2_{\rm sample}\left( z , k \right) + \left( 0.2 \, \Delta^2_{21 } \left( z , k \right)_{\rm fid}\right)^2 ~,
\end{equation}
where $\sigma_{\rm exp}$ denotes the experimental error,\footnote{The experimental sensitivities $\sigma_{\rm exp}$ are explicitly dependent on the power spectrum of the fiducial astrophysical model, as shown explicitly in eq.~(15) of ref.~\cite{Pober:2012zz}. \label{Footnote:sense}} $\sigma_{\rm sample}$ is the error introduced by Poisson noise due to cosmic variance, and the third term accounts for a model uncertainty budget of $20 \%$, in line with the  error budget computed for {\tt 21cmFAST} and propagated for {\tt Zeus21}~\cite{Zahn:2010yw, Munoz:2023kkg}. We compute the experimental error for each of our telescope configurations, and for each of the three fiducial astrophysical scenarios using \texttt{21cmSense}~\cite{Murray:2024the, Pober:2012zz, Pober:2013}, assuming the default system temperature given by 
\begin{equation}\label{eq:Tsys}
    T_{\rm sys} = 100 {\ \rm K} + 260 {\ \rm K} \left( \frac{\nu(z)}{150 {\rm \ MHz}} \right)^{-2.6} ~,
\end{equation}
where $\nu(z)$ is the redshifted 21-cm line frequency. Modeling $T_{\rm sys}$ as in eq.~\eqref{eq:Tsys}, accounts for a receiver temperature of $100 {\ \rm K} $, and includes a sky temperature consistent with the measured diffuse radio background at $\sim 100$~MHz~\cite{Mozdzen:2017}.\footnote{As pointed out in ref.~\cite{Facchinetti:2023slb}, other works, including ref.~\cite{Mason:2022obt}, considered the temperature given in eq.~\eqref{eq:Tsys} to correspond to their pessimistic scenario.} We use the same system temperature for all the experimental configurations considered in this work. In this work, we focus our analysis on producing bounds for {\tt SKA}, but for completeness include a discussion on the experimental sensitivity of {\tt HERA}. 

For {\tt SKA}, we consider two configurations, the initial {\tt SKA AA*} configuration and the final {\tt SKA AA4} configuration (sometimes equivalently referred to in the literature as {\tt SKA1-LOW} and {\tt SKA2-LOW}, respectively~\cite{Koopmans:2015sua}). Both follow a spiral layout, where {\tt AA*} has 307 stations, while {\tt AA4} has 512 stations. The {\tt SKA-LOW} stations are $\sim 38$~m in diameter, but can also be divided into sub-stations of 12~m and 18~m. These sub-stations provide an increased field of view and access to shorter baselines. We assume a Gaussian beam for each baseline, and that the experiment is located at a latitude of $30.7^\circ$. We refer the interested reader to ref.~\cite{SKA_MEMO:2025} for more details and follow their specifications. Both {\tt SKA} configurations are explicitly implemented in the {\tt SKA\_forecast} notebook provided in {\tt 21cmSense}. For our sensitivities, we use a deep survey, where we set the number of tracking hours to be the same as the number of observation hours per day.  We use a fixed bandwidth of 8 MHz (corresponding to our redshift spacing), and assume an observation time of 1080 hrs (6 hrs per day for 180 days). 

For {\tt HERA}, we use a hexagonal antenna layout comprising 331 antennas (11 on each side), with a dish size of 14~m, and separation of 12.12~m. We assume a Gaussian beam for each baseline, and that the experiment is located at a latitude of $30.8^\circ$. For our sensitivities, we perform a drift-scan, where the number of tracking hours is equal to the beam-crossing time, similarly to refs.~\cite{Mason:2022obt, Facchinetti:2023slb}. Similarly to {\tt SKA}, we use a fixed bandwidth of 8~MHz, and assume an observation time of 1080~hrs (6~hrs per day for 180~days).

\subsection{Marginalization}
\label{ssec:Marginalization}

To properly account for degeneracies between accreting PBHs and astrophysical parameters, a marginalization should be performed. In the context of 21-cm studies, this is commonly performed using  MCMC methods~\cite{Greig:2015qca, Mena:2019nhm}, which efficiently sample the posterior without evaluating the full parameter volume, or with Fisher forecasts~\cite{Mason:2022obt, Facchinetti:2023slb, Sun:2023acy}, which provide an estimate of parameter correlations when likelihood evaluations are computationally expensive. 

In our case, computations of the likelihood take $\sim 1 $s, making a full parameter-scan feasible. As such, we run a full 4-dimensional parameter scan over the three parameters in \autoref{tab:Astroparameters}, together with the PBH parameter $f_{\rm PBH}$. We do this for the PBH masses $M_{\rm PBH} = \{1,10,100,1000 \} \ M_{\odot}$,\footnote{We compute a few additional masses within this range to more precisely find the PBH mass where the bound vanishes at $f_{\rm PBH} = 1$, to avoid unphysical features in~\autoref{fig:bound}.} keeping the rest of the astrophysical parameters fixed to the values given in \autoref{tab:All_Astroparams} of \autoref{App:Fixed Astro Params}. The grid has dimension $20^4$, corresponding to $20$ log-spaced points per parameters. We verified \textit{a posteriori} that the grid adequately covers the relevant parameter space, such that the marginalized likelihood vanishes at physically meaningful values, while not missing the region of maximum likelihood. 

From the resulting 4-dimensional likelihood, we perform a marginalization over the astrophysical parameters by interpolating the grid, and then integrating using the trapezoidal rule. The resulting 1-dimensional posterior is then used to construct a highest-density interval, with the bounds stated at the 95\% probability value.\footnote{We compute the bound assuming a flat prior on $f_{\rm PBH}$ over the range $(0,1)$, in line with the constraint derived from the CMB in ref.~\cite{Agius:2024ecw}.}

\section{Results}
\label{sec:Results}

In this section, we show the potential sensitivity of future 21-cm data to accreting PBHs. We begin in \autoref{ssec:Results-Mock-data}, by using the BHL accretion model as an example, to show that the constraining power of future experiments on any exotic energy-injection mechanism, whose 21-cm signal is governed by its heating and ionizing impact on the IGM, depends sensitively on the choice of fiducial model. We then show in \autoref{ssec:Results-PR-vs-BHL} how the 21-cm sensitivity to PBHs varies between the BHL and PR accretion models. In \autoref{ssec:Results-Forecast_and_discussion}, we combine these elements to present the projected exclusion limits on the PBH abundance, $f_{\rm PBH}$, illustrating how both systematics impact the projected constraints.

\begin{figure}
    \centering
    \textsc{less-constraining Scenario:} $N_\alpha = 3\times10^3, \, L_X = 10^{41}\, {\rm erg} \, {\rm s}^{-1} \,M_\odot^{-1} \,{\rm yr}, \, T_{\rm vir} = 10^4\, \text{K}$. \\
    \begin{minipage}{0.49\textwidth}
        \centering
        \includegraphics[width=\linewidth]{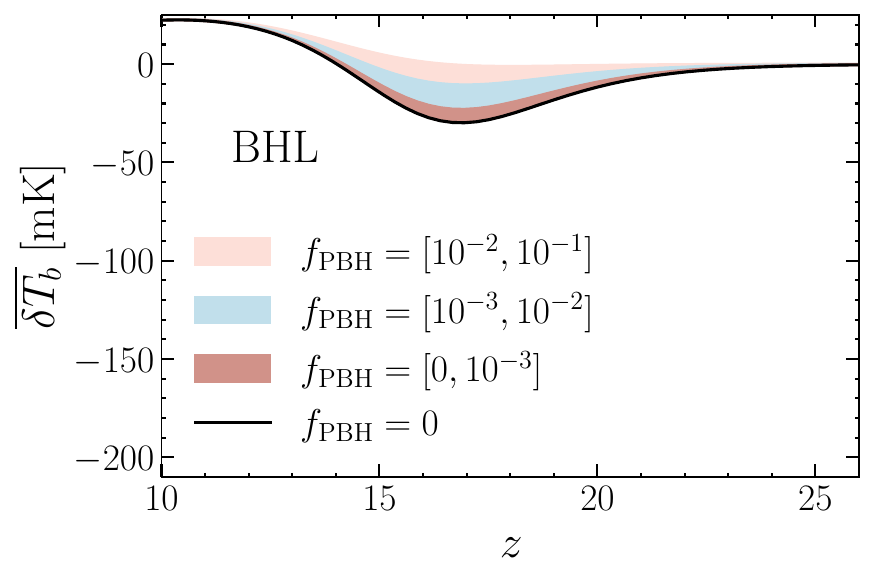}
    \end{minipage}
    \hfill
    \begin{minipage}{0.49\textwidth}
        \centering
        \includegraphics[width=\linewidth]{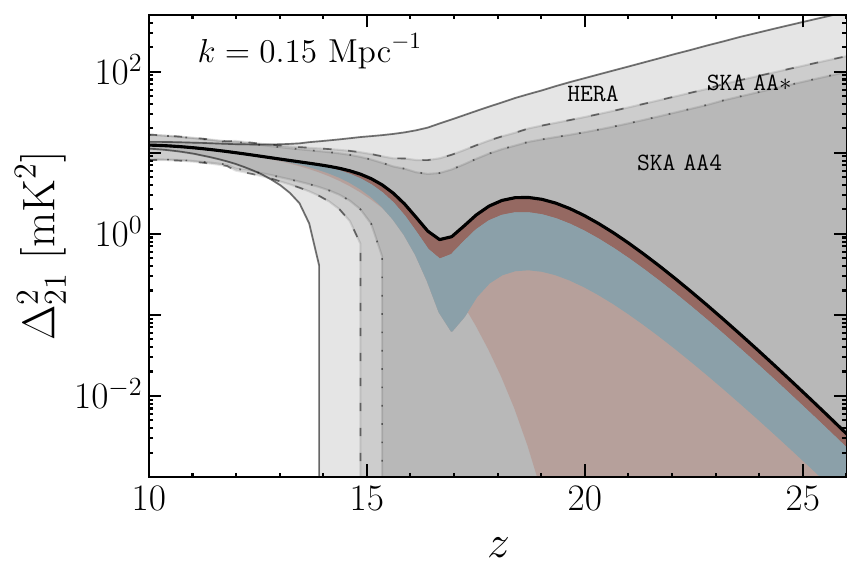}
    \end{minipage}
        \vspace{0.5cm} 

    \textsc{benchmark Scenario:} $N_\alpha = 9690, \, L_X = 10^{40.5}\, {\rm erg} \, {\rm s}^{-1} \,M_\odot^{-1} \,{\rm yr}, \, T_{\rm vir} = 10^4\, \text{K}$. \\
    
    \begin{minipage}{0.49\textwidth}
        \centering
        \includegraphics[width=\linewidth]{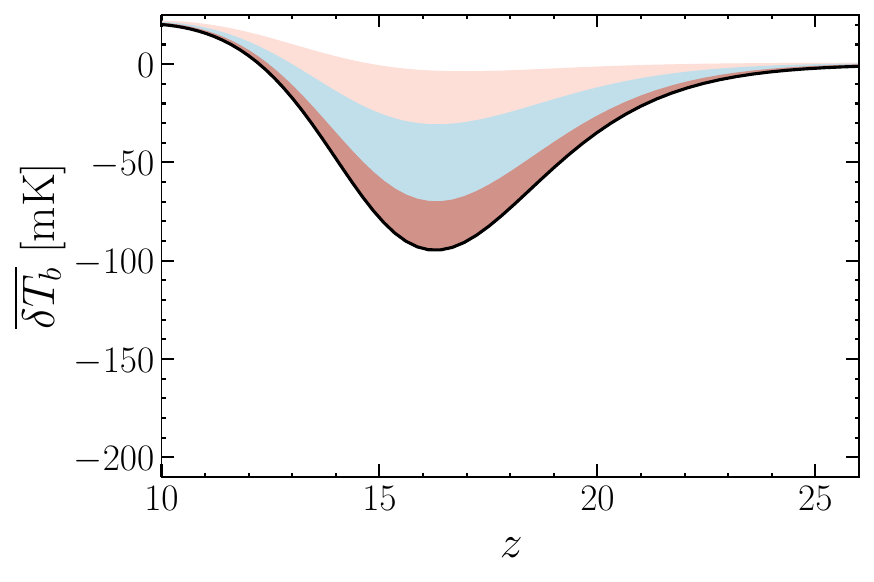}
    \end{minipage}
    \hfill
    \begin{minipage}{0.49\textwidth}
        \centering
        \includegraphics[width=\linewidth]{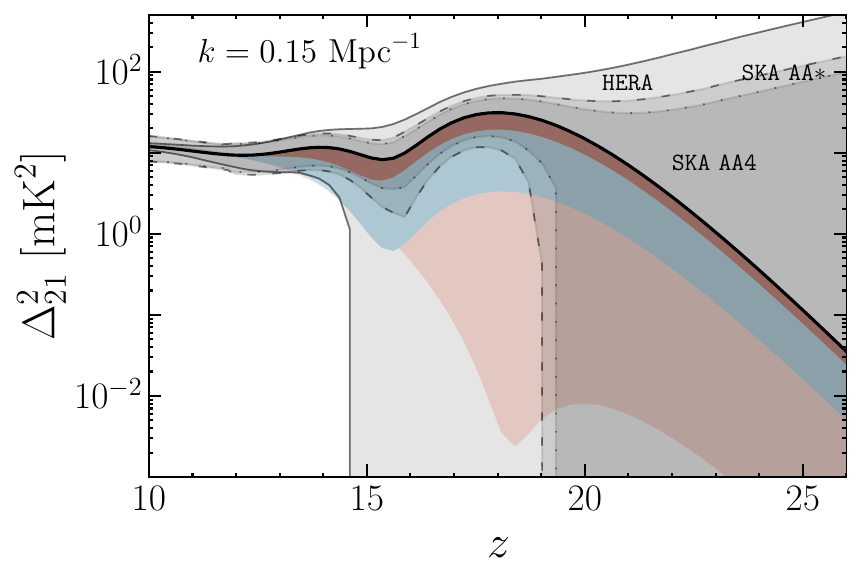}
    \end{minipage}
        \vspace{0.5cm}

    \textsc{More-Constraining Scenario:} $N_\alpha = 4\times10^4, \, L_X = 10^{39.5}\, {\rm erg} \, {\rm s}^{-1} \,M_\odot^{-1} \,{\rm yr}, \, T_{\rm vir} = 10^5 \, \text{K}$. \\
    
    \begin{minipage}{0.49\textwidth}
        \centering
        \includegraphics[width=\linewidth]{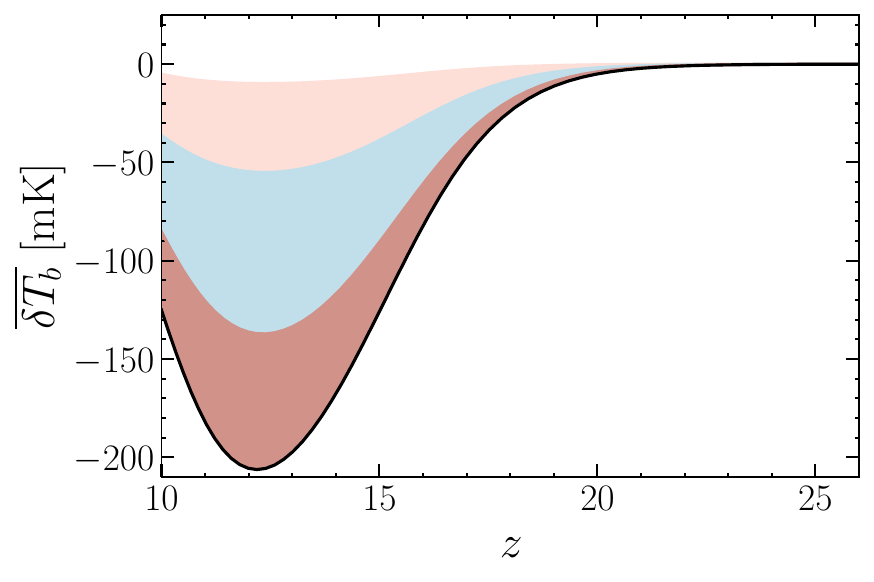}
    \end{minipage}
    \hfill
    \begin{minipage}{0.49\textwidth}
        \centering
        \includegraphics[width=\linewidth]{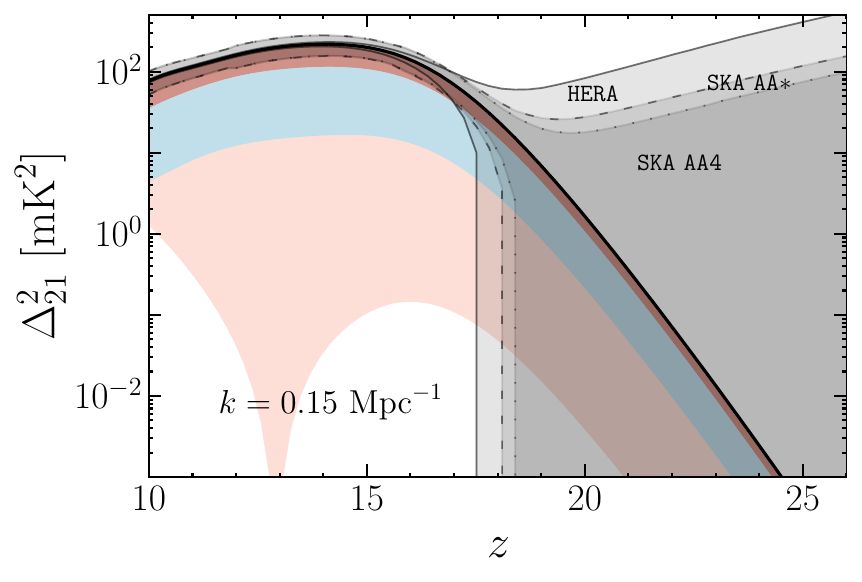}
    \end{minipage}
    
    \caption{Predictions of the 21-cm global signal (left panels) and the power spectrum (right panels), as a function of $z$, for the \textsc{less-constraining} (first row), \textsc{benchmark} (middle row) and \textsc{more-constraining} (bottom row) astrophysical scenarios. We assume the BHL accretion model and PBHs of mass $M_{\rm PBH} = 10^2  \, M_\mathbf{\odot}$, with the colors corresponding to different values of $f_{\rm PBH}$ as indicated in the legend of the top left panel; for the power spectrum, we fix  $k = 0.15 \, {\rm Mpc}^{-1}$.  We also show the $1\sigma$ experimental sensitivities, $\sigma_{\rm exp}\left( z , k \right)$, from the {\tt HERA} experiment, and the {\tt SKA AA*} and {\tt SKA AA4} configurations, computed using {\tt 21cmSense}~\cite{Murray:2024the, Pober:2012zz, Pober:2013}. The astrophysical parameters for each row are given in the title, with the definitions provided in \autoref{tab:Astroparameters}. The remaining astrophysical parameters are fixed to those shown in \autoref{tab:All_Astroparams}.}
    \label{fig:pess_fid_opt}
\end{figure}

\subsection{The importance of the fiducial model}
\label{ssec:Results-Mock-data}

We analyze three distinct astrophysical scenarios by varying the three parameters  ($N_\alpha$, $L_{X}$, $T_{\rm vir}$) defined in \autoref{ssec:Null_hypothesis}. These three distinct scenarios are defined in \autoref{tab:Astroparameters}: a \textsc{benchmark} model based on default parameters in \texttt{Zeus21}, a \textsc{less-constraining} model, and a \textsc{more-constraining} model. The constraining power of each scenario is determined by how prominently the baseline 21-cm signal stands out against experimental noise and how sensitive it is to additional energy injection, where this sensitivity can be understood using simple physical principles, discussed below. 
\begin{itemize}
    \item The \textsc{less-constraining} scenario combines a low flux of Lyman-$\alpha$ photons per baryon ($N_\alpha$), high X-ray luminosity ($L_X$), and a low minimum halo virial temperature ($T_{\rm vir}$). The combination of weak Lyman-$\alpha$ coupling and strong X-ray heating produces a shallow absorption trough in the global signal. This is because a low Lyman-$\alpha$ coupling implies a weak coupling between the spin temperature and the gas temperature, compounded with an increase in the gas temperature, so that the difference between $T_{\rm CMB}$ and $T_{k}$ is smaller. Moreover, a low $T_{\rm vir}$ allows star formation to occur in smaller halos and thus at earlier times, shifting the features in the signal to earlier times (lower frequencies) where the expected experimental sensitivity is weaker~\cite{Lopez-Honorez:2016sur}. The resulting  signal, shown in the top row of \autoref{fig:pess_fid_opt}, provides a poor baseline for constraining power with respect to any DM candidate that contributes to the heating of the IGM in this epoch.

    \item The \textsc{more-constraining} scenario assumes the opposite: high $N_\alpha$, low $L_X$, and a high $T_{\rm vir}$. Strong Lyman-$\alpha$ coupling and minimal X-ray heating create a deep, prominent absorption signal, and thus a larger power spectrum. A high $T_{\rm vir}$ causes star formation to occur only at later times in more massive halos, enhancing the power spectrum fluctuations. As shown in the bottom row of \autoref{fig:pess_fid_opt}, this scenario produces a deep global signal, and a power spectrum that is higher in magnitude than the other cases. As such, this scenario is more sensitive to deviations caused by exotic energy injection, allowing for more stringent constraints to be set, as we will quantify in \autoref{ssec:Results-Forecast_and_discussion}. 
    \item The \textsc{benchmark} scenario assumes an intermediate value of each of these parameters, and thus an intermediate constraining power compared to the other two scenarios. It was chosen based on the default parameters defined in {\tt Zeus21}, in line with the discussion presented in \autoref{ssec:Null_hypothesis}, and is roughly consistent with previous fiducial benchmarks~\cite{Park:2018ljd, Mena:2019nhm, Munoz:2023kkg, Cruz:2024fsv}. 
\end{itemize}

We characterize and visualize each of these three astrophysical scenarios in \autoref{fig:pess_fid_opt}. The figure shows the global signal (left panels) and the power spectrum (right panels) as a function of redshift, evaluated at a representative spatial scale $k = 0.15$ Mpc$^{-1}$. Each plot also shows the impact of a sub-dominant population of PBHs  with relative abundance $f_{\rm PBH}$ and assuming BHL accretion. We include the $1\sigma$ experimental sensitivity associated with the experimental configurations that we consider here, overlaid on the power spectrum plots. As shown in the plot, the sensitivity bands increasingly tighten the constraints for each fiducial scenario from top to bottom. We emphasize that the experimental sensitivity depends on fiducial model's power spectrum (see footnote~\ref{Footnote:sense}). This dependence explains the shift in the intersection of the {\tt SKA} sensitivity curve with the $x$-axis between the \textsc{benchmark} and \textsc{more-constraining} scenario, and it occurs since the fiducial power spectrum in the \textsc{benchmark} scenario is greater than the \textsc{more-constraining} in the narrow redshift range of $z \simeq 17 - 20$.

As shown, the \textsc{less-constraining}  scenario features a shallower global absorption dip and a suppressed power spectrum, especially at higher redshift. In this scenario, the $1\sigma$ experimental sensitivity band encompasses the signal, making it difficult to provide constraints on the PBH hypothesis, under our accretion modeling assumptions. This result generalizes to other exotic sources, where the experimental sensitivity for this fiducial scenario is of the same order as the signal, so distinguishing between different injections proves difficult, as has already been shown in the context of DM injections~\cite{Lopez-Honorez:2016sur, Decant:2024bpg}. In contrast, the \textsc{more-constraining} scenario plotted in the bottom row clearly shows a very pronounced absorption dip and enhanced fluctuations. As a consequence, even a small amount of exotic injection is expected to alter the signal in a detectable way. The \textsc{benchmark} scenario is also depicted in the middle row for comparison, which shows an intermediate regime for distinguishability. We quantify the implications of these three scenarios on the resulting PBH bounds in \autoref{ssec:Results-Forecast_and_discussion}. 

We remark here that by choosing these three fiducial astrophysical scenarios, we choose an agnostic approach as far as the correlations between the parameters are concerned. This choice avoids relying on a quantitative model that attempts to describe the underlying astrophysical quantities and their evolution amid large uncertainties, and in the absence of complementary observations that fully constrain the properties of the first stars.

\subsection{Uncertainty in accretion model}
\label{ssec:Results-PR-vs-BHL}

\begin{figure}
    \centering
    
    \begin{minipage}{0.49\textwidth}
        \centering
        \includegraphics[width=\linewidth]{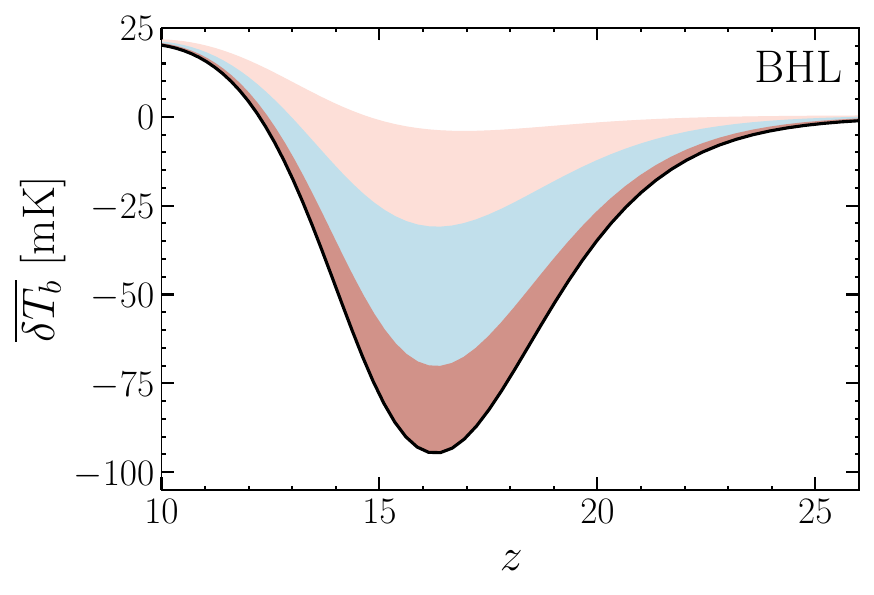}
    \end{minipage}
    \hfill
    \begin{minipage}{0.49\textwidth}
        \centering
        \includegraphics[width=\linewidth]{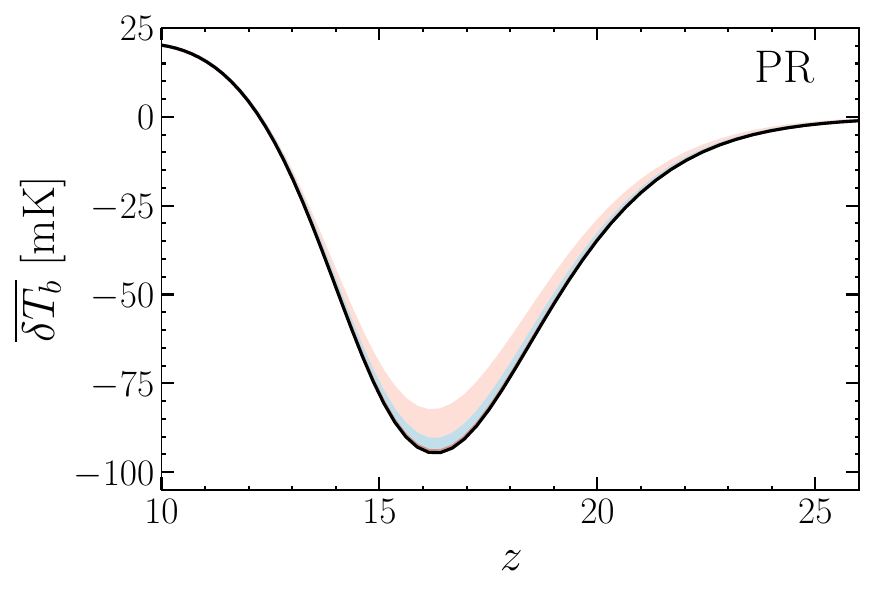}
    \end{minipage}
    
    \vspace{0.0cm} 
    
    \begin{minipage}{0.49\textwidth}
        \centering
        \includegraphics[width=\linewidth, trim={-1mm 0 0 0}]{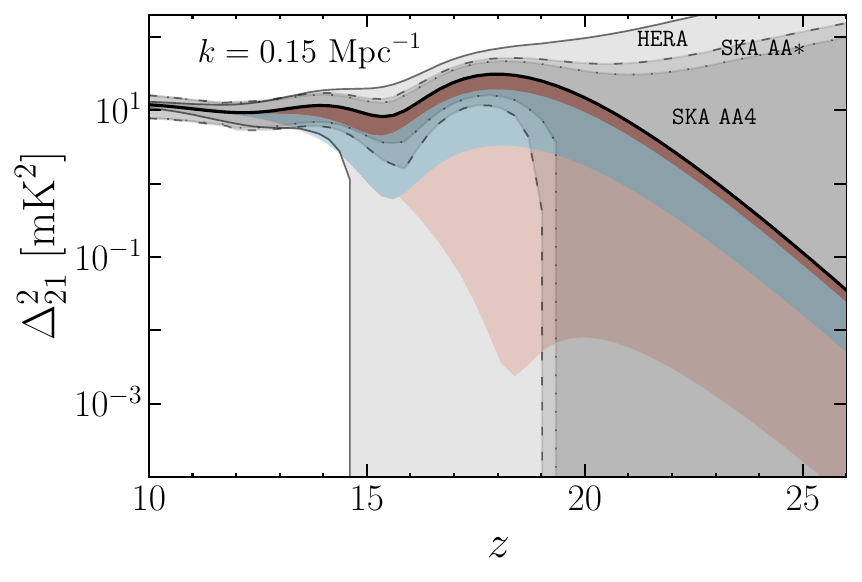}
    \end{minipage}
    \hfill
    \begin{minipage}{0.49\textwidth}
        \centering
        \includegraphics[width=\linewidth, trim={-1mm 0 0 0}]{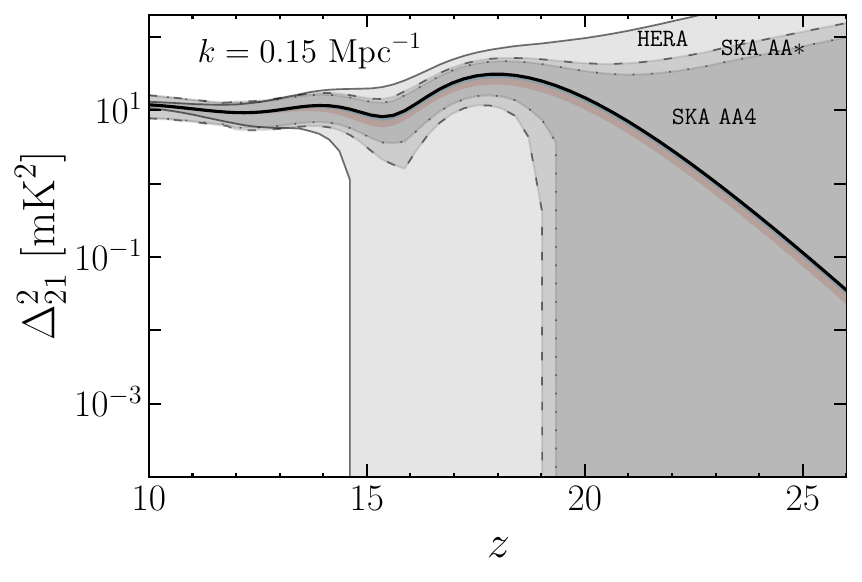}
    \end{minipage}
    
    \vspace{0.0cm}
    
    \begin{minipage}{0.49\textwidth}
        \centering
        \includegraphics[width=\linewidth, trim={-4mm 0 0 0}, clip]{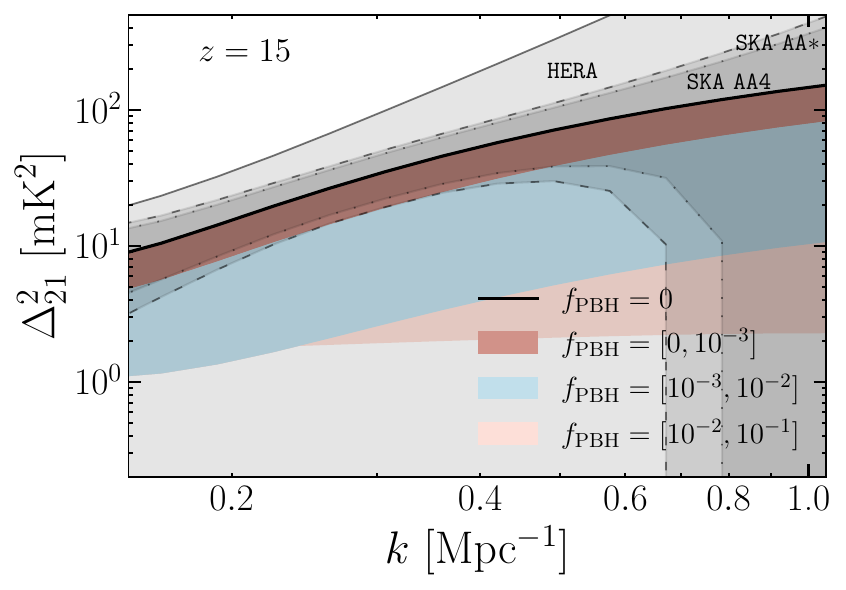}
    \end{minipage}
    \hfill
    \begin{minipage}{0.49\textwidth}
        \centering
        \includegraphics[width=\linewidth, trim={-4mm 0 0 0}, clip]{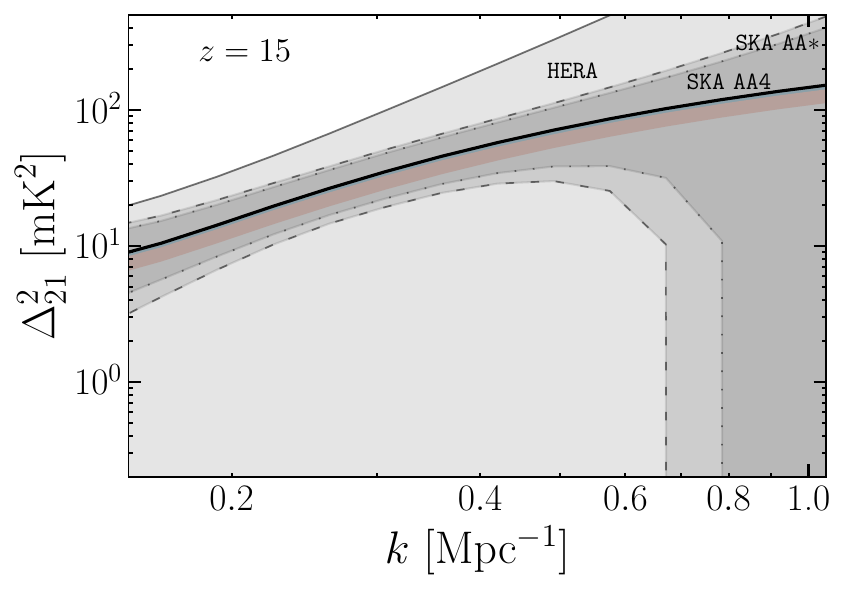}
    \end{minipage}
    
    \caption{Predictions of the 21-cm signal for PBHs for the BHL accretion model (left column) and the PR accretion model (right column), and for a PBH mass of ${M_{\rm PBH} = 10^2~M_\odot}$ and for a range of fractional PBH abundance $f_{\rm PBH}$. \textit{Top row:} impact on the global differential brightness temperature, $\delta T_b$, as a function of redshift. \textit{Middle row:} impact of accreting PBHs on the 21-cm power spectrum, $\Delta^2_{21}$, at a fixed scale $k = 0.15~\textrm{Mpc}^{-1}$, as a function of redshift. The $1\sigma$ sensitivities, $\sigma_{\rm exp}\left( z , k \right)$, are shown for the {\tt HERA} telescope, and for the {\tt SKA AA*} and {\tt SKA AA4} configurations, computed using {\tt 21cmSense}~\cite{Murray:2024the, Pober:2012zz, Pober:2013}. \textit{Bottom row:} same as the middle row, but now showing the power spectrum, $\Delta^2_{21}$, at a fixed redshift $z = 15$, as a function of $k$ instead. In each plot we assume the benchmark astrophysics scenario, corresponding to the middle row in \autoref{fig:pess_fid_opt}, and given in \autoref{tab:Astroparameters}. The legend provided in the bottom left panel applies to all panels.}
    \label{fig:BHL_vs_PR}
\end{figure}

The sensitivity of the 21-cm signal to accreting PBHs is determined by their model-dependent energy injection into the IGM. Of the two accretion models we consider, the BHL accretion model produces a substantially larger impact on the ionization fraction and gas temperature at late times, compared to the PR model, as shown in \autoref{fig:xe_Tk}. In \autoref{fig:BHL_vs_PR}, we show how altered properties of the IGM translate into the 21-cm global signal and power spectrum.

\begin{figure}[t]
    \centering
    \includegraphics[width=0.8\linewidth]{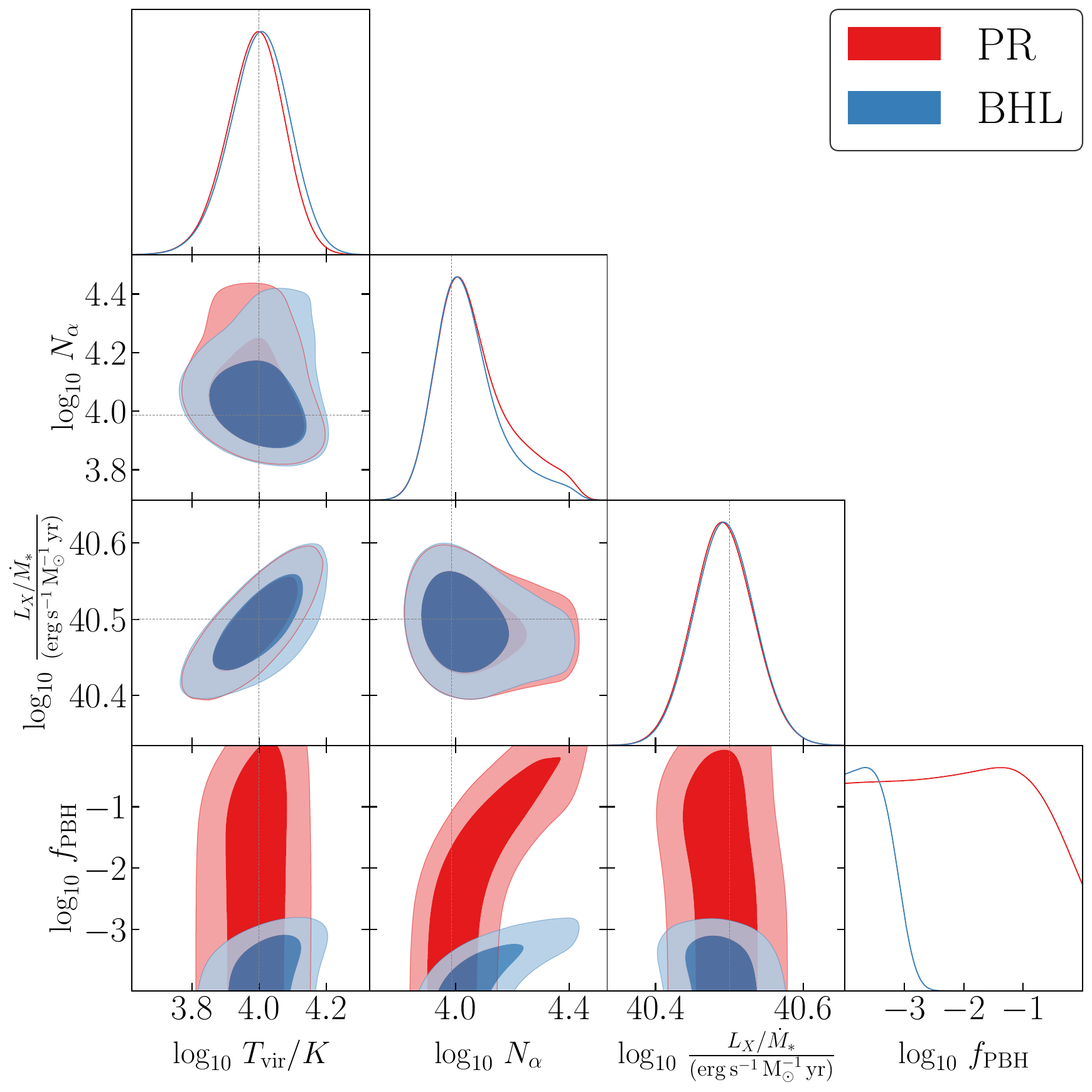}
    \caption{Comparison of the posteriors for our astrophysical and PBH parameters at 68\% and 95\% probability for a PBH mass of $M_{\rm PBH} = 10^2 \, M_{\odot}$ and for two accretion models, BHL (blue lines and contours) and PR (red lines and contours). The fiducial astrophysical model is assumed to be the \textsc{benchmark} scenario in \autoref{tab:Astroparameters}. }
    \label{fig:triangle}
\end{figure}

In particular, in the top row of \autoref{fig:BHL_vs_PR}  we show the global signal, where the impact of accreting PBHs under the BHL prescription is significantly larger. We also show in the middle and bottom panels of \autoref{fig:BHL_vs_PR} the impact of each accretion scenario on the power spectrum, including the $1\sigma$ experimental sensitivities of three different telescope configurations. The middle panel shows the power spectrum as a function of redshift at a fixed scale $k = 0.15 $~Mpc$^{-1}$, and the bottom panel shows the power spectrum as a function of $k$ at a fixed redshift $z = 15 $. It is apparent from these figures that BHL accretion can be much more constrained by the 21-cm power spectrum compared to PR accretion, by comparing the bands describing experimental sensitivities to the signal in the presence of accreting PBHs.

The difference in constraining power between the two accretion models is further described in \autoref{fig:triangle}, where we show the joint posteriors for our astrophysical and PBH parameters at 68\% and 95\% probability at $M_{\rm PBH} = 10^2  \ M_\odot$ for BHL and PR. The underlying fiducial model that we use for this scenario is the benchmark astrophysics scenario in \autoref{tab:Astroparameters}. As shown in the triangle plot, PR accreting PBHs are unconstrained, while the PBH abundance for BHL accreting PBHs is constrained by $f_{\rm PBH} \lesssim  10^{-2.6}$.

In the next section, we show how this substantial difference in the 21-cm power spectrum between accretion models translates into 21-cm bounds on the abundance of accreting PBHs.

\subsection{21-cm forecasts and discussion}
\label{ssec:Results-Forecast_and_discussion}

\begin{figure}[t]
    \centering
    \includegraphics[width=1\linewidth]{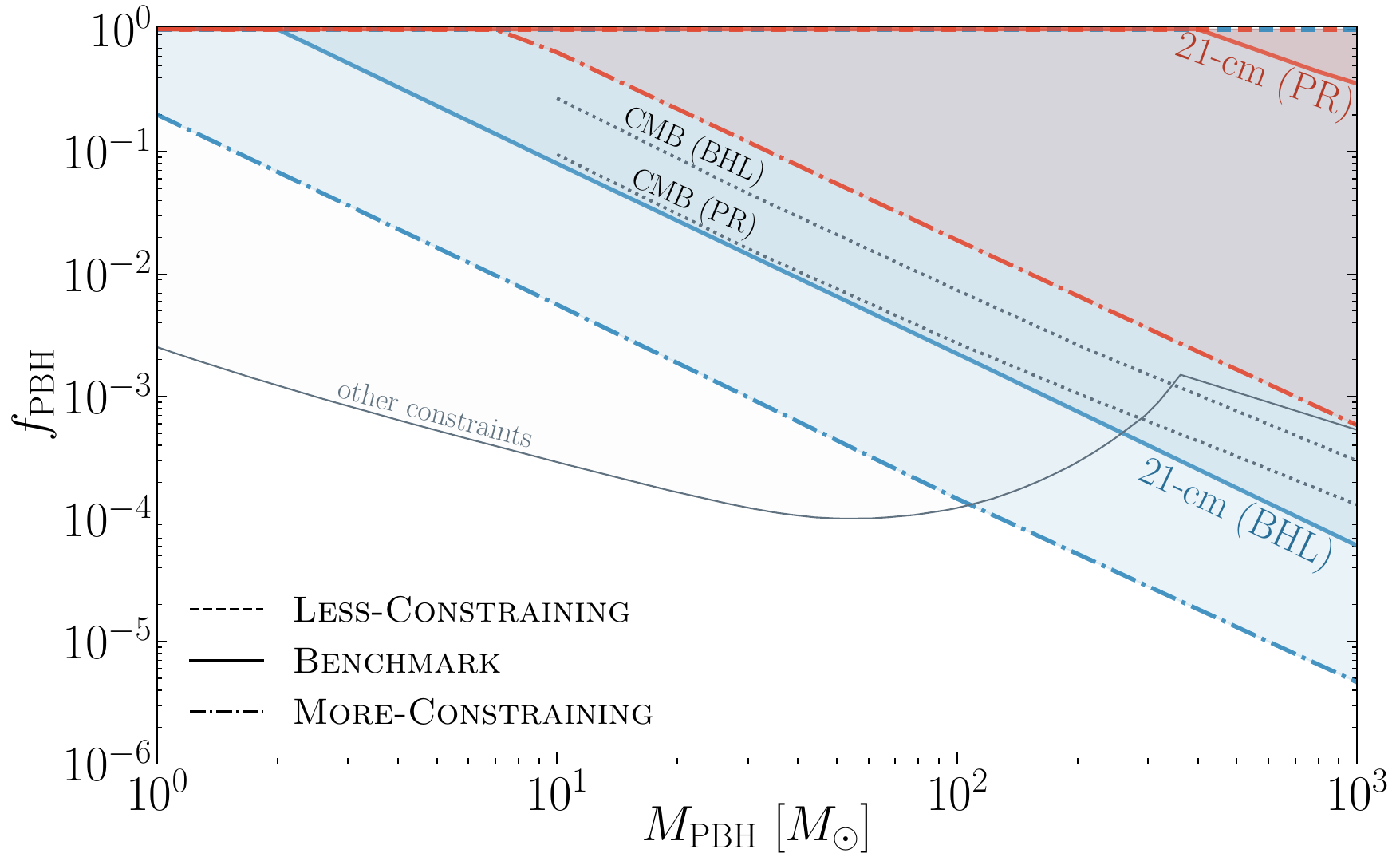}
    \caption{Sensitivity of {\tt SKA AA4} to $\fPBH$, the fractional contribution of PBHs to the DM abundance, at 95\% probability  versus the PBH mass, assuming a monochromatic PBH mass function. Results are shown as a function of the PBH mass, for two accretion models: BHL accretion (blue curves and blue shaded region) and PR accretion (red curves and red shaded region). We show results for three different fiducial astrophysical scenarios: benchmark (solid lines, labeled ``21-cm (BHL)'' and ``21-cm (PR)''), \textsc{more-constraining} (dash-dotted lines), and \textsc{less-constraining} scenarios (dashed lines); the astrophysics parameters for each of these three scenarios are provided in \autoref{tab:Astroparameters}. We remark that the \textsc{less-constraining} scenario is on the $x$-axis with $f_{\rm PBH} = 1$, for both models.  
    For comparison, we also show with a thin gray line (labeled ``other constraints'') the combined constraints from other probes, derived from  gravitational waves from merging events~\cite{Kavanagh:2018ggo, Chen:2019irf,Andres-Carcasona:2024wqk}, radio and X-ray observations~\cite{Manshanden:2018tze}, microlensing~\cite{Oguri:2017ock, Blaineau:2022nhy, Esteban-Gutierrez:2023qcz}, dynamical effects~\cite{Monroy-Rodriguez:2014ula, Brandt:2016aco}, and dwarf galaxy heating~\cite{Lu:2020bmd}. We also show the CMB constraints reported in ref.~\cite{Agius:2024ecw} (dotted gray lines), labeled for both the BHL and PR accretion models.{\protect\footnotemark} We make use of~\cite{Kavanagh2019} for plotting. }
\label{fig:bound}
\end{figure}

The forecasted constraints at 95\% probability on $f_{\rm PBH}$ are presented in \autoref{fig:bound}, which captures the two primary sources of uncertainty investigated in this work: the astrophysical modeling of the cosmic dawn and the PBH accretion prescription. The results demonstrate that these uncertainties have a significant impact on the projected sensitivity of 21-cm experiments.
\footnotetext{The analysis in ref.~\cite{Agius:2024ecw} was focused on the mass range $M_{\rm PBH} \geq10 \, M_{\odot}$, and the bound was not computed for lower masses.}
Our study reveals two key findings. First, the choice of the fiducial astrophysical scenario is critical. As shown by the spread between the \textsc{less-constraining} and \textsc{more-constraining} lines in \autoref{fig:bound}, the projected constraints vary by orders of magnitude. In the  \textsc{less-constraining} scenario (characterized by low $N_{\alpha}$, high $L_X$, and low $T_{\rm vir}$), 21-cm observations may offer no constraining power, whereas the \textsc{more-constraining}  scenario (characterized by high $N_{\alpha}$, low $L_X$, and high $T_{\rm vir}$) combined with the BHL accretion model could yield the most stringent constraints on stellar-mass PBHs. We claim that this result generalizes to any 21-cm constraint on additional heating and ionization of the IGM from exotic sources.

Secondly, the accretion model itself affects the forecast sensitivity. The PR model, which includes radiative feedback, yields constraints that are at least two orders of magnitude weaker than those from the BHL model, for each set of astrophysical parameters that we consider in this work. In particular, these bounds are weaker than existing, independent limits from other probes. Conversely, the widely-used BHL model suggests that 21-cm experiments could probe a large and unconstrained region of the PBH parameter space, particularly for $M_{\rm PBH} \gtrsim 10 \, M_{\odot}$, assuming a \textsc{more-constraining} astrophysical scenario. We remark here, that despite the substantial late time decrease in energy injection from the PR model compared to the BHL model, as shown in \autoref{fig:energy_dep}, where there is suppression in the accretion rate of $\sim$10 orders of magnitude at $z \simeq 20$, a bound can still be obtained. This is due to the significant energy deposition at early times, altering the evolution equations at late times. We can understand this effect as a modification of the initial conditions for later time evolution, effectively increasing the ionization and temperature floor. Thus, despite orders of magnitude lower injection at late times, the PR model still allows for some constraint to be placed, in our \textsc{more-constraining} astrophysical scenario. Counterintuitively, this implies that the 21-cm cosmological probe can be sensitive to early-time injections. 

Therefore, our study highlights that disambiguating the underlying astrophysical model is crucial for the interpretation of any future 21-cm signal, in the context of bounds on exotic energy injection. This strongly motivates independent observational efforts to constrain the properties of the first stars and galaxies. Indeed, a more robust approach than the one given here, where we have bracketed the uncertainty between the \textsc{less-constraining} and \textsc{more-constraining} scenario, would be to use future independent observations to derive posteriors for the astrophysical parameters, and then use these posteriors in a joint 21-cm forecast.\footnote{This would be a natural application of the methodology discussed, e.g., in refs.~\cite{Greig:2015qca, Greig:2016wjs, Park:2018ljd, Katz:2024ayw, Katz:2025sie}, to future data.}

\section{Conclusions and Outlook}
\label{sec:conclusions}

Observations of the redshifted 21-cm signal with next-generation radio interferometers, such as the {\tt SKA}~\cite{Mellema:2012ht}, are expected to open up a new window into the early Universe. The 21-cm signal is highly sensitive to the thermal and ionization history of the IGM and thus provides a potentially powerful probe of exotic sources of energy injection, such as accreting~\cite{Hektor:2018qqw, Mena:2019nhm, Yang:2021idt, Sun:2025ksr} or evaporating~\cite{Clark:2018ghm, Yang:2020egn, Halder:2021jiv, Halder:2021rbq, Mittal:2021egv, Natwariya:2021xki, Cang:2021owu, Saha:2021pqf, Mukhopadhyay:2022jqc, Yang:2022puh, Zhao:2024jad, Zhao:2025ddy, Sun:2025ksr} PBHs and DM annihilations~\cite{Furlanetto:2006wp, Valdes:2007cu, Evoli:2014pva, Lopez-Honorez:2016sur, DAmico:2018sxd, Liu:2018uzy, Qin:2023kkk, Facchinetti:2023slb, Novosyadlyj:2024bie, Zhao:2024jad, Bae:2025uqa, Natwariya:2025jlw, Sun:2025ksr} or decays~\cite{Furlanetto:2006wp, Valdes:2007cu, Poulin:2016anj, Clark:2018ghm, Liu:2018uzy, Mitridate:2018iag, Qin:2023kkk, Sun:2023acy, Novosyadlyj:2024bie, Zhao:2024jad, Zhao:2025ddy}. However, the potential for this signal to constrain new physics is fundamentally limited by the precision with which the standard astrophysical processes that govern the cosmic dawn can be modeled, introducing substantial uncertainty due to our incomplete knowledge of early galaxy formation and emission. 

In this work, we have presented a detailed forecast of the sensitivity of upcoming 21-cm power spectrum measurements to accreting PBHs, quantifying the impact of key systematic uncertainties. By exploring a range of plausible, yet observationally unconstrained, astrophysical scenarios, we demonstrated that the projected constraints on the PBH abundance, $f_{\rm PBH}$, can vary by several orders of magnitude. In a more-constraining astrophysical scenario, characterized by strong Lyman-$\alpha$ coupling, minimal X-ray heating, and a high mass threshold for star formation in halos, 21-cm observations could yield leading bounds on PBHs with mass $M_{\rm PBH} \gtrsim 10 \, M_{_\odot} $. Conversely, for a less-constraining scenario, with weaker Lyman-$\alpha$ coupling, stronger heating, and a lower halo mass threshold, upcoming experiments may offer no improvement over existing limits. Our results highlight that the significant impact of astrophysical uncertainties on constraints of exotic energy injection is a generic feature of any 21-cm forecast, with heating and ionization as the primary channels.

Furthermore, we investigated the impact of the PBH accretion model by comparing the standard Bondi-Hoyle-Lyttleton prescription with the more recent Park-Ricotti model, which incorporates a more comprehensive treatment of radiative feedback. Our analysis shows that the inclusion of feedback, which strongly suppresses the accretion rate at late times, relaxes the forecasted 21-cm constraints by at least two orders of magnitude across both constraining astrophysical scenarios. This result highlights the importance of accurate theoretical modeling of accretion in deriving robust constraints on PBHs from cosmological observables.

Finally, our results emphasize the importance of a multi-probe observational strategy. Synergies with complementary datasets, such as high-redshift UV luminosity functions from {\tt JWST}~\cite{Gardner:2023, Finkelstein2023} and future X-ray data~\cite{Cassano:2018zwm}, will be essential for mitigating the degeneracy between astrophysics and exotic physics~\cite{Fialkov:2016zyq, Lopez-Honorez:2016sur, Cohen:2016jbh, Park:2018ljd, Qin:2020xyh, Munoz:2021psm, Pochinda:2023uom, Katz:2024ayw, Decant:2024bpg, Katz:2025sie, Sims:2025hfm, Dhandha:2025dtn}, thereby enabling the 21-cm signal to reach its full potential as a probe of fundamental physics, and narrowing the range of uncertainties identified in this work.


\acknowledgments
We would like to thank Héctor Cruz and Julián Muñoz for helpful clarifications on {\tt Zeus21}, and James Davies for clarification regarding differences between versions of {\tt 21cmFAST}. We are grateful to Joshua Foster and Yitian Sun for prompt communication regarding their recent work~\cite{Sun:2025ksr}, and to Antonio J. Iovino for illuminating discussions regarding the work in ref.~\cite{Andres-Carcasona:2024wqk}. We extend our thanks also to  Gaétan Facchinetti, Hongwan Liu, Laura Lopez-Honorez, Andrei Mesinger, Diego Redigolo, Justus Schwagereit and Tracy Slatyer for useful discussions. We made use of the SOM Graviton computing infrastructure for this work. 

DA and SPR acknowledge support from the Generalitat Valenciana grants \newline CIGRIS/2021/054 and CIPROM/2022/36, respectively. DA and SPR are also supported by grant PID2023-151418NB-I00, which is funded by MCIU/AEI/10.13039/501100011033/ FEDER, UE, and by the European Union’s Horizon 2020 research and innovation programme under the Marie Skłodowska-Curie grants H2020-MSCA-ITN-2019/860881-HIDDeN and HORIZON-MSCA-2021-SE-01/101086085-ASYMMETRY. 
RE acknowledges support from DOE Grant DE-SC0025309 and Simons Investigator in Physics Awards~623940 and MPS-SIP-00010469. DG acknowledges support from the project ``Theoretical Astroparticle Physics (TAsP)'' funded by INFN. MV acknowledges support from the project ``Theoretical Particle Physics and Cosmology (TPPC)'' funded by INFN. DA thanks INFN Pisa, MIT CTP – a Leinweber Institute, YITP, and University of Amsterdam -- GRAPPA for their hospitality, where part of this work was carried out. 

This work made use of {\tt Zeus21}~\cite{Munoz:2023kkg}, {\tt NumPy}~\cite{Harris:2020xlr}, {\tt SciPy}~\cite{Virtanen:2019joe}, {\tt CLASS}~\cite{Blas:2011rf}, {\tt getdist}~\cite{Lewis:2019xzd}, and any dependencies.


\appendix

\section{Comparison to previous work}

\label{appendix:comparison_to_mena}

There is a previous study of future constraints on the PBH abundance using 21-cm cosmology~\cite{Mena:2019nhm}. In that work, BHL accreting PBHs were considered, and self-consistently implemented in {\tt 21cmFAST}~\cite{Mesinger:2010ne, Murray:2020trn}. The results that we show in \autoref{fig:bound} are consistent within approximately an order of magnitude of those presented in ref.~\cite{Mena:2019nhm}, and we use this section to outline the differences between our work and ref.~\cite{Mena:2019nhm} that could account for this difference. 

Firstly, ref.~\cite{Mena:2019nhm} used a different numerical code, {\tt 21cmFAST}~\cite{Mesinger:2010ne, Murray:2020trn}, for computing the 21-cm signal and power spectrum, and we make use of {\tt Zeus21}. While a comprehensive comparison has been carried out in refs.~\cite{Munoz:2023kkg, Cruz:2024fsv}, where differences in observables were shown to be within the 20\% level typically found between radiative transfer simulations and approximate semi-numeric models~\cite{Zahn:2010yw}, this comparison was carried out for more recent versions of {\tt 21cmFAST}. In particular, ref.~\cite{Mena:2019nhm} used {\tt 21cmFASTv1.2}, while this extensive testing was carried out comparing {\tt Zeus21} to {\tt 21cmFASTv3}. Between versions of {\tt 21cmFAST}, there are improvements that have a significant impact on the resulting power spectrum,\footnote{We remark that it has been pointed out that additional approximations made in {\tt 21cmFAST} may also lead to changes in the power spectra above the 20\% level~\cite{Flitter:2023rzv,Flitter:2024eay}.} where we give a non-exhaustive list below:
\begin{itemize}
 
    \item It was pointed out in {\tt Zeus21}~\cite{Munoz:2023kkg}, that older versions of {\tt 21cmFAST} had underestimated adiabatic fluctuations. This approximation can affect predictions of the temperature fluctuations to differ by approximately one order of magnitude (see Fig. 13 of ref.~\cite{Munoz:2023kkg} for a comparison, and ref.~\cite{Munoz2015} for further details). This has since been updated in {\tt 21cmFAST}, but was not available to the authors of ref.~\cite{Mena:2019nhm}. 
    
    \item Recent versions of {\tt 21cmFAST} introduced several updates between version {\tt 3.0} and {\tt 4.0}, leading to discrepancies exceeding $10\%$ in some cases. These changes include a correction to the adiabatic treatment (as described above), and an updated approach to the construction of density fields, where the cloud-in-cloud method is now preferred over the previous nearest-neighbor method.
    
\end{itemize}

Secondly, in ref.~\cite{Mena:2019nhm}, a minimal set of four astrophysics parameters was used. Namely, the UV ionization efficiency, the number of X-ray photons per solar mass, the minimum virial temperature of halos hosting galaxies, and the number of photons per baryon between Lyman-$\alpha$ and the Lyman limit. In our study, we use the same parameters, but reduce to a minimal model containing three parameters, keeping the UV ionization efficiency fixed, since that parameter does not have an analogue in {\tt Zeus21}. This is due to the UV ionization parameter defining the efficiency at which a grid point in {\tt 21cmFAST} is fully ionized, and since {\tt Zeus21} does not evolve grids, it is not used. We checked that the three parameters that we have chosen have the largest effect on the bound. At the level of statistics, this implies that these parameters have the strongest correlation with $f_{\rm PBH}$. We verified this ourselves, and this correlation is shown in the triangle plot in Fig.~13 of ref.~\cite{Mena:2019nhm}.

Thirdly, ref.~\cite{Mena:2019nhm} uses a different effective velocity in their computation compared to our use of a Maxwell-Boltzmann distribution. In particular, ref.~\cite{Mena:2019nhm} chooses to apply an effective velocity that assumes a radiative efficiency $\epsilon \propto \dot{M}^a_{\rm PBH}$ with $a = 1$, despite using a radiative efficiency with an arbitrary power-law dependence as in their ADAF model. This is discussed in Sec.~III.A of ref.~\cite{Mena:2019nhm}, and the general power-law dependence is shown in their eq.~(3.6). Despite all the caveats discussed there, it was chosen to follow previous works and adopt the expression for the effective velocity assuming the $a = 1$ result. In taking a Maxwell-Boltzmann velocity distribution as described in \autoref{sssec:Environment_properties}, we are using the general power law dependence and not fixing $a = 1$. We checked that the effect of using $a = 1$, leads to a larger injection and thus a more stringent constraint by a factor of $\sim 2$.

Fourthly, in the three astrophysical scenarios that we consider, the benchmark case is most similar, but not an exact reproduction of the global signal and power spectrum in ref.~\cite{Mena:2019nhm}. Finding a set of parameters in {\tt Zeus21} that exactly reproduced that global signal and power spectrum proved unsuccessful, given the reasons stipulated above. Phenomenologically, observe that in figure~3 of ref.~\cite{Mena:2019nhm}, the benchmark global signal has an absorption trough at $z \sim 15$, with depth $ \overline{\delta T_b} \sim -150$~mK. The corresponding trough for our fiducial scenario has an absorption trough at $z\sim 16$, with depth $\overline{\delta T_b}\sim -80$~mK. Since experiments provide more sensitivity at earlier times, and the depth of this trough translates to the amplitude of the power spectrum (see \autoref{fig:pess_fid_opt} for visualization of both effects), we would expect the previous study to correspond to more stringent bounds, which is actually reported there. Those limits roughly correspond to our optimistic scenario, with the caveat that we use a different velocity distribution, as detailed in the previous paragraph. 

Finally, since {\tt Zeus21} breaks down at $z \sim 10$~\cite{Munoz:2023kkg},\footnote{Since {\tt Zeus21} does not evolve reionization, the 21-cm signal in this setup follows the underlying density field by construction.} we restrict to redshifts $z > 10$, with explicit binning given in \autoref{ssec:statistic}. In contrast, ref.~\cite{Mena:2019nhm} used 9 log-spaced measurements in redshift in the interval $z \sim 8.3 - 19.5$. As described above, experiments have increased sensitivity at later times, so including these redshift bins leads to an increase in the constraining power, further explaining those more stringent bounds for the BHL model.

\section{Fixed astrophysical parameters}
\label{App:Fixed Astro Params}

\begin{table}[t]
\centering
\begin{tabular}{|ll|ll|ll|}
\hline
Parameter & Value & Parameter & Value & Parameter & Value \\
\hline
$\alpha_\star$ & 0.5 & $M_{\rm esc}$ & $10^{10}~M_\odot$ & \bm{$N_\alpha$} & \textbf{9690} \\
$\beta_\star$ & $-0.5$ & $\alpha_{\rm esc}$ & 0.0 & $N_{\rm LW}$ & 6200 \\
$\epsilon_\star$ & $10^{-1}$ & \bm{$\log_{10}L_{X} / \dot{M}_*$} & \textbf{40.5} $\boldsymbol{{\rm erg} \, {\rm s}^{-1} \,M_\odot^{-1} \,{\rm yr}}$ & $A_{\rm LW}$ & 2.0 \\
$M_{\rm pivot}$ & $3 \times 10^{11}~M_\odot$ & $E_{0,X}$ & 500~eV & $\beta_{\rm LW}$ & 0.6 \\
\bm{$T_{\rm vir}$} & \textbf{10\textsuperscript{4}~K} & $E_{\rm max,X}$ & 2000~eV & $A_{\rm vcb}$ & 1.0 \\
$f_{\rm esc,0}$ & $10^{-1}$ & $\alpha_X$ & $-1.0$ & $\beta_{\rm vcb}$ & 1.8 \\
\hline
\end{tabular}
\caption{Parameters for Pop II sources in the fiducial {\tt Zeus21} model. Parameters that vary for different astrophysical scenarios are shown in bold. See \autoref{tab:Astroparameters} for the ranges that we vary over, and ref.~\cite{Cruz:2024fsv} for definitions of these parameters.}
\label{tab:All_Astroparams}
\end{table}

Modeling the cosmic dawn and the formation of the first stars necessitates the choice of many astrophysical parameters, making various assumptions on the properties of the first stars. In \autoref{tab:All_Astroparams}, we provide a list of the Pop II parameters that we assume to be fixed in this work. Furthermore, we list other flags in {\tt Zeus21} that we use, to allow for reproducibility of our results. In particular, we use the flags: {\tt USE\_RELATIVE\_VELOCITIES = True}, {\tt USE\_LW\_FEEDBACK = True}, { \tt kmax\_CLASS = 500}, and {\tt zmax\_CLASS =  50}. Furthermore, we use the {\tt Planck2018} parameters~\cite{Planck:2018vyg}: $\Omega_c = 0.11933$, $\Omega_b =  0.02242$, $h =  0.6766$, $ A_s= \exp(3.047) \times 10^{-10}$, $n_s =  0.9665$ and $z_{\rm reio} =  7.82$.

\bibliographystyle{JHEP}
\bibliography{bib_21-cm.bib}

@article{Furlanetto:2006jb,
    author = "Furlanetto, Steven and Oh, S. Peng and Briggs, Frank",
    title = "{Cosmology at low frequencies: The 21 cm transition and the high-redshift Universe}",
    eprint = "astro-ph/0608032",
    archivePrefix = "arXiv",
    doi = "10.1016/j.physrep.2006.08.002",
    journal = "Phys. Rept.",
    volume = "433",
    pages = "181--301",
    year = "2006"
}

@article{Pritchard:2011xb,
    author = "Pritchard, Jonathan R. and Loeb, Abraham",
    title = "{21-cm cosmology in the 21st century}",
    eprint = "1109.6012",
    archivePrefix = "arXiv",
    primaryClass = "astro-ph.CO",
    doi = "10.1088/0034-4885/75/8/086901",
    journal = "Rept. Prog. Phys.",
    volume = "75",
    pages = "086901",
    year = "2012"
}

@incollection{Furlanetto:2015apc,
	author      = "Furlanetto, Steven R.",
	title          = "{The 21-cm line as a probe of reionization}",
	editor       = "Mesinger, Andrei",
	booktitle   = "Understanding the Epoch of Cosmic Reionization: Challenges and Progress",
	publisher  = "Springer International Publishing",
	volume     = "423",
	year          = "2016",	
	pages       = "247-280",
	eprint         = "1511.01131",
	archivePrefix  = "arXiv",
	primaryClass   = "astro-ph.CO",
	SLACcitation   = "%%CITATION = ARXIV:1511.01131;%%"
}

@phdthesis{Villanueva-Domingo:2021vbi,
    author = "Villanueva-Domingo, Pablo",
    title = "{Shedding light on dark matter through 21 cm cosmology and reionization constraints}",
    eprint = "2112.08201",
    archivePrefix = "arXiv",
    primaryClass = "astro-ph.CO",
    school = "U. Valencia",
    year = "2021"
}

@article{Poulin:2017bwe,
    author = "Poulin, Vivian and Serpico, Pasquale D. and Calore, Francesca and Clesse, Sebastien and Kohri, Kazunori",
    title = "{CMB bounds on disk-accreting massive primordial black holes}",
    eprint = "1707.04206",
    archivePrefix = "arXiv",
    primaryClass = "astro-ph.CO",
    reportNumber = "LAPTH-022-17, TTK-17-22",
    doi = "10.1103/PhysRevD.96.083524",
    journal = "Phys. Rev. D",
    volume = "96",
    number = "8",
    pages = "083524",
    year = "2017"
}

@article{Jangra:2024sif,
    author = "Jangra, Pratibha and Gaggero, Daniele and Kavanagh, Bradley J. and Diego, J. M.",
    title = "{The cosmic history of primordial black hole accretion and its uncertainties}",
    eprint = "2412.11921",
    archivePrefix = "arXiv",
    primaryClass = "astro-ph.CO",
    doi = "10.1088/1475-7516/2025/08/006",
    journal = "JCAP",
    volume = "08",
    pages = "006",
    year = "2025"
}

@article{Agius:2024ecw,
    author = "Agius, Dominic and Essig, Rouven and Gaggero, Daniele and Scarcella, Francesca and Suczewski, Gregory and Valli, Mauro",
    title = "{Feedback in the dark: a critical examination of CMB bounds on primordial black holes}",
    eprint = "2403.18895",
    archivePrefix = "arXiv",
    primaryClass = "hep-ph",
    doi = "10.1088/1475-7516/2024/07/003",
    journal = "JCAP",
    volume = "07",
    pages = "003",
    year = "2024"
}

@article{Mellema:2012ht,
    author = "Mellema, Garrelt and others",
    title = "{Reionization and the Cosmic Dawn with the Square Kilometre Array}",
    eprint = "1210.0197",
    archivePrefix = "arXiv",
    primaryClass = "astro-ph.CO",
    doi = "10.1007/s10686-013-9334-5",
    journal = "Exper. Astron.",
    volume = "36",
    pages = "235--318",
    year = "2013"
}

@article{Shchekinov:2006eb,
    author = "Shchekinov, Yuri A. and Vasiliev, E. O.",
    title = "{Particle decay in the early {U}niverse: predictions for 21 cm}",
    eprint = "astro-ph/0604231",
    archivePrefix = "arXiv",
    doi = "10.1111/j.1365-2966.2007.11715.x",
    journal = "Mon. Not. Roy. Astron. Soc.",
    volume = "379",
    pages = "1003--1010",
    year = "2007"
}

@article{Valdes:2007cu,
    author = "Valdes, Marcos and Ferrara, A. and Mapelli, M. and Ripamonti, E.",
    title = "{Constraining DM through 21 cm observations}",
    eprint = "astro-ph/0701301",
    archivePrefix = "arXiv",
    doi = "10.1111/j.1365-2966.2007.11594.x",
    journal = "Mon. Not. Roy. Astron. Soc.",
    volume = "377",
    pages = "245--252",
    year = "2007"
}

@article{HERA:2016,
    author = "DeBoer, David R. and others",
    title = "{Hydrogen Epoch of Reionization Array (HERA)}",
    eprint = "1606.07473",
    archivePrefix = "arXiv",
    primaryClass = "astro-ph.IM",
    doi = "10.1088/1538-3873/129/974/045001",
    journal = "Publ. Astron. Soc. Pac.",
    volume = "129",
    number = "974",
    pages = "045001",
    year = "2017"
}

@article{Park:2011rf,
    author = "Park, KwangHo and Ricotti, Massimo",
    title = "{Accretion onto black holes from large scales regulated by radiative feedback. II. Growth rate and duty cycle}",
    eprint = "1110.4634",
    archivePrefix = "arXiv",
    primaryClass = "astro-ph.CO",
    doi = "10.1088/0004-637X/747/1/9",
    journal = "Astrophys. J.",
    volume = "747",
    pages = "9",
    year = "2012"
}

@article{Park:2012cr,
    author = "Park, KwangHo and Ricotti, Massimo",
    title = "{Accretion onto black holes from large scales regulated by radiative feedback. III. Enhanced luminosity of intermediate mass black holes moving at supersonic speeds}",
    eprint = "1211.0542",
    archivePrefix = "arXiv",
    primaryClass = "astro-ph.CO",
    doi = "10.1088/0004-637X/767/2/163",
    journal = "Astrophys. J.",
    volume = "767",
    pages = "163",
    year = "2013"
}

@article{Park:2010yh,
    author = "Park, KwangHo and Ricotti, Massimo",
    title = "{Accretion onto intermediate mass black holes regulated by radiative feedback I. Parametric study for spherically symmetric accretion}",
    eprint = "1006.1302",
    archivePrefix = "arXiv",
    primaryClass = "astro-ph.CO",
    doi = "10.1088/0004-637X/739/1/2",
    journal = "Astrophys. J.",
    volume = "739",
    pages = "2",
    year = "2011"
}

@article{Pober:2012zz,
    author = "Pober, Jonathan C. and Parsons, Aaron R. and DeBoer, David R. and McDonald, Patrick and McQuinn, Matthew and Aguirre, James E. and Ali, Zaki and Bradley, Richard F. and Chang, Tzu-Ching and Morales, Miguel F.",
    title = "{The baryon acoustic oscillation broadband and broad-beam array: design overview and sensitivity forecasts}",
    eprint = "1210.2413",
    archivePrefix = "arXiv",
    primaryClass = "astro-ph.CO",
    doi = "10.1088/0004-6256/145/3/65",
    journal = "Astron. J.",
    volume = "145",
    pages = "65",
    year = "2013"
}

@article{Murray:2024the,
    author = "Murray, Steven G. and Pober, Jonathan and Kolopanis, Matthew",
    title = "{21cmSense v2: a modular, open-source 21cm sensitivity calculator}",
    eprint = "2406.02415",
    archivePrefix = "arXiv",
    primaryClass = "astro-ph.CO",
    doi = "10.21105/joss.06501",
    journal = "J. Open Source Softw.",
    volume = "9",
    pages = "6501",
    year = "2024"
}

@article{Pober:2013,
    author = "Pober, Jonathan C. and others",
    title = "{What next-generation 21 cm power spectrum measurements can teach us about the epoch of reionization}",
    eprint = "1310.7031",
    archivePrefix = "arXiv",
    primaryClass = "astro-ph.CO",
    doi = "10.1088/0004-637X/782/2/66",
    journal = "Astrophys. J.",
    volume = "782",
    pages = "66",
    year = "2014"
}

@article{Oh:2001,
    author = "Oh, S. Peng and Haiman, Zoltan",
    title = "{Second-generation objects in the universe: radiative cooling and collapse of halos with virial temperatures above 10\textasciicircum{}4 kelvin}",
    eprint = "astro-ph/0108071",
    archivePrefix = "arXiv",
    doi = "10.1086/339393",
    journal = "Astrophys. J.",
    volume = "569",
    pages = "558",
    year = "2002"
}

@article{Evoli:2014pva,
    author = "Evoli, Carmelo and Mesinger, Andrei and Ferrara, Andrea",
    title = "{Unveiling the nature of dark matter with high redshift 21 cm line experiments}",
    eprint = "1408.1109",
    archivePrefix = "arXiv",
    primaryClass = "astro-ph.HE",
    doi = "10.1088/1475-7516/2014/11/024",
    journal = "JCAP",
    volume = "11",
    pages = "024",
    year = "2014"
}

@article{Lopez-Honorez:2016sur,
    author = "Lopez-Honorez, Laura and Mena, Olga and Molin{\'e}, {\'A}ngeles and Palomares-Ruiz, Sergio and Vincent, Aaron C.",
    title = "{The 21 cm signal and the interplay between dark matter annihilations and astrophysical processes}",
    eprint = "1603.06795",
    archivePrefix = "arXiv",
    primaryClass = "astro-ph.CO",
    reportNumber = "CFTP-16-007, IFIC-16-16, IPPP-16-20",
    doi = "10.1088/1475-7516/2016/08/004",
    journal = "JCAP",
    volume = "08",
    pages = "004",
    year = "2016"
}

@article{Xie:2012rs,
    author = "Xie, Fu-Guo and Yuan, Feng",
    title = "{The radiative efficiency of hot accretion flows}",
    eprint = "1207.3113",
    archivePrefix = "arXiv",
    primaryClass = "astro-ph.HE",
    doi = "10.1111/j.1365-2966.2012.22030.x",
    journal = "Mon. Not. Roy. Astron. Soc.",
    volume = "427",
    pages = "1580",
    year = "2012"
}

@article{Stocker:2018avm,
    author = {St{\"o}cker, Patrick and Kr{\"a}mer, Michael and Lesgourgues, Julien and Poulin, Vivian},
    title = "{Exotic energy injection with ExoCLASS: Application to the Higgs portal model and evaporating black holes}",
    eprint = "1801.01871",
    archivePrefix = "arXiv",
    primaryClass = "astro-ph.CO",
    doi = "10.1088/1475-7516/2018/03/018",
    journal = "JCAP",
    volume = "03",
    pages = "018",
    year = "2018"
}

@article{Slatyer:2015kla,
    author = "Slatyer, Tracy R.",
    title = "{Indirect dark matter signatures in the cosmic dark ages II. Ionization, heating and photon production from arbitrary energy injections}",
    eprint = "1506.03812",
    archivePrefix = "arXiv",
    primaryClass = "astro-ph.CO",
    reportNumber = "MIT-CTP-4683",
    doi = "10.1103/PhysRevD.93.023521",
    journal = "Phys. Rev. D",
    volume = "93",
    number = "2",
    pages = "023521",
    year = "2016"
}

@article{HERA:2021noe,
    author = "Abdurashidova, Zara and others",
    collaboration = "HERA",
    title = "{HERA phase I limits on the cosmic 21 cm signal: constraints on astrophysics and cosmology during the epoch of reionization}",
    eprint = "2108.07282",
    archivePrefix = "arXiv",
    primaryClass = "astro-ph.CO",
    doi = "10.3847/1538-4357/ac2ffc",
    journal = "Astrophys. J.",
    volume = "924",
    number = "2",
    pages = "51",
    year = "2022"
}

@article{Lehmer:2012ak,
    author = "Lehmer, B. D. and others",
    title = "{The 4 Ms Chandra deep field-south number counts apportioned by source class: pervasive active galactic nuclei and the ascent of normal galaxies}",
    eprint = "1204.1977",
    archivePrefix = "arXiv",
    primaryClass = "astro-ph.CO",
    doi = "10.1088/0004-637X/752/1/46",
    journal = "Astrophys. J.",
    volume = "752",
    pages = "46",
    year = "2012"
}

@article{HERA:2022wmy,
    author = "Abdurashidova, Zara and others",
    collaboration = "HERA",
    title = "{Improved constraints on the 21 cm EoR power spectrum and the X-ray heating of the IGM with HERA phase I observations}",
    eprint = "2210.04912",
    archivePrefix = "arXiv",
    primaryClass = "astro-ph.CO",
    doi = "10.3847/1538-4357/acaf50",
    journal = "Astrophys. J.",
    volume = "945",
    number = "2",
    pages = "124",
    year = "2023"
}

@article{Delos:2024poq,
    author = "Delos, M. Sten and Rantala, Antti and Young, Sam and Schmidt, Fabian",
    title = "{Structure formation with primordial black holes: collisional dynamics, binaries, and gravitational waves}",
    eprint = "2410.01876",
    archivePrefix = "arXiv",
    primaryClass = "astro-ph.CO",
    doi = "10.1088/1475-7516/2024/12/005",
    journal = "JCAP",
    volume = "12",
    pages = "005",
    year = "2024"
}

@article{Koopmans:2015sua,
    author = "Koopmans, L. V. E. and others",
    editor = "Bourke, Tyler L. and others",
    title = "{The Cosmic Dawn and Epoch of Reionization with the Square Kilometre Array}",
    eprint = "1505.07568",
    archivePrefix = "arXiv",
    primaryClass = "astro-ph.CO",
    doi = "10.22323/1.215.0001",
    journal = "PoS",
    volume = "AASKA14",
    pages = "001",
    year = "2015"
}

@article{Planck:2018vyg,
    author = "Aghanim, N. and others",
    collaboration = "Planck",
    title = "{Planck 2018 results. VI. Cosmological parameters}",
    eprint = "1807.06209",
    archivePrefix = "arXiv",
    primaryClass = "astro-ph.CO",
    doi = "10.1051/0004-6361/201833910",
    journal = "Astron. Astrophys.",
    volume = "641",
    pages = "A6",
    year = "2020",
    note = "[Erratum: Astron.Astrophys. 652, C4 (2021)]"
}

@article{Liu2019,
    author = "Liu, Hongwan and Ridgway, Gregory W. and Slatyer, Tracy R.",
    title = "{Code package for calculating modified cosmic ionization and thermal histories with dark matter and other exotic energy injections}",
    eprint = "1904.09296",
    archivePrefix = "arXiv",
    primaryClass = "astro-ph.CO",
    doi = "10.1103/PhysRevD.101.023530",
    journal = "Phys. Rev. D",
    volume = "101",
    number = "2",
    pages = "023530",
    year = "2020"
}

@Article{Shull:1985,
	author   = "Shull, J. M. and van Steenberg, M. E.",
	title   = "{X-ray secondary heating and ionization in quasar emission-line clouds}",
	journal   = "Astrophys. J.",
	year   = "1985",
	volume   = "298",
	pages   = "268-274",
	doi   = "10.1086/163605"
}

@article{Bird:2016dcv,
    author = {Bird, Simeon and others},
    title = "{Did LIGO detect dark matter?}",
    eprint = "1603.00464",
    archivePrefix = "arXiv",
    primaryClass = "astro-ph.CO",
    doi = "10.1103/PhysRevLett.116.201301",
    journal = "Phys. Rev. Lett.",
    volume = "116",
    number = "20",
    pages = "201301",
    year = "2016"
}

@article{LIGOScientific:2016aoc,
    author = "Abbott, B. P. and others",
    collaboration = "LIGO Scientific, Virgo",
    title = "{Observation of gravitational waves from a binary black hole merger}",
    eprint = "1602.03837",
    archivePrefix = "arXiv",
    primaryClass = "gr-qc",
    reportNumber = "LIGO-P150914",
    doi = "10.1103/PhysRevLett.116.061102",
    journal = "Phys. Rev. Lett.",
    volume = "116",
    number = "6",
    pages = "061102",
    year = "2016"
}

@article{Mason:2022obt,
    author = "Mason, Charlotte A. and Mu{\~n}oz, Julian B. and Greig, Bradley and Mesinger, Andrei and Park, Jaehong",
    title = "{21cmfish: Fisher-matrix framework for fast parameter forecasts from the cosmic 21-cm signal}",
    eprint = "2212.09797",
    archivePrefix = "arXiv",
    primaryClass = "astro-ph.CO",
    doi = "10.1093/mnras/stad2145",
    journal = "Mon. Not. Roy. Astron. Soc.",
    volume = "524",
    number = "3",
    pages = "4711--4728",
    year = "2023"
}

@article{Decant:2024bpg,
    author = "Decant, Quentin and Dimitriou, Androniki and Honorez, Laura Lopez and Zaldivar, Bryan",
    title = "{Simulation-based inference on warm dark matter from HERA forecasts}",
    eprint = "2412.10310",
    archivePrefix = "arXiv",
    primaryClass = "astro-ph.CO",
    reportNumber = "ULB-TH/24-17",
    doi = "10.1088/1475-7516/2025/07/004",
    journal = "JCAP",
    volume = "07",
    pages = "004",
    year = "2025"
}

@article{Blas:2011rf,
    author = "Blas, Diego and Lesgourgues, Julien and Tram, Thomas",
    title = "{The Cosmic Linear Anisotropy Solving System (CLASS) II: approximation schemes}",
    eprint = "1104.2933",
    archivePrefix = "arXiv",
    primaryClass = "astro-ph.CO",
    reportNumber = "CERN-PH-TH-2011-082, LAPTH-010-11",
    doi = "10.1088/1475-7516/2011/07/034",
    journal = "JCAP",
    volume = "07",
    pages = "034",
    year = "2011"
}

@article{Kavanagh:2018ggo,
    author = "Kavanagh, Bradley J. and Gaggero, Daniele and Bertone, Gianfranco",
    title = "{Merger rate of a subdominant population of primordial black holes}",
    eprint = "1805.09034",
    archivePrefix = "arXiv",
    primaryClass = "astro-ph.CO",
    doi = "10.1103/PhysRevD.98.023536",
    journal = "Phys. Rev. D",
    volume = "98",
    number = "2",
    pages = "023536",
    year = "2018"
}

@article{Niemeyer:1997mt,
    author = "Niemeyer, Jens C. and Jedamzik, K.",
    title = "{Near-critical gravitational collapse and the initial mass function of primordial black holes}",
    eprint = "astro-ph/9709072",
    archivePrefix = "arXiv",
    doi = "10.1103/PhysRevLett.80.5481",
    journal = "Phys. Rev. Lett.",
    volume = "80",
    pages = "5481--5484",
    year = "1998"
}

@article{Niemeyer:1999ak,
    author = "Niemeyer, Jens C. and Jedamzik, K.",
    title = "{Dynamics of primordial black hole formation}",
    eprint = "astro-ph/9901292",
    archivePrefix = "arXiv",
    doi = "10.1103/PhysRevD.59.124013",
    journal = "Phys. Rev. D",
    volume = "59",
    pages = "124013",
    year = "1999"
}

@article{Musco:2004ak,
    author = "Musco, Ilia and Miller, John C. and Rezzolla, Luciano",
    title = "{Computations of primordial black hole formation}",
    eprint = "gr-qc/0412063",
    archivePrefix = "arXiv",
    doi = "10.1088/0264-9381/22/7/013",
    journal = "Class. Quant. Grav.",
    volume = "22",
    pages = "1405--1424",
    year = "2005"
}

@article{Escriva:2020tak,
    author = "Escriv{\`a}, Albert and Germani, Cristiano and Sheth, Ravi K.",
    title = "{Analytical thresholds for black hole formation in general cosmological backgrounds}",
    eprint = "2007.05564",
    archivePrefix = "arXiv",
    primaryClass = "gr-qc",
    reportNumber = "ICCUB-20-016",
    doi = "10.1088/1475-7516/2021/01/030",
    journal = "JCAP",
    volume = "01",
    pages = "030",
    year = "2021"
}

@article{Musco:2020jjb,
    author = "Musco, Ilia and De Luca, Valerio and Franciolini, Gabriele and Riotto, Antonio",
    title = "{Threshold for primordial black holes. II. A simple analytic prescription}",
    eprint = "2011.03014",
    archivePrefix = "arXiv",
    primaryClass = "astro-ph.CO",
    doi = "10.1103/PhysRevD.103.063538",
    journal = "Phys. Rev. D",
    volume = "103",
    number = "6",
    pages = "063538",
    year = "2021"
}

@article{Carr:1975qj,
    author = "Carr, Bernard J.",
    title = "{The primordial black hole mass spectrum}",
    doi = "10.1086/153853",
    journal = "Astrophys. J.",
    volume = "201",
    pages = "1--19",
    year = "1975"
}

@article{Ivanov:1994pa,
	author         = "Ivanov, P. and Naselsky, P. and Novikov, I.",
	title          = "{Inflation and primordial black holes as dark matter}",
	journal        = "Phys. Rev.",
	volume         = "D50",
	year           = "1994",
	pages          = "7173-7178",
	doi            = "10.1103/PhysRevD.50.7173",
	reportNumber   = "NORDITA-94-12-A",
	SLACcitation   = "%%CITATION = PHRVA,D50,7173;%%"
}

@article{Yokoyama:1995ex,
	author         = "Yokoyama, Junichi",
	title          = "{Formation of MACHO primordial black holes in inflationary cosmology}",
	journal        = "Astron. Astrophys.",
	volume         = "318",
	year           = "1997",
	pages          = "673",
	eprint         = "astro-ph/9509027",
	archivePrefix  = "arXiv",
	primaryClass   = "astro-ph",
	reportNumber   = "YITP-U-95-26",
	SLACcitation   = "%%CITATION = ASTRO-PH/9509027;%%"
}

@article{Yokoyama:1998qw,
	author         = "Yokoyama, Jun'ichi",
	title          = "{Formation of primordial black holes in the inflationary universe}",
	journal        = "Phys. Rept.",
	volume         = "307",
	year           = "1998",
	pages          = "133-139",
	doi            = "10.1016/S0370-1573(98)00044-1",
	reportNumber   = "YITP-98-40",
	SLACcitation   = "%%CITATION = PRPLC,307,133;%%"
}

@article{Shibata:1999zs,
	author         = "Shibata, Masaru and Sasaki, Misao",
	title          = "{Black hole formation in the Friedmann universe: Formulation and computation in numerical relativity}",
	journal        = "Phys. Rev.",
	volume         = "D60",
	year           = "1999",
	pages          = "084002",
	doi            = "10.1103/PhysRevD.60.084002",
	eprint         = "gr-qc/9905064",
	archivePrefix  = "arXiv",
	primaryClass   = "gr-qc",
	reportNumber   = "OU-TAP-93",
	SLACcitation   = "%%CITATION = GR-QC/9905064;%%"
}

@article{Young:2014ana,
	author         = "Young, Sam and Byrnes, Christian T. and Sasaki, Misao",
	title          = "{Calculating the mass fraction of primordial black holes}",
	journal        = "JCAP",
	volume         = "1407",
	year           = "2014",
	pages          = "045",
	doi            = "10.1088/1475-7516/2014/07/045",
	eprint         = "1405.7023",
	archivePrefix  = "arXiv",
	primaryClass   = "gr-qc",
	SLACcitation   = "%%CITATION = ARXIV:1405.7023;%%"
}

@article{Chen:2019irf,
    author = "Chen, Zu-Cheng and Huang, Qing-Guo",
    title = "{Distinguishing primordial black holes from astrophysical black holes by Einstein Telescope and Cosmic Explorer}",
    eprint = "1904.02396",
    archivePrefix = "arXiv",
    primaryClass = "astro-ph.CO",
    doi = "10.1088/1475-7516/2020/08/039",
    journal = "JCAP",
    volume = "08",
    pages = "039",
    year = "2020"
}

@article{Manshanden:2018tze,
    author = "Manshanden, Julien and Gaggero, Daniele and Bertone, Gianfranco and Connors, Riley M. T. and Ricotti, Massimo",
    title = "{Multi-wavelength astronomical searches for primordial black holes}",
    eprint = "1812.07967",
    archivePrefix = "arXiv",
    primaryClass = "astro-ph.HE",
    doi = "10.1088/1475-7516/2019/06/026",
    journal = "JCAP",
    volume = "06",
    pages = "026",
    year = "2019"
}

@article{Oguri:2017ock,
    author = "Oguri, Masamune and Diego, Jose M. and Kaiser, Nick and Kelly, Patrick L. and Broadhurst, Tom",
    title = "{Understanding caustic crossings in giant arcs: characteristic scales, event rates, and constraints on compact dark matter}",
    eprint = "1710.00148",
    archivePrefix = "arXiv",
    primaryClass = "astro-ph.CO",
    doi = "10.1103/PhysRevD.97.023518",
    journal = "Phys. Rev. D",
    volume = "97",
    number = "2",
    pages = "023518",
    year = "2018"
}

@article{Blaineau:2022nhy,
    author = "Blaineau, T. and others",
    title = "{New limits from microlensing on Galactic black holes in the mass range 10 M{\ensuremath{\odot}} {\ensuremath{<}} M {\ensuremath{<}} 1000 M{\ensuremath{\odot}}}",
    eprint = "2202.13819",
    archivePrefix = "arXiv",
    primaryClass = "astro-ph.GA",
    doi = "10.1051/0004-6361/202243430",
    journal = "Astron. Astrophys.",
    volume = "664",
    pages = "A106",
    year = "2022"
}

@article{Esteban-Gutierrez:2023qcz,
    author = "Esteban-Guti{\'e}rrez, A. and Mediavilla, E. and Jim{\'e}nez-Vicente, J. and Mu{\~n}oz, J. A.",
    title = "{Constraints on the abundance of primordial black holes from X-ray quasar microlensing observations: Substellar to planetary mass range}",
    eprint = "2307.07473",
    archivePrefix = "arXiv",
    primaryClass = "astro-ph.CO",
    doi = "10.3847/1538-4357/ace62f",
    journal = "Astrophys. J.",
    volume = "954",
    number = "2",
    pages = "172",
    year = "2023"
}

@article{Monroy-Rodriguez:2014ula,
    author = "Monroy-Rodr{\'\i}guez, Miguel A. and Allen, Christine",
    title = "{The end of the MACHO era revisited: new limits on MACHO masses from halo wide binaries}",
    eprint = "1406.5169",
    archivePrefix = "arXiv",
    primaryClass = "astro-ph.GA",
    doi = "10.1088/0004-637X/790/2/159",
    journal = "Astrophys. J.",
    volume = "790",
    number = "2",
    pages = "159",
    year = "2014"
}

@article{Brandt:2016aco,
    author = "Brandt, Timothy D.",
    title = "{Constraints on MACHO dark matter from compact stellar systems in ultra-faint dwarf galaxies}",
    eprint = "1605.03665",
    archivePrefix = "arXiv",
    primaryClass = "astro-ph.GA",
    doi = "10.3847/2041-8205/824/2/L31",
    journal = "Astrophys. J. Lett.",
    volume = "824",
    number = "2",
    pages = "L31",
    year = "2016"
}

@article{Lu:2020bmd,
    author = "Lu, Philip and Takhistov, Volodymyr and Gelmini, Graciela B. and Hayashi, Kohei and Inoue, Yoshiyuki and Kusenko, Alexander",
    title = "{Constraining primordial black holes with dwarf galaxy heating}",
    eprint = "2007.02213",
    archivePrefix = "arXiv",
    primaryClass = "astro-ph.CO",
    reportNumber = "IPMU20-0076, RIKEN-iTHEMS-Report-20",
    doi = "10.3847/2041-8213/abdcb6",
    journal = "Astrophys. J. Lett.",
    volume = "908",
    number = "2",
    pages = "L23",
    year = "2021"
}

@misc{Kavanagh2019,
  author    = {Kavanagh, Bradley J.},
  title     = {bradkav/PBHbounds: Release version},
  year      = {2019},
  copyright = {Open Access},
  doi       = {10.5281/zenodo.3538999},
  publisher = {Zenodo},
}

@article{Munoz2015,
    author = {Mu{\~n}oz, Julian B. and Ali-Ha{\"\i}moud, Yacine and Kamionkowski, Marc},
    title = "{Primordial non-gaussianity from the bispectrum of 21-cm fluctuations in the dark ages}",
    eprint = "1506.04152",
    archivePrefix = "arXiv",
    primaryClass = "astro-ph.CO",
    doi = "10.1103/PhysRevD.92.083508",
    journal = "Phys. Rev. D",
    volume = "92",
    number = "8",
    pages = "083508",
    year = "2015"
}

@article{Sheth:1999su,
    author = "Sheth, Ravi K. and Mo, H. J. and Tormen, Giuseppe",
    title = "{Ellipsoidal collapse and an improved model for the number and spatial distribution of dark matter haloes}",
    eprint = "astro-ph/9907024",
    archivePrefix = "arXiv",
    doi = "10.1046/j.1365-8711.2001.04006.x",
    journal = "Mon. Not. Roy. Astron. Soc.",
    volume = "323",
    pages = "1",
    year = "2001"
}

@article{Dhandha:2025dtn,
    author = "Dhandha, Jiten and others",
    title = "{Narrowing the discovery space of the cosmological 21-cm signal using multi-wavelength constraints}",
    eprint = "2508.13761",
    archivePrefix = "arXiv",
    primaryClass = "astro-ph.CO",
    doi = "10.1093/mnras/staf1736",
    journal = "Mon. Not. Roy. Astron. Soc.",
    volume = "544",
    number = "2",
    pages = "1608–1626",
    year = "2025"
}

@article{Sims:2025hfm,
    author = "Sims, Peter H. and others",
    title = "{Rapid and late cosmic reionization driven by massive galaxies: a joint analysis of constraints from 21-cm, Lyman line {\&} CMB data sets}",
    eprint = "2504.09725",
    archivePrefix = "arXiv",
    primaryClass = "astro-ph.CO",
    doi = "10.1093/mnras/staf1864",
    journal = "Mon. Not. Roy. Astron. Soc.",
    volume  = "544",
    number = "4",
    pages = "3856-3882",
    year = "2025"
}

@article{Madau:1996yh,
    author = "Madau, Piero and Ferguson, H. C. and Dickinson, M. E. and Giavalisco, M. and Steidel, C. C. and Fruchter, A.",
    title = "{High redshift galaxies in the hubble deep field. color selection and star formation history to z=4}",
    eprint = "astro-ph/9607172",
    archivePrefix = "arXiv",
    doi = "10.1093/mnras/283.4.1388",
    journal = "Mon. Not. Roy. Astron. Soc.",
    volume = "283",
    pages = "1388--1404",
    year = "1996"
}

@article{Gnedin:2014uta,
    author = "Gnedin, Nickolay Y.",
    title = "{Cosmic reionization on computers I. Design and calibration of simulations}",
    eprint = "1403.4245",
    archivePrefix = "arXiv",
    primaryClass = "astro-ph.CO",
    reportNumber = "FERMILAB-PUB-14-066-A",
    doi = "10.1088/0004-637X/793/1/29",
    journal = "Astrophys. J.",
    volume = "793",
    pages = "29",
    year = "2014"
}

@article{Greig:2016wjs,
    author = "Greig, Bradley and Mesinger, Andrei",
    title = "{The global history of reionization}",
    eprint = "1605.05374",
    archivePrefix = "arXiv",
    primaryClass = "astro-ph.CO",
    doi = "10.1093/mnras/stw3026",
    journal = "Mon. Not. Roy. Astron. Soc.",
    volume = "465",
    number = "4",
    pages = "4838--4852",
    year = "2017"
}

@article{LIGOScientific:2016dsl,
    author = "Abbott, B. P. and others",
    collaboration = "LIGO Scientific, Virgo",
    title = "{Binary black hole mergers in the first advanced LIGO observing run}",
    eprint = "1606.04856",
    archivePrefix = "arXiv",
    primaryClass = "gr-qc",
    reportNumber = "LIGO-P1600088",
    doi = "10.1103/PhysRevX.6.041015",
    journal = "Phys. Rev. X",
    volume = "6",
    number = "4",
    pages = "041015",
    year = "2016",
    note = "[Erratum: Phys.Rev.X 8, 039903 (2018)]"
}

@article{Liu:2019awk,
    author = "Liu, Adrian and Shaw, J. Richard",
    title = "{Data analysis for precision 21 cm cosmology}",
    eprint = "1907.08211",
    archivePrefix = "arXiv",
    primaryClass = "astro-ph.IM",
    doi = "10.1088/1538-3873/ab5bfd",
    journal = "Publ. Astron. Soc. Pac.",
    volume = "132",
    number = "1012",
    pages = "062001",
    year = "2020"
}

@article{Chen:2003gz,
    author = "Chen, Xue-Lei and Kamionkowski, Marc",
    title = "{Particle decays during the cosmic dark ages}",
    eprint = "astro-ph/0310473",
    archivePrefix = "arXiv",
    doi = "10.1103/PhysRevD.70.043502",
    journal = "Phys. Rev. D",
    volume = "70",
    pages = "043502",
    year = "2004"
}

@article{Slatyer:2009yq,
    author = "Slatyer, Tracy R. and Padmanabhan, Nikhil and Finkbeiner, Douglas P.",
    title = "{CMB constraints on WIMP annihilation: energy absorption during the recombination epoch}",
    eprint = "0906.1197",
    archivePrefix = "arXiv",
    primaryClass = "astro-ph.CO",
    doi = "10.1103/PhysRevD.80.043526",
    journal = "Phys. Rev. D",
    volume = "80",
    pages = "043526",
    year = "2009"
}

@article{LIGOScientific:2017bnn,
    author = "Abbott, Benjamin P. and others",
    collaboration = "LIGO Scientific, VIRGO",
    title = "{GW170104: Observation of a 50-solar-mass binary black hole coalescence at redshift 0.2}",
    eprint = "1706.01812",
    archivePrefix = "arXiv",
    primaryClass = "gr-qc",
    reportNumber = "LIGO-P170104",
    doi = "10.1103/PhysRevLett.118.221101",
    journal = "Phys. Rev. Lett.",
    volume = "118",
    number = "22",
    pages = "221101",
    year = "2017",
    note = "[Erratum: Phys.Rev.Lett. 121, 129901 (2018)]"
}

@article{LIGOScientific:2016sjg,
    author = "Abbott, B. P. and others",
    collaboration = "LIGO Scientific, Virgo",
    title = "{GW151226: Observation of gravitational waves from a 22-solar-mass binary black hole coalescence}",
    eprint = "1606.04855",
    archivePrefix = "arXiv",
    primaryClass = "gr-qc",
    reportNumber = "LIGO-P151226",
    doi = "10.1103/PhysRevLett.116.241103",
    journal = "Phys. Rev. Lett.",
    volume = "116",
    number = "24",
    pages = "241103",
    year = "2016"
}

@article{LIGOScientific:2017ycc,
    author = "Abbott, B. P. and others",
    collaboration = "LIGO Scientific, Virgo",
    title = "{GW170814: A three-detector observation of gravitational waves from a binary black hole coalescence}",
    eprint = "1709.09660",
    archivePrefix = "arXiv",
    primaryClass = "gr-qc",
    doi = "10.1103/PhysRevLett.119.141101",
    journal = "Phys. Rev. Lett.",
    volume = "119",
    number = "14",
    pages = "141101",
    year = "2017"
}

@article{Barkana:2000fd,
    author = "Barkana, Rennan and Loeb, Abraham",
    title = "{In the beginning: the first sources of light and the reionization of the Universe}",
    eprint = "astro-ph/0010468",
    archivePrefix = "arXiv",
    doi = "10.1016/S0370-1573(01)00019-9",
    journal = "Phys. Rept.",
    volume = "349",
    pages = "125--238",
    year = "2001"
}

@article{Hirata:2005mz,
    author = "Hirata, Christopher M.",
    title = "{Wouthuysen-Field coupling strength and application to high-redshift 21 cm radiation}",
    eprint = "astro-ph/0507102",
    archivePrefix = "arXiv",
    doi = "10.1111/j.1365-2966.2005.09949.x",
    journal = "Mon. Not. Roy. Astron. Soc.",
    volume = "367",
    pages = "259--274",
    year = "2006"
}

@article{Cappelluti:2012rd,
    author = "Cappelluti, N. and others",
    title = "{The nature of the unresolved extragalactic soft CXB}",
    eprint = "1208.4105",
    archivePrefix = "arXiv",
    primaryClass = "astro-ph.CO",
    doi = "10.1111/j.1365-2966.2012.21867.x",
    journal = "Mon. Not. Roy. Astron. Soc.",
    volume = "427",
    pages = "651",
    year = "2012"
}

@article{Pochinda:2023uom,
    author = "Pochinda, S. and others",
    title = "{Constraining the properties of Population III galaxies with multiwavelength observations}",
    eprint = "2312.08095",
    archivePrefix = "arXiv",
    primaryClass = "astro-ph.CO",
    doi = "10.1093/mnras/stae1185",
    journal = "Mon. Not. Roy. Astron. Soc.",
    volume = "531",
    number = "1",
    pages = "1113--1132",
    year = "2024"
}

@article{Tseliakhovich:2010bj,
    author = "Tseliakhovich, Dmitriy and Hirata, Christopher",
    title = "{Relative velocity of dark matter and baryonic fluids and the formation of the first structures}",
    eprint = "1005.2416",
    archivePrefix = "arXiv",
    primaryClass = "astro-ph.CO",
    doi = "10.1103/PhysRevD.82.083520",
    journal = "Phys. Rev. D",
    volume = "82",
    pages = "083520",
    year = "2010"
}

@article{Dvorkin:2013cea,
    author = "Dvorkin, Cora and Blum, Kfir and Kamionkowski, Marc",
    title = "{Constraining dark matter-baryon scattering with linear cosmology}",
    eprint = "1311.2937",
    archivePrefix = "arXiv",
    primaryClass = "astro-ph.CO",
    doi = "10.1103/PhysRevD.89.023519",
    journal = "Phys. Rev. D",
    volume = "89",
    number = "2",
    pages = "023519",
    year = "2014"
}

@article{Serpico2020,
  author        = {Serpico, Pasquale D. and Poulin, Vivian and Inman, Derek and Kohri, Kazunori},
  journal       = {Phys. Rev. Res.},
  title         = {{Cosmic microwave background bounds on primordial black holes including dark matter halo accretion}},
  year          = {2020},
  number        = {2},
  pages         = {023204},
  volume        = {2},
  archiveprefix = {arXiv},
  doi           = {10.1103/PhysRevResearch.2.023204},
  eprint        = {2002.10771},
  primaryclass  = {astro-ph.CO},
  reportnumber  = {LAPTH-005/20, KEK-Cosmo-248, KEK-TH-2198, IPMU20-0021},
}

@article{Mack:2006gz,
    author = "Mack, Katherine J. and Ostriker, Jeremiah P. and Ricotti, Massimo",
    title = "{Growth of structure seeded by primordial black holes}",
    eprint = "astro-ph/0608642",
    archivePrefix = "arXiv",
    doi = "10.1086/518998",
    journal = "Astrophys. J.",
    volume = "665",
    pages = "1277--1287",
    year = "2007"
}

@article{Slatyer:2012yq,
    author = "Slatyer, Tracy R.",
    title = "{Energy injection and absorption in the cosmic dark ages}",
    eprint = "1211.0283",
    archivePrefix = "arXiv",
    primaryClass = "astro-ph.CO",
    doi = "10.1103/PhysRevD.87.123513",
    journal = "Phys. Rev. D",
    volume = "87",
    number = "12",
    pages = "123513",
    year = "2013"
}

@article{Leitherer:1999rq,
    author = "Leitherer, Claus and Schaerer, Daniel and Goldader, Jeffrey D. and Gonzalez Delgado, Rosa M. and Robert, Carmelle and Kune, Denis Foo and Mello, Duilia F. de and Devost, Daniel and Heckman, Timothy M.",
    title = "{Starburst99: synthesis models for galaxies with active star formation}",
    eprint = "astro-ph/9902334",
    archivePrefix = "arXiv",
    doi = "10.1086/313233",
    journal = "Astrophys. J. Suppl.",
    volume = "123",
    pages = "3--40",
    year = "1999"
}

@article{Greig:2018hja,
    author = "Greig, Bradley and Mesinger, Andrei",
    title = "{21CMMC with a 3D light-cone: the impact of the co-evolution approximation on the astrophysics of reionization and cosmic dawn}",
    eprint = "1801.01592",
    archivePrefix = "arXiv",
    primaryClass = "astro-ph.CO",
    doi = "10.1093/mnras/sty796",
    journal = "Mon. Not. Roy. Astron. Soc.",
    volume = "477",
    number = "3",
    pages = "3217--3229",
    year = "2018"
}

@article{Mesinger:2010ne,
    author = "Mesinger, Andrei and Furlanetto, Steven and Cen, Renyue",
    title = "{21cmFAST: a fast, semi-numerical simulation of the high-redshift 21-cm signal}",
    eprint = "1003.3878",
    archivePrefix = "arXiv",
    primaryClass = "astro-ph.CO",
    doi = "10.1111/j.1365-2966.2010.17731.x",
    journal = "Mon. Not. Roy. Astron. Soc.",
    volume = "411",
    pages = "955",
    year = "2011"
}

@article{Natarajan:2009bm,
    author = "Natarajan, Aravind and Schwarz, Dominik J.",
    title = "{Dark matter annihilation and its effect on CMB and Hydrogen 21 cm observations}",
    eprint = "0903.4485",
    archivePrefix = "arXiv",
    primaryClass = "astro-ph.CO",
    reportNumber = "BI-TP-2009-08",
    doi = "10.1103/PhysRevD.80.043529",
    journal = "Phys. Rev. D",
    volume = "80",
    pages = "043529",
    year = "2009"
}

@article{Furlanetto:2006wp,
    author = "Furlanetto, Steven R. and Oh, S. Peng and Pierpaoli, Elena",
    title = "{The effects of dark matter decay and annihilation on the high-redshift 21 cm background}",
    eprint = "astro-ph/0608385",
    archivePrefix = "arXiv",
    doi = "10.1103/PhysRevD.74.103502",
    journal = "Phys. Rev. D",
    volume = "74",
    pages = "103502",
    year = "2006"
}

@article{Murray:2020trn,
    author = "Murray, Steven G. and Greig, Bradley and Mesinger, Andrei and Mu{\~n}oz, Julian B. and Qin, Yuxiang and Park, Jaehong and Watkinson, Catherine A.",
    title = "{21cmFAST v3: a Python-integrated C code for generating 3D realizations of the cosmic 21cm signal}",
    eprint = "2010.15121",
    archivePrefix = "arXiv",
    primaryClass = "astro-ph.IM",
    doi = "10.21105/joss.02582",
    journal = "J. Open Source Softw.",
    volume = "5",
    number = "54",
    pages = "2582",
    year = "2020"
}

@article{Munoz:2023kkg,
    author = "Mu{\~n}oz, Julian B.",
    title = "{An effective model for the cosmic-dawn 21-cm signal}",
    eprint = "2302.08506",
    archivePrefix = "arXiv",
    primaryClass = "astro-ph.CO",
    doi = "10.1093/mnras/stad1512",
    journal = "Mon. Not. Roy. Astron. Soc.",
    volume = "523",
    number = "2",
    pages = "2587--2607",
    year = "2023"
}

@article{Cruz:2024fsv,
    author = "Cruz, Hector Afonso G. and Mu\~noz, Julian B. and Sabti, Nashwan and Kamionkowski, Marc",
    title = "{Effective model for the 21-cm signal with population III stars}",
    eprint = "2407.18294",
    archivePrefix = "arXiv",
    primaryClass = "astro-ph.CO",
    doi = "10.1103/PhysRevD.111.083503",
    journal = "Phys. Rev. D",
    volume = "111",
    number = "8",
    pages = "083503",
    year = "2025"
}

@article{Fragos:2012vf,
    author = "Fragos, Tassos and others",
    title = "{X-ray binary evolution across cosmic time}",
    eprint = "1206.2395",
    archivePrefix = "arXiv",
    primaryClass = "astro-ph.HE",
    doi = "10.1088/0004-637X/764/1/41",
    journal = "Astrophys. J.",
    volume = "764",
    pages = "41",
    year = "2013"
}

@article{Lehmer:2016dxd,
    author = "Lehmer, B. D. and others",
    title = "{The evolution of normal galaxy X-ray emission through cosmic history: Constraints from the 6 Ms Chandra deep field-south}",
    eprint = "1604.06461",
    archivePrefix = "arXiv",
    primaryClass = "astro-ph.GA",
    doi = "10.3847/0004-637X/825/1/7",
    journal = "Astrophys. J.",
    volume = "825",
    number = "1",
    pages = "7",
    year = "2016"
}

@Article{Hoyle1939effect,
  author    = {Hoyle, F. and Lyttleton, R. A.},
  journal   = {Math. Proc. Camb. Philos. Soc.},
  title     = {The effect of interstellar matter on climatic variation},
  year      = {1939},
  issn      = {1469-8064},
  number    = {3},
  pages     = {405--415},
  volume    = {35},
  doi       = {10.1017/s0305004100021150},
  publisher = {Cambridge University Press (CUP)},
}

@article{Hoyle1940accretion,
  author={Hoyle, Fred and Lyttleton, R. A.},
  title={On the accretion of interstellar matter by stars},
  doi   = "10.1017/S0305004100017369",
  journal={Math. Proc. Camb. Philos. Soc.},
  volume={36},
  number={3},
  pages={325--330},
  year={1940},
  organization={Cambridge University Press}
}

@article{Hoyle1940physical,
  author={Hoyle, F and Lyttleton, R. A.},
  title={On the physical aspects of accretion by stars},
  doi   = "10.1017/S0305004100017461",
  journal={Math. Proc. Camb. Philos. Soc.},
  volume={36},
  number={4},
  pages={424--437},
  year={1940},
  organization={Cambridge University Press}
}

@article{Hoyle1941accretion,
  author={Hoyle, Fred and Lyttleton, R. A.},
  title={On the accretion theory of stellar evolution},
  doi   = "10.1093/mnras/101.4.227",
  journal={Mon. Not. Roy. Astron. Soc.},
  volume={101},
  pages={227},
  year={1941}
}

@article{Bondi:1944rnk,
    author = "Bondi, H. and Hoyle, F.",
    title = "{On the mechanism of accretion by stars}",
    doi = "10.1093/mnras/104.5.273",
    journal = "Mon. Not. Roy. Astron. Soc.",
    volume = "104",
    number = "5",
    pages = "273--282",
    year = "1944"
}

@article{Bondi:1952ni,
    author = "Bondi, H.",
    title = "{On spherically symmetrical accretion}",
    doi = "10.1093/mnras/112.2.195",
    journal = "Mon. Not. Roy. Astron. Soc.",
    volume = "112",
    pages = "195",
    year = "1952"
}

@article{Bogdan:2023ilu,
    author = "Bogdan, Akos and others",
    title = "{Evidence for heavy-seed origin of early supermassive black holes from a z{\,}{\ensuremath{\approx}}{\,}10 X-ray quasar}",
    eprint = "2305.15458",
    archivePrefix = "arXiv",
    primaryClass = "astro-ph.GA",
    doi = "10.1038/s41550-023-02111-9",
    journal = "Nature Astron.",
    volume = "8",
    number = "1",
    pages = "126--133",
    year = "2024"
}

@article{Greig:2017jdj,
    author = "Greig, Bradley and Mesinger, Andrei",
    title = "{Simultaneously constraining the astrophysics of reionization and the epoch of heating with 21CMMC}",
    eprint = "1705.03471",
    archivePrefix = "arXiv",
    primaryClass = "astro-ph.CO",
    doi = "10.1093/mnras/stx2118",
    journal = "Mon. Not. Roy. Astron. Soc.",
    volume = "472",
    number = "3",
    pages = "2651--2669",
    year = "2017"
}

@article{Tegmark:1996yt,
    author = "Tegmark, Max and Silk, Joseph and Rees, Martin J. and Blanchard, Alain and Abel, Tom and Palla, Francesco",
    title = "{How small were the first cosmological objects?}",
    eprint = "astro-ph/9603007",
    archivePrefix = "arXiv",
    doi = "10.1086/303434",
    journal = "Astrophys. J.",
    volume = "474",
    pages = "1--12",
    year = "1997"
}

@article{Inman:2019wvr,
    author = {Inman, Derek and Ali-Ha{\"\i}moud, Yacine},
    title = "{Early structure formation in primordial black hole cosmologies}",
    eprint = "1907.08129",
    archivePrefix = "arXiv",
    primaryClass = "astro-ph.CO",
    doi = "10.1103/PhysRevD.100.083528",
    journal = "Phys. Rev. D",
    volume = "100",
    number = "8",
    pages = "083528",
    year = "2019"
}

@article{Koulen:2025xjq,
    author = "Koulen, Julia Monika and Profumo, Stefano and Smyth, Nolan",
    title = "{Primordial black holes and the first stars}",
    eprint = "2506.06171",
    archivePrefix = "arXiv",
    primaryClass = "astro-ph.CO",
    doi = "10.1103/n4b8-cymr",
    journal = "Phys. Rev. D",
    volume = "112",
    number = "4",
    pages = "043044",
    year = "2025"
}

@article{Cole:2019zhu,
    author = "Cole, Philippa S. and Silk, Joseph",
    title = "{Small-scale primordial fluctuations in the 21 cm Dark Ages signal}",
    eprint = "1912.02171",
    archivePrefix = "arXiv",
    primaryClass = "astro-ph.CO",
    doi = "10.1093/mnras/staa3638",
    journal = "Mon. Not. Roy. Astron. Soc.",
    volume = "501",
    number = "2",
    pages = "2627--2634",
    year = "2021"
}

@article{Turner_2022,
   title={The Road to Precision Cosmology},
   volume={72},
   ISSN={1545-4134},
   url={http://dx.doi.org/10.1146/annurev-nucl-111119-041046},
   DOI={10.1146/annurev-nucl-111119-041046},
   number={1},
   journal={Ann. Rev. Nucl. Part. Sci.},
   publisher={Annual Reviews},
   author={Turner, Michael S.},
   year={2022},
   pages={1–35},
   eprint = "2201.04741",
    archivePrefix = "arXiv",
    primaryClass = "astro-ph.CO"
}

@article{Trac:2006vr,
    author = "Trac, Hy and Cen, Renyue",
    title = "{Radiative transfer simulations of cosmic reionization. 1. Methodology and initial results}",
    eprint = "astro-ph/0612406",
    archivePrefix = "arXiv",
    doi = "10.1086/522566",
    journal = "Astrophys. J.",
    volume = "671",
    pages = "1",
    year = "2007"
}

@article{Mondal:2023xjx,
    author = "Mondal, Rajesh and Barkana, Rennan",
    title = "{Prospects for precision cosmology with the 21{\,}cm signal from the dark ages}",
    eprint = "2305.08593",
    archivePrefix = "arXiv",
    primaryClass = "astro-ph.CO",
    doi = "10.1038/s41550-023-02057-y",
    journal = "Nature Astron.",
    volume = "7",
    number = "9",
    pages = "1025--1030",
    year = "2023"
}

@article{McGreer:2014qwa,
    author = "McGreer, Ian and Mesinger, Andrei and D'Odorico, Valentina",
    title = "{Model-independent evidence in favour of an end to reionization by $z \approx$ 6}",
    eprint = "1411.5375",
    archivePrefix = "arXiv",
    primaryClass = "astro-ph.CO",
    doi = "10.1093/mnras/stu2449",
    journal = "Mon. Not. Roy. Astron. Soc.",
    volume = "447",
    number = "1",
    pages = "499--505",
    year = "2015"
}

@article{Poulin:2015pna,
    author = "Poulin, Vivian and Serpico, Pasquale D. and Lesgourgues, Julien",
    title = "{Dark matter annihilations in halos and high-redshift sources of reionization of the universe}",
    eprint = "1508.01370",
    archivePrefix = "arXiv",
    primaryClass = "astro-ph.CO",
    doi = "10.1088/1475-7516/2015/12/041",
    journal = "JCAP",
    volume = "12",
    pages = "041",
    year = "2015"
}

@article{Poulin:2016anj,
    author = "Poulin, Vivian and Lesgourgues, Julien and Serpico, Pasquale D.",
    title = "{Cosmological constraints on exotic injection of electromagnetic energy}",
    eprint = "1610.10051",
    archivePrefix = "arXiv",
    primaryClass = "astro-ph.CO",
    doi = "10.1088/1475-7516/2017/03/043",
    journal = "JCAP",
    volume = "03",
    pages = "043",
    year = "2017"
}

@misc{SKA_MEMO:2025,
  author       = {Seethapuram Sridhar, Sarrvesh and
                  Williams, Wendy and
                  Price, Danny and
                  Breen, shari and
                  Ball, Lewis},
  title        = "{SKA Low and Mid subarray templates}",
  year         = 2025,
  howpublished = {SKAO},
  version      = {SKAO-TEL-0002380, Revision 02},
  doi          = {10.5281/zenodo.16951088},
  url          = {https://doi.org/10.5281/zenodo.16951088},
}

@article{Mozdzen:2017,
       author = {{Mozdzen}, T.~J. and {Bowman}, J.~D. and {Monsalve}, R.~A. and {Rogers}, A.~E.~E.},
        title = "{Improved measurement of the spectral index of the diffuse radio background between 90 and 190 MHz}",
      journal = "Mon. Not. Roy. Astron. Soc.",
         year = 2017,
       volume = {464},
       number = {4},
        pages = {4995-5002},
          doi = {10.1093/mnras/stw2696},
archivePrefix = {arXiv},
       eprint = {1609.08705},
 primaryClass = {astro-ph.IM},
       adsurl = {https://ui.adsabs.harvard.edu/abs/2017MNRAS.464.4995M}
}

@article{Zahn:2010yw,
    author = "Zahn, Oliver and Mesinger, Andrei and McQuinn, Matthew and Trac, Hy and Cen, Renyue and Hernquist, Lars E.",
    title = "{Comparison of reionization models: Radiative transfer simulations and approximate, semi-numeric models}",
    eprint = "1003.3455",
    archivePrefix = "arXiv",
    primaryClass = "astro-ph.CO",
    doi = "10.1111/j.1365-2966.2011.18439.x",
    journal = "Mon. Not. Roy. Astron. Soc.",
    volume = "414",
    pages = "727",
    year = "2011"
}

@article{Bouwens:2021,
    author = {{Bouwens}, R.~J. and others},
    title = "{New determinations of the UV luminosity functions from $z \sim 9$ to $z \sim 2$ show a remarkable consistency with halo growth and a constant star formation efficiency}",
    journal = "Astrophys. J.",
    year = 2021,
    volume = {162},
    number = {2},
    eid = {47},
    pages = {47},
    doi = {10.3847/1538-3881/abf83e},
    archivePrefix = {arXiv},
    eprint = {2102.07775},
    primaryClass = {astro-ph.GA}
}

@article{Chen:2003gc,
    author = "Chen, Xue-Lei and Miralda-Escude, Jordi",
    title = "{The spin - kinetic temperature coupling and the heating rate due to Lyman - alpha scattering before reionization: Predictions for 21cm emission and absorption}",
    eprint = "astro-ph/0303395",
    archivePrefix = "arXiv",
    doi = "10.1086/380829",
    journal = "Astrophys. J.",
    volume = "602",
    pages = "1--11",
    year = "2004"
}

@article{Chen:2006zr,
    author = "Chen, Xue-Lei and Miralda-Escude, Jordi",
    title = "{The 21cm signature of the first stars}",
    eprint = "astro-ph/0605439",
    archivePrefix = "arXiv",
    doi = "10.1086/528941",
    journal = "Astrophys. J.",
    volume = "684",
    pages = "18--33",
    year = "2008"
}

@article{Flitter:2023rzv,
    author = "Flitter, Jordan and Kovetz, Ely D.",
    title = "{New tool for 21-cm cosmology. II. Investigating the effect of early linear fluctuations}",
    eprint = "2309.03948",
    archivePrefix = "arXiv",
    primaryClass = "astro-ph.CO",
    doi = "10.1103/PhysRevD.109.043513",
    journal = "Phys. Rev. D",
    volume = "109",
    number = "4",
    pages = "043513",
    year = "2024"
}

@article{Flitter:2024eay,
    author = "Flitter, Jordan and Libanore, Sarah and Kovetz, Ely D.",
    title = "{More careful treatment of matter density fluctuations in 21-cm simulations}",
    eprint = "2411.00089",
    archivePrefix = "arXiv",
    primaryClass = "astro-ph.CO",
    doi = "10.1103/zc4v-2yx4",
    journal = "Phys. Rev. D",
    volume = "112",
    number = "2",
    pages = "023537",
    year = "2025"
}

@article{Bouwens:2014fua,
    author = "Bouwens, R. J. and others",
    title = "{UV luminosity functions at redshifts $z \sim$4 to $z \sim$10: 10000 galaxies from HST legacy fields}",
    eprint = "1403.4295",
    archivePrefix = "arXiv",
    primaryClass = "astro-ph.CO",
    doi = "10.1088/0004-637X/803/1/34",
    journal = "Astrophys. J.",
    volume = "803",
    number = "1",
    pages = "34",
    year = "2015"
}

@article{Barkana:2004vb,
    author = "Barkana, Rennan and Loeb, Abraham",
    title = "{Detecting the earliest galaxies through two new sources of 21cm fluctuations}",
    eprint = "astro-ph/0410129",
    archivePrefix = "arXiv",
    doi = "10.1086/429954",
    journal = "Astrophys. J.",
    volume = "626",
    pages = "1--11",
    year = "2005"
}

@article{Mena:2019nhm,
    author = "Mena, Olga and Palomares-Ruiz, Sergio and Villanueva-Domingo, Pablo and Witte, Samuel J.",
    title = "{Constraining the primordial black hole abundance with 21-cm cosmology}",
    eprint = "1906.07735",
    archivePrefix = "arXiv",
    primaryClass = "astro-ph.CO",
    doi = "10.1103/PhysRevD.100.043540",
    journal = "Phys. Rev. D",
    volume = "100",
    number = "4",
    pages = "043540",
    year = "2019"
}

@article{Yang:2021idt,
    author = "Yang, Yupeng",
    title = "{Constraints on accreting primordial black holes with the global 21-cm signal}",
    eprint = "2108.11130",
    archivePrefix = "arXiv",
    primaryClass = "astro-ph.CO",
    doi = "10.1103/PhysRevD.104.063528",
    journal = "Phys. Rev. D",
    volume = "104",
    number = "6",
    pages = "063528",
    year = "2021"
}

@article{Mesinger:2012ys,
    author = "Mesinger, Andrei and Ferrara, Andrea and Spiegel, David S.",
    title = "{Signatures of X-rays in the early Universe}",
    eprint = "1210.7319",
    archivePrefix = "arXiv",
    primaryClass = "astro-ph.CO",
    doi = "10.1093/mnras/stt198",
    journal = "Mon. Not. Roy. Astron. Soc.",
    volume = "431",
    pages = "621",
    year = "2013"
}

@article{Cassano:2018zwm,
    author = "Cassano, R. and others",
    title = "{SKA-Athena Synergy White Paper}",
    eprint = "1807.09080",
    archivePrefix = "arXiv",
    primaryClass = "astro-ph.HE",
    month = "7",
    year = "2018"
}

@article{Madau:1996cs,
    author = "Madau, Piero and Meiksin, Avery and Rees, Martin J.",
    title = "{21-cm tomography of the intergalactic medium at high redshift}",
    eprint = "astro-ph/9608010",
    archivePrefix = "arXiv",
    doi = "10.1086/303549",
    journal = "Astrophys. J.",
    volume = "475",
    pages = "429",
    year = "1997"
}

@article{Barkana:2016nyr,
    author = "Barkana, Rennan",
    title = "{The rise of the first stars: Supersonic streaming, radiative feedback, and 21-cm cosmology}",
    eprint = "1605.04357",
    archivePrefix = "arXiv",
    primaryClass = "astro-ph.CO",
    doi = "10.1016/j.physrep.2016.06.006",
    journal = "Phys. Rept.",
    volume = "645",
    pages = "1--59",
    year = "2016"
}

@article{Carr:1981pko,
    author = "Carr, B. J.",
    title = "{Pregalactic black hole accretion and the thermal history of the Universe}",
    doi = "10.1093/mnras/194.3.639",
    journal = "Mon. Not. Roy. Astron. Soc.",
    volume = "194",
    number = "3",
    pages = "639--668",
    year = "1981"
}

@article{Facchinetti:2022kbg,
    author = "Facchinetti, Ga{\'e}tan and Lucca, Matteo and Clesse, S{\'e}bastien",
    title = "{Relaxing CMB bounds on primordial black holes: The role of ionization fronts}",
    eprint = "2212.07969",
    archivePrefix = "arXiv",
    primaryClass = "astro-ph.CO",
    reportNumber = "ULB-TH/22-27",
    doi = "10.1103/PhysRevD.107.043537",
    journal = "Phys. Rev. D",
    volume = "107",
    number = "4",
    pages = "043537",
    year = "2023"
}

@article{Mesinger:2016ddl,
    author = "Mesinger, Andrei and Greig, Bradley and Sobacchi, Emanuele",
    title = "{The evolution of 21 cm structure (EOS): public, large-scale simulations of Cosmic Dawn and Reionization}",
    eprint = "1602.07711",
    archivePrefix = "arXiv",
    primaryClass = "astro-ph.CO",
    doi = "10.1093/mnras/stw831",
    journal = "Mon. Not. Roy. Astron. Soc.",
    volume = "459",
    number = "3",
    pages = "2342--2353",
    year = "2016"
}

@article{Qin:2023kkk,
    author = "Qin, Wenzer and Munoz, Julian B. and Liu, Hongwan and Slatyer, Tracy R.",
    title = "{Birth of the first stars amidst decaying and annihilating dark matter}",
    eprint = "2308.12992",
    archivePrefix = "arXiv",
    primaryClass = "astro-ph.CO",
    reportNumber = "MIT-CTP/5596",
    doi = "10.1103/PhysRevD.109.103026",
    journal = "Phys. Rev. D",
    volume = "109",
    number = "10",
    pages = "103026",
    year = "2024"
}

@article{Facchinetti:2023slb,
    author = "Facchinetti, Ga{\'e}tan and Lopez-Honorez, Laura and Qin, Yuxiang and Mesinger, Andrei",
    title = "{21cm signal sensitivity to dark matter decay}",
    eprint = "2308.16656",
    archivePrefix = "arXiv",
    primaryClass = "astro-ph.CO",
    reportNumber = "ULB-TH/23-09",
    doi = "10.1088/1475-7516/2024/01/005",
    journal = "JCAP",
    volume = "01",
    pages = "005",
    year = "2024"
}

@article{Sasaki:2018dmp,
    author = "Sasaki, Misao and Suyama, Teruaki and Tanaka, Takahiro and Yokoyama, Shuichiro",
    title = "{Primordial black holes{\textemdash}perspectives in gravitational wave astronomy}",
    eprint = "1801.05235",
    archivePrefix = "arXiv",
    primaryClass = "astro-ph.CO",
    doi = "10.1088/1361-6382/aaa7b4",
    journal = "Class. Quant. Grav.",
    volume = "35",
    number = "6",
    pages = "063001",
    year = "2018"
}

@article{DAmico:2018sxd,
    author = "D'Amico, Guido and Panci, Paolo and Strumia, Alessandro",
    title = "{Bounds on dark matter annihilations from 21 cm data}",
    eprint = "1803.03629",
    archivePrefix = "arXiv",
    primaryClass = "astro-ph.CO",
    reportNumber = "CERN-TH-2018-052, IFUP-TH-2018",
    doi = "10.1103/PhysRevLett.121.011103",
    journal = "Phys. Rev. Lett.",
    volume = "121",
    number = "1",
    pages = "011103",
    year = "2018"
}

@article{Greig:2015qca,
    author = "Greig, Bradley and Mesinger, Andrei",
    title = "{21CMMC: an MCMC analysis tool enabling astrophysical parameter studies of the cosmic 21 cm signal}",
    eprint = "1501.06576",
    archivePrefix = "arXiv",
    primaryClass = "astro-ph.CO",
    doi = "10.1093/mnras/stv571",
    journal = "Mon. Not. Roy. Astron. Soc.",
    volume = "449",
    number = "4",
    pages = "4246--4263",
    year = "2015"
}

@article{Witte:2018itc,
    author = "Witte, Samuel and Villanueva-Domingo, Pablo and Gariazzo, Stefano and Mena, Olga and Palomares-Ruiz, Sergio",
    title = "{EDGES result versus CMB and low-redshift constraints on ionization histories}",
    eprint = "1804.03888",
    archivePrefix = "arXiv",
    primaryClass = "astro-ph.CO",
    reportNumber = "IFIC/18-15, IFIC-18-15",
    doi = "10.1103/PhysRevD.97.103533",
    journal = "Phys. Rev. D",
    volume = "97",
    number = "10",
    pages = "103533",
    year = "2018"
}

@article{Clark:2018ghm,
    author = "Clark, Steven and Dutta, Bhaskar and Gao, Yu and Ma, Yin-Zhe and Strigari, Louis E.",
    title = "{21 cm limits on decaying dark matter and primordial black holes}",
    eprint = "1803.09390",
    archivePrefix = "arXiv",
    primaryClass = "astro-ph.HE",
    reportNumber = "MI-TH-1879",
    doi = "10.1103/PhysRevD.98.043006",
    journal = "Phys. Rev. D",
    volume = "98",
    number = "4",
    pages = "043006",
    year = "2018"
}

@article{Liu:2018uzy,
    author = "Liu, Hongwan and Slatyer, Tracy R.",
    title = "{Implications of a 21-cm signal for dark matter annihilation and decay}",
    eprint = "1803.09739",
    archivePrefix = "arXiv",
    primaryClass = "astro-ph.CO",
    reportNumber = "MIT-CTP/4996, MIT-CTP-4996",
    doi = "10.1103/PhysRevD.98.023501",
    journal = "Phys. Rev. D",
    volume = "98",
    number = "2",
    pages = "023501",
    year = "2018"
}

@article{Mitridate:2018iag,
    author = "Mitridate, Andrea and Podo, Alessandro",
    title = "{Bounds on dark matter decay from 21 cm line}",
    eprint = "1803.11169",
    archivePrefix = "arXiv",
    primaryClass = "hep-ph",
    doi = "10.1088/1475-7516/2018/05/069",
    journal = "JCAP",
    volume = "05",
    pages = "069",
    year = "2018"
}

@article{Yang:2020egn,
    author = "Yang, Yupeng",
    title = "{Constraints on primordial black holes and curvature perturbations from the global 21-cm signal}",
    eprint = "2009.11547",
    archivePrefix = "arXiv",
    primaryClass = "astro-ph.CO",
    doi = "10.1103/PhysRevD.102.083538",
    journal = "Phys. Rev. D",
    volume = "102",
    number = "8",
    pages = "083538",
    year = "2020"
}

@article{Halder:2021jiv,
    author = "Halder, Ashadul and Pandey, Madhurima",
    title = "{Probing the effects of primordial black holes on 21-cm EDGES signal along with interacting dark energy and dark matter{\textendash}baryon scattering}",
    eprint = "2101.05228",
    archivePrefix = "arXiv",
    primaryClass = "astro-ph.CO",
    doi = "10.1093/mnras/stab2795",
    journal = "Mon. Not. Roy. Astron. Soc.",
    volume = "508",
    number = "3",
    pages = "3446--3454",
    year = "2021"
}

@article{Halder:2021rbq,
    author = "Halder, Ashadul and Banerjee, Shibaji",
    title = "{Bounds on abundance of primordial black hole and dark matter from EDGES 21-cm signal}",
    eprint = "2102.00959",
    archivePrefix = "arXiv",
    primaryClass = "astro-ph.CO",
    doi = "10.1103/PhysRevD.103.063044",
    journal = "Phys. Rev. D",
    volume = "103",
    number = "6",
    pages = "063044",
    year = "2021"
}

@article{Mittal:2021egv,
    author = "Mittal, Shikhar and Ray, Anupam and Kulkarni, Girish and Dasgupta, Basudeb",
    title = "{Constraining primordial black holes as dark matter using the global 21-cm signal with X-ray heating and excess radio background}",
    eprint = "2107.02190",
    archivePrefix = "arXiv",
    primaryClass = "astro-ph.CO",
    reportNumber = "TIFR/TH/21-6",
    doi = "10.1088/1475-7516/2022/03/030",
    journal = "JCAP",
    volume = "03",
    pages = "030",
    year = "2022"
}

@article{Natwariya:2021xki,
    author = "Natwariya, Pravin Kumar and Nayak, Alekha C. and Srivastava, Tripurari",
    title = "{Constraining spinning primordial black holes with global 21-cm signal}",
    eprint = "2107.12358",
    archivePrefix = "arXiv",
    primaryClass = "astro-ph.CO",
    doi = "10.1093/mnras/stab3754",
    journal = "Mon. Not. Roy. Astron. Soc.",
    volume = "510",
    pages = "4236",
    year = "2021"
}

@article{Cang:2021owu,
    author = "Cang, Junsong and Gao, Yu and Ma, Yin-Zhe",
    title = "{21-cm constraints on spinning primordial black holes}",
    eprint = "2108.13256",
    archivePrefix = "arXiv",
    primaryClass = "astro-ph.CO",
    doi = "10.1088/1475-7516/2022/03/012",
    journal = "JCAP",
    volume = "03",
    number = "03",
    pages = "012",
    year = "2022"
}

@article{Saha:2021pqf,
    author = "Saha, Akash Kumar and Laha, Ranjan",
    title = "{Sensitivities on nonspinning and spinning primordial black hole dark matter with global 21-cm troughs}",
    eprint = "2112.10794",
    archivePrefix = "arXiv",
    primaryClass = "astro-ph.CO",
    doi = "10.1103/PhysRevD.105.103026",
    journal = "Phys. Rev. D",
    volume = "105",
    number = "10",
    pages = "103026",
    year = "2022"
}

@article{Mukhopadhyay:2022jqc,
    author = "Mukhopadhyay, Upala and Majumdar, Debasish and Halder, Ashadul",
    title = "{Constraining PBH mass distributions from 21cm brightness temperature results and an analytical mapping between probability distribution of 21cm signal and PBH masses}",
    eprint = "2203.13008",
    archivePrefix = "arXiv",
    primaryClass = "astro-ph.CO",
    doi = "10.1088/1475-7516/2022/10/099",
    journal = "JCAP",
    volume = "10",
    pages = "099",
    year = "2022"
}

@article{Yang:2022puh,
    author = "Yang, Yupeng",
    title = "{Impact of radiation from primordial black holes on the 21-cm angular-power spectrum in the dark ages}",
    eprint = "2209.00851",
    archivePrefix = "arXiv",
    primaryClass = "astro-ph.CO",
    doi = "10.1103/PhysRevD.106.123508",
    journal = "Phys. Rev. D",
    volume = "106",
    number = "12",
    pages = "123508",
    year = "2022"
}

@article{Novosyadlyj:2024bie,
    author = "Novosyadlyj, Bohdan and Kulinich, Yu. and Koval, Danylo",
    title = "{Global signal in the redshifted hydrogen 21-cm line from the dark ages and cosmic dawn: Dependence on the nature of dark matter and modeling of first light}",
    eprint = "2410.07380",
    archivePrefix = "arXiv",
    primaryClass = "astro-ph.CO",
    doi = "10.1103/PhysRevD.111.083514",
    journal = "Phys. Rev. D",
    volume = "111",
    number = "8",
    pages = "083514",
    year = "2025"
}

@article{Zhao:2024jad,
    author = "Zhao, Meng-Lin and Wang, Sai and Zhang, Xin",
    title = "{Prospects for probing dark matter particles and primordial black holes with the Hongmeng mission using the 21 cm global spectrum at cosmic dawn}",
    eprint = "2412.19257",
    archivePrefix = "arXiv",
    primaryClass = "astro-ph.CO",
    doi = "10.1088/1475-7516/2025/07/039",
    journal = "JCAP",
    volume = "07",
    pages = "039",
    year = "2025"
}

@article{Bae:2025uqa,
    author = "Bae, Hwan and Erickcek, Adrienne L. and Delos, M. Sten and Mu{\~n}oz, Julian B.",
    title = "{21-cm constraints on dark matter annihilation after an early matter-dominated era}",
    eprint = "2502.08719",
    archivePrefix = "arXiv",
    primaryClass = "astro-ph.CO",
    doi = "10.1103/6rkn-tlrh",
    journal = "Phys. Rev. D",
    volume = "112",
    number = "8",
    pages = "083013",
    year = "2025"
}

@article{Sun:2023acy,
    author = "Sun, Yitian and Foster, Joshua W. and Liu, Hongwan and Mu{\~n}oz, Julian B. and Slatyer, Tracy R.",
    title = "{Inhomogeneous energy injection in the 21-cm power spectrum: Sensitivity to dark matter decay}",
    eprint = "2312.11608",
    archivePrefix = "arXiv",
    primaryClass = "hep-ph",
    reportNumber = "MIT-CTP/5657, FERMILAB-PUB-23-0816-T-V",
    doi = "10.1103/PhysRevD.111.043015",
    journal = "Phys. Rev. D",
    volume = "111",
    number = "4",
    pages = "043015",
    year = "2025"
}

@article{Zhao:2025ddy,
    author = "Zhao, Meng-Lin and Shao, Yue and Wang, Sai and Zhang, Xin",
    title = "{Prospects for probing dark matter particles and primordial black holes with the Square Kilometre Array using the 21 cm power spectrum at cosmic dawn}",
    eprint = "2507.02651",
    archivePrefix = "arXiv",
    primaryClass = "astro-ph.CO",
    doi = "10.1088/1674-1137/ae1375",
    journal = "Chin. Phys. C",
    volume = "50",
    pages = "025101",
    year = "2026"
}

@article{Natwariya:2025jlw,
    author = "Natwariya, Pravin Kumar and Kadota, Kenji and Nishizawa, Atsushi J.",
    title = "{Sensitivity toward dark matter annihilation imprints on the 21-cm signal with SKA-Low: A convolutional neural network approach}",
    eprint = "2508.08251",
    archivePrefix = "arXiv",
    primaryClass = "astro-ph.CO",
    doi = "10.1103/23hy-p8zq",
    journal = "Phys. Rev. D",
    volume = "113",
    number = "2",
    pages = "023038",
    year = "2026"
}

@article{Sun:2025ksr,
    author = "Sun, Yitian and Foster, Joshua W. and Mu{\~n}oz, Julian B.",
    title = "{Constraining inhomogeneous energy injection from annihilating dark matter and primordial black holes with 21-cm cosmology}",
    eprint = "2509.22772",
    archivePrefix = "arXiv",
    primaryClass = "hep-ph",
    month = "9",
    year = "2025"
}

@article{Leitherer:2014,
    author = {{Leitherer}, Claus and {Ekstr{\"o}m}, Sylvia and {Meynet}, Georges and {Schaerer}, Daniel and {Agienko}, Katerina B. and {Levesque}, Emily M.},
    title = "{The effects of stellar rotation. II. a comprehensive set of Starburst99 models}",
    journal = "Astrophys. J. Suppl.",
    year = 2014,
    volume = {212},
    number = {1},
    eid = {14},
    pages = {14},
    doi = {10.1088/0067-0049/212/1/14},
    archivePrefix = {arXiv},
    eprint = {1403.5444},
    primaryClass = {astro-ph.GA}
}

@article{Leitherer:2010,
    author = {{Leitherer}, Claus and others},
    title = "{A library of theoretical ultraviolet spectra of massive, hot stars for evolutionary synthesis}",
    journal = "Astrophys. J. Suppl.",
    year = 2010,
    volume = {189},
    number = {2},
    pages = {309-335},
    doi = {10.1088/0067-0049/189/2/309},
    archivePrefix = {arXiv},
    eprint = {1006.5624},
     primaryClass = {astro-ph.SR}
}

@article{Munoz:2021psm,
    author = "Mu{\~n}oz, Julian B. and Qin, Yuxiang and Mesinger, Andrei and Murray, Steven G. and Greig, Bradley and Mason, Charlotte",
    title = "{The impact of the first galaxies on cosmic dawn and reionization}",
    eprint = "2110.13919",
    archivePrefix = "arXiv",
    primaryClass = "astro-ph.CO",
    doi = "10.1093/mnras/stac185",
    journal = "Mon. Not. Roy. Astron. Soc.",
    volume = "511",
    number = "3",
    pages = "3657--3681",
    year = "2022"
}

@article{Fialkov:2016zyq,
    author = "Fialkov, Anastasia and Cohen, Aviad and Barkana, Rennan and Silk, Joseph",
    title = "{Constraining the redshifted 21-cm signal with the unresolved soft X-ray background}",
    eprint = "1602.07322",
    archivePrefix = "arXiv",
    primaryClass = "astro-ph.CO",
    doi = "10.1093/mnras/stw2540",
    journal = "Mon. Not. Roy. Astron. Soc.",
    volume = "464",
    number = "3",
    pages = "3498--3508",
    year = "2017"
}

@article{Afshordi:2003zb,
    author = "Afshordi, N. and McDonald, P. and Spergel, D. N.",
    title = "{Primordial black holes as dark matter: the power spectrum and evaporation of early structures}",
    eprint = "astro-ph/0302035",
    archivePrefix = "arXiv",
    doi = "10.1086/378763",
    journal = "Astrophys. J. Lett.",
    volume = "594",
    pages = "L71--L74",
    year = "2003"
}

@article{Cohen:2016jbh,
    author = "Cohen, Aviad and Fialkov, Anastasia and Barkana, Rennan and Lotem, Matan",
    title = "{Charting the parameter space of the global 21-cm signal}",
    eprint = "1609.02312",
    archivePrefix = "arXiv",
    primaryClass = "astro-ph.CO",
    doi = "10.1093/mnras/stx2065",
    journal = "Mon. Not. Roy. Astron. Soc.",
    volume = "472",
    number = "2",
    pages = "1915--1931",
    year = "2017"
}

@article{Qin:2021gkn,
    author = "Qin, Yuxiang and Mesinger, Andrei and Bosman, Sarah E. I. and Viel, Matteo",
    title = "{Reionization and galaxy inference from the high-redshift Ly{\,}{\ensuremath{\alpha}} forest}",
    eprint = "2101.09033",
    archivePrefix = "arXiv",
    primaryClass = "astro-ph.CO",
    doi = "10.1093/mnras/stab1833",
    journal = "Mon. Not. Roy. Astron. Soc.",
    volume = "506",
    number = "2",
    pages = "2390--2407",
    year = "2021"
}

@article{AliHamoud:2011,
    author = "Ali-Haimoud, Yacine and Hirata, Christopher M.",
    title = "{HyRec: A fast and highly accurate primordial hydrogen and helium recombination code}",
    eprint = "1011.3758",
    archivePrefix = "arXiv",
    primaryClass = "astro-ph.CO",
    doi = "10.1103/PhysRevD.83.043513",
    journal = "Phys. Rev. D",
    volume = "83",
    pages = "043513",
    year = "2011"
}

@article{Park:2018ljd,
    author = "Park, Jaehong and Mesinger, Andrei and Greig, Bradley and Gillet, Nicolas",
    title = "{Inferring the astrophysics of reionization and cosmic dawn from galaxy luminosity functions and the 21-cm signal}",
    eprint = "1809.08995",
    archivePrefix = "arXiv",
    primaryClass = "astro-ph.GA",
    doi = "10.1093/mnras/stz032",
    journal = "Mon. Not. Roy. Astron. Soc.",
    volume = "484",
    number = "1",
    pages = "933--949",
    year = "2019"
}

@article{Qin:2020xyh,
    author = "Qin, Yuxiang and Mesinger, Andrei and Park, Jaehong and Greig, Bradley and Mu{\~n}oz, Julian B.",
    title = "{A tale of two sites {\textendash} I. Inferring the properties of minihalo-hosted galaxies from current observations}",
    eprint = "2003.04442",
    archivePrefix = "arXiv",
    primaryClass = "astro-ph.CO",
    doi = "10.1093/mnras/staa1131",
    journal = "Mon. Not. Roy. Astron. Soc.",
    volume = "495",
    number = "1",
    pages = "123--140",
    year = "2020"
}

@article{Katz:2024ayw,
    author = "Katz, Omer Zvi and Outmezguine, Nadav and Redigolo, Diego and Volansky, Tomer",
    title = "{Probing new physics at cosmic dawn with 21-cm cosmology}",
    eprint = "2401.10978",
    archivePrefix = "arXiv",
    primaryClass = "hep-ph",
    doi = "10.1016/j.nuclphysb.2024.116502",
    journal = "Nucl. Phys. B",
    volume = "1003",
    pages = "116502",
    year = "2024"
}

@article{Katz:2025sie,
    author = "Katz, Omer Zvi and Redigolo, Diego and Volansky, Tomer",
    title = "{Closing in on Pop-III stars: constraints and predictions across the spectrum}",
    eprint = "2502.03525",
    archivePrefix = "arXiv",
    primaryClass = "astro-ph.CO",
    doi = "10.1088/1475-7516/2025/10/047",
    journal = "JCAP",
    volume = "10",
    pages = "047",
    year = "2025"
}

@Article{Slatyer2024,
  author    = {Slatyer, Tracy R.},
  journal   = {Nucl. Phys. B},
  title     = {What does cosmology teach us about non-gravitational properties of dark matter?},
  year      = {2024},
  issn      = {0550-3213},
  pages     = {116468},
  volume    = {1003},
  doi       = {10.1016/j.nuclphysb.2024.116468},
  publisher = {Elsevier BV},
}

@Article{Cirelli2024,
title={{Dark matter}},
	author={Marco Cirelli and Alessandro Strumia and Jure Zupan},
	journal={SciPost Phys. Rev.},
	pages={1},
	year={2026},
	publisher={SciPost},
	doi={10.21468/SciPostPhysRev.1},
	url={https://scipost.org/10.21468/SciPostPhysRev.1},
    eprint = "2406.01705",
    archivePrefix = "arXiv",
    primaryClass = "hep-ph",
}

@article{Gardner:2023,
   author={Gardner, Jonathan P. and others},
   title={The James Webb Space Telescope Mission},
   volume={135},
   ISSN={1538-3873},
   url={http://dx.doi.org/10.1088/1538-3873/acd1b5},
   DOI={10.1088/1538-3873/acd1b5},
   number={1048},
   journal={Publ. Astron. Soc. Pac.},
   publisher={IOP Publishing},
   year={2023},
  pages={068001} 
}

@article{Finkelstein2023,
    author = {{Finkelstein}, Steven L. and others},
    title = "{CEERS Key Paper. I. An Early Look into the First 500 Myr of Galaxy Formation with JWST}",
    journal = {Astrophys. J. Lett.},
    year = 2023,
    volume = {946},
    number = {1},
    eid = {L13},
    pages = {L13},
    doi = {10.3847/2041-8213/acade4},
archivePrefix = {arXiv},
    eprint = {2211.05792},
 primaryClass = {astro-ph.GA}
}

@article{Fender:2013ei,
    author = "Fender, Rob and Maccarone, Tom and Heywood, Ian",
    title = "{The closest black holes}",
    eprint = "1301.1341",
    archivePrefix = "arXiv",
    primaryClass = "astro-ph.HE",
    doi = "10.1093/mnras/sts688",
    journal = "Mon. Not. Roy. Astron. Soc.",
    volume = "430",
    pages = "1538",
    year = "2013"
}

@article{Perna:2003ck,
    author = "Perna, Rosalba and Narayan, Ramesh and Rybicki, George and Stella, Luigi and Treves, Aldo",
    title = "{Bondi accretion and the problem of the missing isolated neutron stars}",
    eprint = "astro-ph/0305421",
    archivePrefix = "arXiv",
    doi = "10.1086/377091",
    journal = "Astrophys. J.",
    volume = "594",
    pages = "936--942",
    year = "2003"
}

@article{Pellegrini:2005pi,
    author = "Pellegrini, Silvia",
    title = "{Nuclear accretion in galaxies of the local Universe: Clues from Chandra observations}",
    eprint = "astro-ph/0502035",
    archivePrefix = "arXiv",
    doi = "10.1086/429267",
    journal = "Astrophys. J.",
    volume = "624",
    pages = "155--161",
    year = "2005",
    note = "[Erratum: Astrophys.J. 636, 564 (2005)]"
}

@Article{Wang2013,
  author        = {Wang, Q. D. and others},
  journal       = {Science},
  title         = {{Dissecting X-ray-emitting gas around the center of our galaxy}},
  year          = {2013},
  pages         = {981},
  volume        = {341},
  archiveprefix = {arXiv},
  doi           = {10.1126/science.1240755},
  eprint        = {1307.5845},
  primaryclass  = {astro-ph.HE},
}

@article{Ricotti:2007jk,
    author = "Ricotti, Massimo",
    title = "{Bondi accretion in the early universe}",
    eprint = "0706.0864",
    archivePrefix = "arXiv",
    primaryClass = "astro-ph",
    doi = "10.1086/516562",
    journal = "Astrophys. J.",
    volume = "662",
    pages = "53--61",
    year = "2007"
}

@article{Ricotti:2007au,
    author = "Ricotti, Massimo and Ostriker, Jeremiah P. and Mack, Katherine J.",
    title = "{Effect of primordial black holes on the cosmic microwave background and cosmological parameter estimates}",
    eprint = "0709.0524",
    archivePrefix = "arXiv",
    primaryClass = "astro-ph",
    doi = "10.1086/587831",
    journal = "Astrophys. J.",
    volume = "680",
    pages = "829",
    year = "2008"
}

@article{Horowitz:2016lib,
    author = "Horowitz, Benjamin",
    title = "{Revisiting primordial black holes constraints from ionization history}",
    eprint = "1612.07264",
    archivePrefix = "arXiv",
    primaryClass = "astro-ph.CO",
    month = "12",
    year = "2016"
}

@article{LuisBernal:2017fmf,
    author = "Luis Bernal, Jos{\'e} and Bellomo, Nicola and Raccanelli, Alvise and Verde, Licia",
    title = "{Cosmological implications of primordial black holes}",
    eprint = "1709.07465",
    archivePrefix = "arXiv",
    primaryClass = "astro-ph.CO",
    doi = "10.1088/1475-7516/2017/10/052",
    journal = "JCAP",
    volume = "10",
    pages = "052",
    year = "2017"
}

@article{Klessen:2023qmc,
    author = "Klessen, Ralf S. and Glover, Simon C. O.",
    title = "{The first stars: formation, properties, and impact}",
    eprint = "2303.12500",
    archivePrefix = "arXiv",
    primaryClass = "astro-ph.CO",
    doi = "10.1146/annurev-astro-071221-053453",
    journal = "Ann. Rev. Astron. Astrophys.",
    volume = "61",
    pages = "65--130",
    year = "2023"
}

@article{Tashiro:2012qe,
    author = "Tashiro, Hiroyuki and Sugiyama, Naoshi",
    title = "{The effect of primordial black holes on 21 cm fluctuations}",
    eprint = "1207.6405",
    archivePrefix = "arXiv",
    primaryClass = "astro-ph.CO",
    doi = "10.1093/mnras/stt1493",
    journal = "Mon. Not. Roy. Astron. Soc.",
    volume = "435",
    pages = "3001",
    year = "2013"
}

@article{Hektor:2018qqw,
    author = {Hektor, Andi and H{\"u}tsi, Gert and Marzola, Luca and Raidal, Martti and Vaskonen, Ville and Veerm{\"a}e, Hardi},
    title = "{Constraining primordial black holes with the EDGES 21-cm absorption signal}",
    eprint = "1803.09697",
    archivePrefix = "arXiv",
    primaryClass = "astro-ph.CO",
    reportNumber = "CERN-TH-2018-073",
    doi = "10.1103/PhysRevD.98.023503",
    journal = "Phys. Rev. D",
    volume = "98",
    number = "2",
    pages = "023503",
    year = "2018"
}

@article{Jester:2009dw,
    author = "Jester, Sebastian and Falcke, Heino",
    title = "{Science with a lunar low-frequency array: from the dark ages of the Universe to nearby exoplanets}",
    eprint = "0902.0493",
    archivePrefix = "arXiv",
    primaryClass = "astro-ph.CO",
    doi = "10.1016/j.newar.2009.02.001",
    journal = "New Astron. Rev.",
    volume = "53",
    pages = "1--26",
    year = "2009"
}

@article{Wouthuysen:1952,
	author = {{Wouthuysen}, S.~A.},
	title = "{On the excitation mechanism of the 21-cm (radio-frequency) interstellar hydrogen emission line.}",
	journal = {Astrophys. J.},
	year = 1952,
	volume = 57,
	pages = {31-32},
	doi = {10.1086/106661},
	adsurl = {http://adsabs.harvard.edu/abs/1952AJ.....57R..31W}
}

@article{Field:1958,
	author = {{Field}, G.~B.},
	title = "{Excitation of the hydrogen 21-cm line}",
	journal = {Proc. Inst. Radio Eng.},
	year = 1958,
	volume = 46,
	pages = {240-250},
	doi = {10.1109/JRPROC.1958.286741},
	adsurl = {http://adsabs.harvard.edu/abs/1958PIRE...46..240F}
}

@article{Piga:2022ysp,
    author = "Piga, Lorenzo and Lucca, Matteo and Bellomo, Nicola and Bosch-Ramon, Valent{\`\i} and Matarrese, Sabino and Raccanelli, Alvise and Verde, Licia",
    title = "{The effect of outflows on CMB bounds from Primordial Black Hole accretion}",
    eprint = "2210.14934",
    archivePrefix = "arXiv",
    primaryClass = "astro-ph.CO",
    reportNumber = "ULB-TH/22-14, UTWI-11-2022",
    doi = "10.1088/1475-7516/2022/12/016",
    journal = "JCAP",
    volume = "12",
    pages = "016",
    year = "2022"
}

@article{Farside:2019,
	author = "Burns, Jack and others",
	title = "{FARSIDE}: A low radio frequency interferometric array on the lunar farside",
    eprint = "1907.05407",
archivePrefix = "arXiv",
primaryClass = "astro-ph.IM",
	journal = "Bull. Am. Astron. Soc.",
	volume = "51",
    pages = "(7)",
    year = "2019",
	note = "https://baas.aas.org/pub/2020n7i178"
}

@article{Andres-Carcasona:2024wqk,
    author = {Andr{\'e}s-Carcasona, M. and Iovino, A. J. and Vaskonen, V. and Veerm{\"a}e, H. and Mart{\'\i}nez, M. and Pujol{\`a}s, O. and Mir, Ll. M.},
    title = "{Constraints on primordial black holes from LIGO-Virgo-KAGRA O3 events}",
    eprint = "2405.05732",
    archivePrefix = "arXiv",
    primaryClass = "astro-ph.CO",
    doi = "10.1103/PhysRevD.110.023040",
    journal = "Phys. Rev. D",
    volume = "110",
    number = "2",
    pages = "023040",
    year = "2024"
}

@article{Burns_2020,
   author={Burns, Jack O.},
   title={Transformative science from the lunar farside: observations of the dark ages and exoplanetary systems at low radio frequencies},
   volume={379},
   ISSN={1471-2962},
   url={http://dx.doi.org/10.1098/rsta.2019.0564},
   DOI={10.1098/rsta.2019.0564},
   number={2188},
   journal={Phil. Trans. A. Math. Phys. Eng. Sci.},
   publisher={The Royal Society},
   pages ={20190564},
   year={2020},
   archivePrefix = {arXiv},
       eprint = {2003.06881},
 primaryClass = {astro-ph.IM},
}

@article{Bowman:2018yin,
    author = "Bowman, Judd D. and Rogers, Alan E. E. and Monsalve, Raul A. and Mozdzen, Thomas J. and Mahesh, Nivedita",
    title = "{An absorption profile centred at 78 megahertz in the sky-averaged spectrum}",
    eprint = "1810.05912",
    archivePrefix = "arXiv",
    primaryClass = "astro-ph.CO",
    doi = "10.1038/nature25792",
    journal = "Nature",
    volume = "555",
    number = "7694",
    pages = "67--70",
    year = "2018"
}

@article{deLeraAcedo:2022kiu,
    author = "de Lera Acedo, E. and others",
    title = "{The REACH radiometer for detecting the 21-cm hydrogen signal from redshift z{\,}{\ensuremath{\approx}}{\,}7.5{\textendash}28}",
    eprint = "2210.07409",
    archivePrefix = "arXiv",
    primaryClass = "astro-ph.CO",
    doi = "10.1038/s41550-022-01817-6",
    journal = "Nature Astron.",
    volume = "6",
    number = "7",
    pages = "998",
    year = "2022"
}

@article{Singh:2021mxo,
    author = "Singh, Saurabh and Nambissan T., Jishnu and Subrahmanyan, Ravi and Udaya Shankar, N. and Girish, B. S. and Raghunathan, A. and Somashekar, R. and Srivani, K. S. and Sathyanarayana Rao, Mayuri",
    title = "{On the detection of a cosmic dawn signal in the radio background}",
    eprint = "2112.06778",
    archivePrefix = "arXiv",
    primaryClass = "astro-ph.CO",
    doi = "10.1038/s41550-022-01610-5",
    journal = "Nature Astron.",
    volume = "6",
    number = "5",
    pages = "607--617",
    year = "2022"
}

@article{Polidan:2024kkh,
    author = "Polidan, Ronald S. and Burns, Jack O. and Ignatiev, Alex and Hegedus, Alex and Pober, Jonathan and Mahesh, Nivedita and Chang, Tzu-Ching and Hallinan, Gregg and Ning, Yuhong and Bowman, Judd",
    title = "{FarView: An in-situ manufactured lunar far side radio array concept for 21-cm Dark Ages cosmology}",
    eprint = "2404.03840",
    archivePrefix = "arXiv",
    primaryClass = "astro-ph.IM",
    doi = "10.1016/j.asr.2024.04.008",
    journal = "Adv. Space Res.",
    volume = "74",
    pages = "528--546",
    year = "2024"
}

@article{Silk:2025znp,
    author = "Silk, Joseph",
    title = "{The limits of cosmology}",
    eprint = "2509.08066",
    archivePrefix = "arXiv",
    primaryClass = "astro-ph.CO",
    journal = "Gen. Rel. Grav.",
    volume = "57",
    pages = "127",
    year = "2025"
}

@article{Harris:2020xlr,
    author = "Harris, Charles R. and others",
    title = "{Array programming with NumPy}",
    eprint = "2006.10256",
    archivePrefix = "arXiv",
    primaryClass = "cs.MS",
    doi = "10.1038/s41586-020-2649-2",
    journal = "Nature",
    volume = "585",
    number = "7825",
    pages = "357--362",
    year = "2020"
}

@article{LuseeNight:2023,
       author = {{Bale}, Stuart D. and others},
        title = "{LuSEE `Night': The Lunar Surface Electromagnetics Experiment}",
         year = 2023,
archivePrefix = {arXiv},
       eprint = {2301.10345},
 primaryClass = {astro-ph.IM},
       adsurl = {https://ui.adsabs.harvard.edu/abs/2023arXiv230110345B}
}

@article{Lewis:2019xzd,
   author = "Lewis, Antony",
   title = "{GetDist: a Python package for analysing Monte Carlo samples}",
   eprint = "1910.13970",
   archivePrefix = "arXiv",
   primaryClass = "astro-ph.IM",
   doi = "10.1088/1475-7516/2025/08/025",
   journal = "JCAP",
   volume = "08",
   pages = "025",
   year = "2025"
}

@article{Virtanen:2019joe,
    author = "Virtanen, Pauli and others",
    title = "{SciPy 1.0--Fundamental algorithms for scientific computing in Python}",
    eprint = "1907.10121",
    archivePrefix = "arXiv",
    primaryClass = "cs.MS",
    doi = "10.1038/s41592-019-0686-2",
    journal = "Nature Meth.",
    volume = "17",
    pages = "261",
    year = "2020"
}

@inbook{DeLuca:2023bcr,
    author = "De Luca, Valerio and Bellomo, Nicola",
    title = "{The accretion, emission, mass and~spin evolution of primordial black holes}",
    eprint = "2312.14097",
    publisher = "Springer",
    year = "2025",
    archivePrefix = "arXiv",
    primaryClass = "astro-ph.CO",
    editor = "Byrnes, Christian and Franciolini, Gabriele and Harada, Tomohiro and Pani, Paolo and Sasaki, Misao",
    booktitle = "{Primordial Black Holes}",
    doi = "10.1007/978-981-97-8887-3",
    isbn = "978-981--978886-6, 978-981--978889-7, 978-981--978887-3",
    series = "Springer Series in Astrophysics and Cosmology",
}

@article{Ali-Haimoud:2016mbv,
    author = {Ali-Ha{\"\i}moud, Yacine and Kamionkowski, Marc},
    title = "{Cosmic microwave background limits on accreting primordial black holes}",
    eprint = "1612.05644",
    archivePrefix = "arXiv",
    primaryClass = "astro-ph.CO",
    doi = "10.1103/PhysRevD.95.043534",
    journal = "Phys. Rev. D",
    volume = "95",
    number = "4",
    pages = "043534",
    year = "2017"
}

@article{Shapiro1976,
  author   = {{Shapiro}, S.~L. and {Lightman}, A.~P.},
  journal  = "Astrophys. J.",
  title    = {{Black holes in X-ray binaries: marginal existence and rotation reversals of accretion disks}},
  year     = {1976},
  pages    = {555--560},
  volume   = {204},
  doi      = {10.1086/154203}
}

@article{Agius:2025nfz,
    author = "Agius, Dominic and Slatyer, Tracy Robyn",
    title = "{Boosting the cosmic 21-cm signal with exotic Lyman-$\alpha$ from dark matter}",
    eprint = "2510.26791",
    archivePrefix = "arXiv",
    primaryClass = "astro-ph.CO",
    reportNumber = "MIT-CTP/5952",
    month = "10",
    year = "2025"
}

@article{Burns_2017,
   title={A Space-based Observational Strategy for Characterizing the First Stars and Galaxies Using the Redshifted 21 cm Global Spectrum},
   volume={844},
   ISSN={1538-4357},
   url={http://dx.doi.org/10.3847/1538-4357/aa77f4},
   DOI={10.3847/1538-4357/aa77f4},
   number={1},
   journal={Astrophys. J.},
   publisher={American Astronomical Society},
   author={Burns, Jack O. and Bradley, Richard and Tauscher, Keith and Furlanetto, Steven and Mirocha, Jordan and Monsalve, Raul and Rapetti, David and Purcell, William and Newell, David and Draper, David and MacDowall, Robert and Bowman, Judd and Nhan, Bang and Wollack, Edward J. and Fialkov, Anastasia and Jones, Dayton and Kasper, Justin C. and Loeb, Abraham and Datta, Abhirup and Pritchard, Jonathan and Switzer, Eric and Bicay, Michael},
   year={2017},
pages={33},
archivePrefix = {arXiv},
       eprint = {1704.02651},
 primaryClass = {astro-ph.IM}
 }

@article{Tauscher:2018uxi,
    author = "Tauscher, Keith and Rapetti, David and Burns, Jack O.",
    title = "{A new goodness-of-fit statistic and its application to 21-cm cosmology}",
    eprint = "1810.00076",
    archivePrefix = "arXiv",
    primaryClass = "astro-ph.CO",
    doi = "10.1088/1475-7516/2018/12/015",
    journal = "JCAP",
    volume = "12",
    pages = "015",
    year = "2018"
}

@article{Chen:2024tvn,
    author = "Chen, Xuelei and others",
    title = "{Large-scale Array for Radio Astronomy on the Farside}",
    eprint = "2403.16409",
    archivePrefix = "arXiv",
    primaryClass = "astro-ph.IM",
    doi = "10.1098/rsta.2023.0094",
    journal = "Phil. Trans. Roy. Soc. Lond. A",
    volume = "382",
    pages = "20230094",
    year = "2024"
}

@inproceedings{Chen:2019xvd,
    author = "Chen, Xuelei and others",
    title = "{Discovering the Sky at the Longest Wavelengths with Small Satellite Constellations}",
    booktitle = "{ISSI-BJ Forum}: {Discover the Sky by Longest Wavelength with Small Satellite Constellation}",
    eprint = "1907.10853",
    archivePrefix = "arXiv",
    primaryClass = "astro-ph.IM",
    reportNumber = "ISSI-BJ Taikong-14",
    month = "7",
    year = "2019"
}

@article{Chen:2020lok,
    author = "Chen, Xuelei and Yan, Jingye and Deng, Li and Wu, Fengquan and Wu, Lin and Xu, Yidong and Zhou, Li",
    title = "{Discovering the Sky at the Longest wavelengths with a lunar orbit array}",
    eprint = "2007.15794",
    archivePrefix = "arXiv",
    primaryClass = "astro-ph.IM",
    doi = "10.1098/rsta.2019.0566",
    journal = "Phil. Trans. Roy. Soc. Lond. A",
    volume = "379",
    pages = "20190566",
    year = "2020"
}

@article{Shi:2022zdx,
    author = "Shi, Yuan and Deng, Furen and Xu, Yidong and Wu, Fengquan and Yan, Qisen and Chen, Xuelei",
    title = "{Lunar Orbit Measurement of the Cosmic Dawns 21 cm Global Spectrum}",
    eprint = "2203.01124",
    archivePrefix = "arXiv",
    primaryClass = "astro-ph.IM",
    doi = "10.3847/1538-4357/ac5965",
    journal = "Astrophys. J.",
    volume = "929",
    number = "1",
    pages = "32",
    year = "2022"
}

@article{2023ChJSS..43...43C,
       author = {{Chen}, Xuelei and {Yan}, Jingye and {Xu}, Yidong and {Deng}, Li and {Wu}, Fengquan and {Wu}, Lin and {Zhou}, Li and {Zhang}, Xiaofeng and {Zhu}, Xiaocheng and {Yang}, Zhongguang and {Wu}, Ji},
        title = "{Discovering the Sky at the Longest Wavelength Mission - A Pathfinder for Exploring the Cosmic Dark Ages}",
      journal = {Chin. J. Space Sci.},
         year = 2023,
        
       volume = {43},
       number = {1},
        pages = {43-59},
          doi = {10.11728/cjss2023.01.220104001},
       adsurl = {https://ui.adsabs.harvard.edu/abs/2023ChJSS..43...43C}
}

@article{Lopez-Honorez:2020lno,
    author = "Lopez-Honorez, Laura and Mena, Olga and Palomares-Ruiz, Sergio and Villanueva-Domingo, Pablo and Witte, Samuel J.",
    title = "{Variations in fundamental constants at the cosmic dawn}",
    eprint = "2004.00013",
    archivePrefix = "arXiv",
    primaryClass = "astro-ph.CO",
    reportNumber = "ULB-TH/20-03",
    doi = "10.1088/1475-7516/2020/06/026",
    journal = "JCAP",
    volume = "06",
    pages = "026",
    year = "2020"
}

\end{document}